\documentclass[epj-spec]{svjour}
\usepackage{graphics}
\usepackage{amsmath}
\usepackage{color}
\newcommand{\TGaseq}[1]{(\ref{TGas:eq:#1})}
\newcommand{\TGasEq}[1]{Eq.~(\ref{TGas:eq:#1})}
\newcommand{\TGaseqs}[2]{(\ref{TGas:eq:#1})--(\ref{TGas:eq:#2})}

\newcommand{\TGasFig}[1]{Fig.~\ref{TGas:fig:#1}}
\newcommand{\TGasfig}[1]{\ref{TGas:fig:#1}}
\newcommand{\TGasSection}[1]{Section~\ref{TGas:sec:#1}}
\newcommand{\TGasSect}[1]{Sect.~\ref{TGas:sec:#1}}
\newcommand{\TGassect}[1]{\ref{TGas:sec:#1}}

\newcommand{\TGaspdiff}[1]{\partial_{#1}}
\newcommand{\TGaspdiffer}[2]{\frac{\partial #1}{\partial #2}}
\newcommand{\TGaspDiff}[1]{\mathcal D_{#1}}
\newcommand{\TGasF}[1]{\vec{#1}}
\newcommand{\TGasunit}[1]{\mathrm{#1}}
\begin{document}
\title{Ultracold gases far from equilibrium}
\author{Thomas Gasenzer\thanks{\email{t.gasenzer@thphys.uni-heidelberg.de}}}
\institute{Institut f\"ur Theoretische Physik, Philosophenweg 16, 69120 Heidelberg, Germany}
\abstract{
Ultracold atomic quantum gases belong to the most exciting challenges of modern physics. 
Their theoretical description has drawn much from (semi-\nolinebreak) classical field equations.
These mean-field approximations are in general reliable for dilute gases in which the atoms collide only rarely with each other, and for situations where the gas is not too far from thermal equilibrium. 
With present-day technology it is, however, possible to drive and observe a system far away from equilibrium.
Functional quantum field theory provides powerful tools to achieve both, analytical understanding and numerical computability, also in higher dimensions, of  far-from-equilibrium quantum many-body dynamics.
In the article, an outline of these approaches is given, including methods based on the two-particle irreducible effective action as well as on renormalisation-group theory.
Their relation to near-equilibrium kinetic theory is discussed, and the distinction between quantum and classical statistical fluctuations is shown to naturally emerge from the functional-integral description.
Example applications to the evolution of an ultracold atomic Bose gas in one spatial dimension underline the power of the methods.
} 
\maketitle
%
\section{Introduction}
\label{TGas:sec:Intro}
Five millimeters of mere nothing separate the micrometer-scale ultracold cloud of sodium atoms from the glass walls of its surrounding vacuum cell in a typical experiment in the basement of maybe the reader's research institution.
Five millimeters between the glass at 293 K and the Bose-Einstein condensed gas at a few nanokelvin.
These eleven orders of magnitude in temperature are to be compared to the eight orders the temperature in a supernova is higher than that in our office.
And the vacuum pressure of $10^{-10}\,$Pa which is quantifying "mere nothing" is similar to the atmospheric pressure at the surface of the moon.

Bose-Einstein condensation, a phenomenon predicted over 70 years ago by Albert Einstein 
\cite{TGas:Einstein1924a} 
on the basis of Bose's new statistical formulation of a photon gas \cite{TGas:Bose1924a}, has revolutionised atomic physics since its ground breaking experimental achievement in dilute alkali gases at JILA (Boulder) \cite{TGas:Anderson1995a}, MIT \cite{TGas:Davis1995b}, and Rice University \cite{TGas:Bradley1997a}.
By today, more than 80 groups worldwide dispose of techniques to produce ultracold and Bose-Einstein condensed gases, mostly of alkalis like $^{87}$Rb, $^{23}$Na, $^{7}$Li but also including metastable $^4$He, as well as $^{41}$K, $^{52}$Cr, $^{85}$Rb, $^{133}$Cs, $^{174}$Yb.
Another 100 groups need to be included when counting the wider range of experiments studying cold atomic gases, atom optics, trapping, cooling, and many more subjects \cite{TGas:InnsbruckAtomTrapsWWW}.
The past decade has seen an exploding range of experiments studying many different properties of such systems, thereby varying densities, atom numbers, dimensionality, interaction strengths, internal (electronic) state multiplicities, as well as character, geometry, shape, size and temporal behaviour of the external trapping potentials, both, between different runs and during the experiment's timeline.
A range of review articles and monographs on theory and experiment can be consulted \cite{TGas:Anglin2002b,TGas:Pethick2001a,TGas:Dalfovo1999a,TGas:Leggett2001a,TGas:Pitaevskii2003a}.
During the past five years, a growing number of experiments with ultracold Fermi gases has been added to the spectrum of activities, see, e.g., Refs.~\cite{TGas:DeMarco1999a,TGas:Truscott2001a,TGas:Regal2004b,TGas:Zwierlein2004a,TGas:Bartenstein2004a}.
For recent reviews see Ref.~\cite{TGas:Inguscio2008a}.
Today, ultracold fermions are regarded as a promising tool to design many-body quantum systems exhibiting many of the phenomena relevant in solid state systems, in particular in the context of superconductivity, and to explore these beyond the range of parameters realistic for degenerate electron gases.
Moreover, new efforts aim at a cross-fertilisation between high-energy physics, in particular concerning the phase structure of quantumchromodynamics, and the physics of strongly correlated atomic gases.  

The quantum degeneracy of ultracold gases results from the (anti-)symmetrisation principle for the wave function describing indistinguishable particles, together with the statistical behaviour of the many-body system.
A non-interacting gas is characterised, for given mean energy and particle number, by the occupation numbers of the single-particle eigenmodes.
Bose-Einstein condensation emerges as the macroscopic occupation of a single mode, at zero as well as finite temperatures \cite{TGas:Huang1987a}.
Although interactions are not required for the existence of this quantum degeneracy, they are, in a realistic physical system, indispensable for reaching the respective degenerate equilibrated state.
In fact, in experiment, the key last step to degeneracy is induced by evaporative cooling \cite{TGas:Anglin2002b}, where collisional relaxation following a removal of the high-frequency tail of atoms restores an equilibrium distribution.

Trapped atomic gases provide the unique possibility to tune both, the interaction strength between the particles and the external boundary conditions fixed by the trapping potential:
External electromagnetic fields can be used to considerably vary, in particular near (Feshbach) zero-energy scattering resonances, the scattering length which quantifies the collisional interactions \cite{TGas:Burnett2002b}.
Thereby, laser light, combined with elaborate lensing technology, allows for almost arbitrary trapping geometries.
Most of these tuning knobs can be turned so quickly as to excite many-body dynamics far away from a thermal or metastable equilibrium state. 

What do we mean by ``far-from-equilibrium'' dynamics?
Consider a non-integrable many-body system.
Far from equilibrium, in contrast to close to it, there is no longer the notion of a precisely defined spectrum of quasiparticle modes whose damping can be described on the grounds of a linear-response analysis.
The latter generally relies on a perturbative expansion in some small parameter, and analytic relations between the fluctuation and response functions reflect the principle of detailed balance in Boltzmann's kinetic theory \cite{TGas:KadanoffBaym1962a,TGas:Kane1965a}.
In far-from-equilibrium or, as it is also often called, nonequilibrium time evolution there is no such fluctuation-dissipation relation, while microcausality and microreversibility are still conserved.
When studying the transition from far-from- to near-equilibrium dynamics one of the key questions is how the known near-equilibrium features are recovered during the time evolution of the system, given specific interaction properties.
An interesting observation gives rise to the distinction between short- and long-time evolution after a sudden quench of some boundary conditions which produces an initial state far from equilibrium.
In particular long-time many-body evolution and equilibration are demanding and still largely unresolved problems.

An important issue when studying nonequilibrium dynamics is the level of approximation on which interaction effects are taken into account.
As pointed out above, interactions are inextricably linked with nonequilibrium dynamics.
If they are weak, i.e., occur rarely which is the case in a dilute gas, low-order perturbative approximations in the diluteness parameter can provide a reliable description over a certain time.
At large times, however, such perturbative descriptions are expected to break down.
Moreover, if the interactions are sufficiently strong, the time at which this breakdown occurs can be shorter than the time at which near-equilibrium kinetic theory, which in general involves perturbative approximations, sets in to be valid.

For systems with large, i.e., classical occupations of the kinematically relevant modes, quantum fluctuations typically play a minor role and, if classical fluctuations are relevant, Monte Carlo simulations are often the method of choice. 
However, if quantum fluctuations become important, no such methods are at hand, and this is generically the case for long-time evolutions and dense, strongly interacting systems.
(Note that recently, for a certain range of applications, stochastic simulation techniques have been studied \cite{TGas:Berges:2005yt,TGas:Berges:2006xc,TGas:Berges:2006rw,TGas:Berges:2007nr}.)
Functional quantum field theoretical techniques represent a powerful approach to such dynamics.
Moreover, they provide analytical insight and make numerical computations feasible, in particular in more than one spatial dimension.

This article provides an introduction to functional quantum field theoretical methods to describe far-from-equilibrium many-body quantum dynamics.
It is beyond its scope to represent a full review and to provide a satisfactory account of the relevant literature.
In \TGasSection{MF}, we will define the relevant observables and give a brief summary of mean-field theory for the short-time and near-equilibrium evolution of a Bose gas.
In \TGasSection{NEqFT}, we shall then discuss the nonequilibrium two-particle irreducible (2PI) effective action \cite{TGas:Luttinger1960a,TGas:Baym1962a,TGas:Cornwall1974a} in the nonperturbative approximation introduced in Refs.~\cite{TGas:Berges:2001fi,TGas:Aarts:2002dj}.
This nonequilibrium approach has been developed and extensively applied, in the context of relativistic quantum field theories, as reported in Refs.~\cite{TGas:Berges:2001fi,TGas:Aarts:2002dj,TGas:Berges:2002cz,TGas:Mihaila:2000sr,TGas:Cooper:2002qd,TGas:Arrizabalaga:2004iw,TGas:Berges:2002wr,TGas:Berges:2004ce,TGas:Berges:2008wm}.
Their extension to abelian and non-abelian gauge theories \cite{TGas:Berges:2004pu,TGas:Berges:2004yj,TGas:Arrizabalaga2004a,TGas:Aarts:2006pa} is the subject of recent and ongoing research.
We will, furthermore, summarise a new functional renormalisation-group approach to far-from-equilibrium dynamics introduced in \cite{TGas:Gasenzer:2007za}.
For applications, the focus will be set on interacting ultracold atomic Bose gases \cite{TGas:Rey2004a,TGas:Baier:2004hm,TGas:Rey2005a,TGas:Gasenzer:2005ze} and an overview be given of the results first presented in Refs.~\cite{TGas:Gasenzer:2005ze,TGas:Temme2006a,TGas:Berges:2007ym,TGas:Branschadel2008a}. 
Example applications to the long-time evolution of an interacting Bose gas will be discussed.
The relation to near-equilibrium evolution is described in \TGasSection{Kin}. 
Quantum statistical fluctuations and their distinction from their classical counterparts are the subject of \TGasSection{QvsCl}.
A summary will be given in \TGasSection{Summary}.

\section{Mean-field dynamics of Bose-Einstein condensates}
\label{TGas:sec:MF}
In this section we give an introduction to mean-field theory of time-evolving ultracold Bose gases.
Mean-field theory is generically valid for the description of near-equilibrium dynamics and has been applied successfully to understand and predict an enormous variety of experimental observations, see, e.g.~\cite{TGas:Leggett2001a,TGas:Pitaevskii2003a}.
Discussing it allows us to lay the foundations for the later development of far-from-equilibrium dynamics and dynamics of strongly interacting systems.
We will first define the observables of interest in the context of most experiments and then give a concise introduction to the Gross-Pitaevskii and Hartree-Fock-Bogoliubov theories of Bose-Einstein condensates near their ground-state configuration.
We close the section with an outlook beyond mean-field theory, focusing on atomic gases near a Feshbach resonance as well as on an exact method to calculate the nonequilibrium dynamics of an interacting Bose gas in one spatial dimension.

\subsection{Observables}
\label{TGas:sec:Obs}
The application of statistical many-body theory to quantum degenerate states, e.g. of non-interacting Bose-Einstein condensates (BEC), aims at the occupation number distribution over the available energy eigenstates.
Quantum states, however, contain additional information which manifests itself in the phase of the wave function.
This phase gives rise to coherence and interference phenomena which, in a number of experiments with degenerate atomic gases, has been made visible in a macroscopic manner, e.g., in the experiments reported in Refs.~\cite{TGas:Andrews1997a,TGas:Mewes1997a,TGas:Anderson1998a,TGas:Bloch1999a,TGas:Gati2006a}.
Let us discuss in more detail the observables required to describe such properties of the many-body system.

Large occupation numbers, together with the phase of the single-particle wave function, lead to an approximate description of coherent matter in terms of a scalar complex field $\phi(x)$.
Here and in the following time and space variables are included in the four-vector $x\equiv(t,\vec{x})\equiv(x_0,\vec{x})$.
In view of the statistical properties of BEC, $\phi$ forms an order parameter.
With respect to coherence properties as well as large occupation numbers, the matter-wave field $\phi(x)$ exists in full analogy with, e.g., Maxwell's classical electromagnetic field $F_{\mu\nu}(x)$.
Recall that an ideal, single-mode laser can be described by a coherent state which characterises it as a coherent superposition of photon number states,
\begin{equation}
\label{TGas:eq:CohState}
  |\alpha\rangle = e^{-|\alpha|^2/2} \sum_{n=0}^\infty \frac{\alpha^n}{\sqrt{n!}}\,|n\rangle. 
\end{equation}
The expansion coefficients are chosen such that $|\alpha\rangle$ is an eigenstate of the Fock annihilation operator $\hat a$ with eigenvalue $\alpha$.
For photons propagating in vacuum, the time evolution of the coherent state reads $|\alpha(t)\rangle=|\alpha(0)\exp(-i\omega t)\rangle$, $\omega$ being the frequency of the mode.
This corresponds, in spatial representation, to a classical oscillation of the wave packet in the oscillator potential $m\omega^2X^2/2$.
Hence, the expectation value of the electric field operator $\hat{\vec{E}}\propto \vec{\varepsilon}(\hat{a}+\hat{a}^\dagger)$ performs the harmonic oscillations of the macroscopic classical field.  

For an ideal-gas BEC, the field can be written as $\phi(x)=\sqrt{N}\psi(x)$, where $\psi(x)$ has the functional form of the quantum mechanical single-particle wave function whose macroscopic mean occupation number is $N$.
Hence, while $|\phi(x)|^2$ describes the particle density at $x$, its phase gives rise to the same interference phenomena as the wave function $\psi$ does for single particles.
The complex matter-wave field $\phi(x)$ describes the spatial and temporal density and phase distributions of the coherent cloud of atoms in a particular internal state.
This analogy with photons gives rise to the notion of an atom laser in experiments where, e.g., a coherent beam of particles is coupled out from a trapped BEC \cite{TGas:Mewes1997a,TGas:Anderson1998a,TGas:Bloch1999a}.

A remark is in order:
Only compact systems are experimentally relevant.
According to the above picture, $\phi(x)$ is the expectation value of the complex non-relativistic quantum field operator obeying the bosonic equal-time commutation relations $[\Phi(\vec{x},t),\Phi^\dagger(\vec{x}',t)]=\delta(\vec{x}-\vec{x}')$.
In finite, closed, non-relativistic gatherings of atoms, the total number of atoms is a conserved quantity, i.e., the expectation value of this field operator with respect to the reduced density matrix describing any subsystem necessarily vanishes, $\phi=\langle\Phi\rangle\equiv0$.
Therefore, a coherent state can not describe such a system. 
This is closely related to the fact that an isolated system does by definition not interfere with any other system such that its total phase can not be measured.
Phase can only be measured by means of interference effects, and hence only the relative phase between different (sub)systems is a meaningful quantity.

This leads to the concept of the phase coherence of a BEC. 
This coherence manifests itself in the off-diagonal elements of the reduced single-particle density matrix, i.e., the single-time two-point correlation function
\begin{equation}
\label{TGas:eq:SPDM}
  n(\vec{x},\vec{y};t)=\langle\Phi^*(\vec{x},t)\Phi(\vec{y},t)\rangle.
\end{equation}
The local particle number density is given by the diagonal elements, $n(\vec{x};t)=n(\vec{x},\vec{x};t)$.
For an infinite uniform ideal-gas BEC, one finds that the first-order coherence function derived from the off-diagonal elements, $n^{(1)}(s)=n(\vec{R}+\vec{s}/2,\vec{R}-\vec{s}/2)\to const.\not=0$ for $s\to\infty$.
Since the momentum distribution of particles is given by the spatial Fourier transform of $n(\vec{x},\vec{y};t)$ with respect to the relative coordinate $\vec{s}=\vec{x}-\vec{y}$, this asymptotic off-diagonal finiteness implies a macroscopic occupation of the zero mode, i.e., Bose-Einstein condensation.
It is termed Off-Diagonal Long-Range Order (ODLRO) and allows the asymptotic factorisation of the single-particle density matrix,   
\begin{equation}
\label{TGas:eq:ODLRO}
   n(\vec{x},\vec{y};t)\quad \stackrel{\scriptstyle|\vec{x}-\vec{y}|\to\infty}{\longrightarrow}\quad \bar\phi^\dagger(\vec{x},t)\bar\phi(\vec{y},t).
\end{equation}
ODLRO was introduced by Penrose and Onsager as a general criterion for interacting BEC \cite{TGas:Penrose1956a}.
It is applicable also in non-uniform systems alternative to the field expectation value $\phi(x)$ and remains meaningful for number-conserving states.
The asymptotic factorisation in \TGasEq{ODLRO} defines $\bar\phi$ as an in general complex order parameter for BEC.
In finite systems, once ``long-range'' is quantified, it can be taken as a measure for local order.
It clearly expresses the fixed phase relation between distant points in the BEC.
In the thermodynamic limit $\bar\phi$ can be identified with the field expectation value $\phi$.

In summary, the field expectation value $\phi$ is a useful concept for most cases where a system involves macroscopic occupations $N\gg1$.
When comparing a formulation in terms of $\phi$ with that based on a density matrix with fixed total particle number differences in the observables are suppressed at least with a factor $1/N$.

In the following, the classical field 
\begin{equation}
\label{TGas:eq:FEV}
  \phi(x)=\langle\Phi(x)\rangle=\mathrm{Tr}[\hat\rho(t_0)\Phi(x)]
\end{equation}
will be regarded as the non-vanishing expectation value of the field operator evaluated at some time $t$ with respect to the density operator $\hat\rho$ at some initial time $t_0$, where the time dependence of $\Phi$ is implied to include the time evolution from $t_0$ to $t$.
Besides this one-point function we will use, in the following, the time-ordered connected Green function 
\begin{equation}
\label{TGas:eq:Gab}
  G_{ab}(x,y) 
   = \langle{\cal T}\Phi_a(x)\Phi_b(y)\rangle_c
   = \langle{\cal T}\Phi_a(x)\Phi_b(y)\rangle - \phi_a(x)\phi_b(y)
   = \langle{\cal T}\tilde\Phi_a(x)\tilde\Phi_b(y)\rangle,
\end{equation}
where ${\cal T}$ denotes time ordering of the subsequent operators and $\tilde\Phi=\Phi-\phi$ is the operator of fluctuations around the classical field $\phi$.
For the first, the field indices are chosen to number $\Phi_1=\Phi$ and $\Phi_2=\Phi^\dagger$ as independent components, $\{a,b\}\in\{1,2\}$.
The time ordering allows us to write $G$ as
\begin{equation}
\label{TGas:eq:GitoFrho}
  G_{ab}(x,y) = F_{ab}(x,y) -\frac{i}{2}\mathrm{sgn}(x_0-y_0)\rho_{ab}(x,y),
\end{equation}
where $F$ and $\rho$ involve the anticommutator and commutator of the fields, respectively,
\begin{equation}
\label{TGas:eq:Fandrho}
  F_{ab}(x,y)
  =\mbox{$\frac{1}{2}$}\langle\{\Phi_a(x),\Phi_b(y)\}\rangle_c,
  \qquad
  \rho_{ab}(x,y)
  =i\langle[\Phi_a(x),\Phi_b(y)]\rangle_c.
\end{equation}
In the $\Phi$-$\Phi^\dagger$-basis, one has $G_{11}=\langle{\cal T}\Phi^\dagger\Phi\rangle_c$, $G_{21}=\langle{\cal T}\Phi\Phi\rangle_c$, etc., and $G$ is related to the single-particle density matrix $n=n_0+\widetilde n$, $n_0=|\phi|^2$, and to the pair function $\widetilde m(\vec{x},\vec{y};t)=\langle\tilde\Phi(\vec{x},t)\tilde\Phi(\vec{y},t)\rangle$ as follows
\begin{eqnarray}
\label{TGas:eq:nmitoF}
  \widetilde n(\vec{x},\vec{y};t) 
  &=& F_{11}(x,y)|_{x_0=y_0=t}-\frac{1}{2}\delta(\vec{x}-\vec{y}),
  \nonumber\\
  \widetilde m(\vec{x},\vec{y};t) 
  &=& F_{21}(x,y)|_{x_0=y_0=t}.
\end{eqnarray}
As will be discussed in more detail in \TGasSect{Kin}, $F$ is called the statistical correlation function, containing, near equilibrium, information about the occupation of the available states, while the spectral correlation function $\rho$ provides the frequencies and widths of these states.
Near equilibrium, $F$ and $\rho$ are linked by a fluctuation-dissipation relation.
$\widetilde m$ quantifies the amount of pair correlations in the system, as, e.g., the number of atoms bound in pairs or, for fermions, the number of long-range correlated (Cooper) pairs, see \TGasSect{Feshbach}.

Connected $n$-point functions with $n\ge3$ contain information about higher-order correlations and are defined analogously.
In this section we will concentrate on $n\le2$.

\subsection{The Gross-Pitaevskii equation}
\label{TGas:sec:GPE}
This subsection intends to give a concise summary of the non-relativistic classical equation of motion for the field expectation value $\phi(x)$ which represents the leading-order approximation to the full quantum dynamics in the case that $\phi\not=0$.
It was first studied by Gross \cite{TGas:Gross1961a} and Pitaevskii \cite{TGas:Pitaevskii1961a} in the context of vortices in superfluid helium.
The equation for a field $\phi$ describing an ideal-gas BEC according to our above discussion has the same form as the Schr\"odinger equation for the single-particle wave function $\psi$ and is a special case of the Gross-Pitaevskii equation (GPE) which, in addition, includes the leading-order effects of interactions between the particles.

We define the system to be studied by the many-body Hamiltonian for a single-species of non-relativistic particles,
\begin{equation}
\label{TGas:eq:GPH}
  H
  = \int \mathrm{d}\vec{x}\,\Phi^\dagger(\vec{x})H_\mathrm{1B}(\vec{x})
    \Phi(\vec{x}) 
    + \frac{1}{2}\int \mathrm{d}\vec{x} \mathrm{d}\vec{y}\,
    \Phi^\dagger(\vec{x})\Phi^\dagger(\vec{y})V(\vec{x}-\vec{y})
    \Phi(\vec{y})\Phi(\vec{x}),
\end{equation}
with the Hamiltonian for a single particle exposed to an external, e.g., trapping potential $V_\mathrm{ext}$,
\begin{equation}
\label{TGas:eq:H1B}
  H_\mathrm{1B}(\vec{x})
  = -\frac{\hbar^2}{2m}\vec{\nabla}^2+V_\mathrm{ext}(\vec{x}),
\end{equation}
and the two-body potential $V(\vec{x}-\vec{y})$ describing the interactions between the particles.
The interactions are assumed to depend only on the relative coordinate between the collision partners as, e.g., in the Born-Oppenheimer approximation of atom-atom collisions.
The GPE is then obtained from the Liouville equation $i\hbar\partial_t\langle\Phi\rangle_t=\langle[\Phi,H]\rangle_t$, where $\langle\cdot\rangle_t$ denotes the expectation value with respect to the density operator at time $t$, by approximating the 3-point function $\langle\Phi^\dagger\Phi\Phi\rangle$ as a product of field expectation values, $\langle\Phi^\dagger\Phi\Phi\rangle\sim\phi^*\phi\phi$:
\begin{equation}
\label{TGas:eq:GPE}
  i\hbar\partial_t\phi(x)
  = \left[H_\mathrm{1B}(\vec{x})+g|\phi(x)|^2\right]\phi(x).
\end{equation}
Here, the contact-potential approximation has been chosen for $V(r)=g\delta(r)$, with $g=4\pi\hbar^2a/m$, where $a$ is the $s$-wave scattering length.
In \TGasSect{Feshbach} below, we will discuss in more detail the scattering theory behind this parametrisation, and note here only, that this approximation renders the GPE to be a low-energy effective theory:
The typical length scale characterising the motion of atoms in a BEC, the thermal de Broglie wave length $\lambda_T=h/\sqrt{2\pi mk_BT}$, is generally much larger than the scale determining the short-distance behaviour of the interatomic potential.
The atoms therefore only see an averaged interaction potential which at large internuclear distance manifests itself in a scattering phase shift. 
At low energies, given that the potential is sufficiently short ranged, only the $s$-wave scattering amplitude survives which tends to a constant related to the phase shift and equal to the minus the $s$-wave scattering length $a$ at vanishing scattering momentum.
The scattering length also quantifies the spatial extent of the highest excited dimer bound state close to the dissociation threshold and therefore the pair correlation length.
Setting $g=0$, \TGasEq{GPE} has the form of the Schr\"odinger equation.
  
We note that, in order to have Bose condensation in an ideal gas, $\lambda_T$ needs to exceed the interatomic distance.
In turn, for the GPE to be a good approximation, the interatomic distance must be much larger than the scattering length $a$.
This means, the gas needs to be dilute, such that multiple scattering effects play no role.
One says, the gas is weakly interacting.

The quartic interaction term in the Hamiltonian shows, that the local energy density rises with increasing particle density if $a$ is positive.
On the other hand, if $a$ is negative, the interaction part is lowered by increasing the local density of particles.
Therefore, a positive scattering length is said to describe repulsive interactions while negative $a$ corresponds to attractive ones.
Note that despite this, also at positive scattering lengths the interatomic potential can support bound states and therefore exert attractive forces.
Also the trap potential plays a crucial role as it, e.g., can stabilise a BEC of atoms with $a<0$, since the kinetic part of the energy rises with increasing curvature of the field $\phi$ at the density peak, see \TGasEq{GPH}.

We briefly discuss stationary solutions of the GPE in the presence of a trapping potential, with the time dependence $\phi(t,\vec{x})=\phi(0,\vec{x})\exp(-i\mu t/\hbar)$, i.e., solutions of the time-independent GPE
\begin{equation}
\label{TGas:eq:SGPE}
  \left[H_\mathrm{1B}(\vec{x})+g|\phi(\vec{x})|^2\right]\phi(\vec{x})
  =\mu\phi(\vec{x}).
\end{equation}
For positive $a$, the dilute-gas BEC is often termed strongly interacting\footnote{%
``Strongly interacting'' does here not imply that the gas requires a beyond-mean-field description, e.g., by being non-dilute in three or a gas with $\gamma>1$ in one spatial dimension, see \TGasSect{BeyondMF}.}  
if the kinetic part can be neglected within the total energy. 
This is called the Thomas-Fermi (TF) regime, in which the density distribution obtained from \TGaseq{SGPE} reflects directly the shape of the potential:
\begin{equation}
\label{TGas:eq:TFprofile}
  n_0(\vec{x})
  = |\phi(\vec{x})|^2
  = \frac{\mu-V_\mathrm{ext}(\vec{x})}{g}.
\end{equation}
Only at the edge of the cloud, where the density vanishes, the GP approximation breaks down.
A BEC in a harmonic-oscillator trap with frequency $\omega_\mathrm{ho}$ is in the TF regime if the total number of particles times the scattering length is much larger than the oscillator length $l_\mathrm{ho}=(\hbar/m\omega_\mathrm{ho})^{1/2}$, i.e., $\beta\equiv Na/l_\mathrm{ho}\gg 1$.
The TF radius $R_\mathrm{TF}=\beta^{1/5}l_\mathrm{ho}$ characterising the extent of the cloud results from the requirement that $N$ particles fit into the profile \TGaseq{TFprofile}.
Clearly, in the TF regime, the shape and size of the atom cloud is distinctly different from that of a nearly ideal gas, which is, to a good approximation, given by the modulus squared of the single-particle wave function.

Consider, now, the GP dynamics disclosed by the non-linear field equation \TGaseq{GPE}.
Writing the field in terms of density and phase, $\phi(x)=\sqrt{n(x)}\exp(-iS(x)/\hbar)$, one derives, by inserting this into the GPE, the hydrodynamic equations
\begin{eqnarray}
\label{TGas:eq:ContEq}
  &&\partial_tn + \vec{\nabla}\cdot(n\vec{v}) = 0,
\\
\label{TGas:eq:QEuler}
  &&m\partial_t\vec{v} + \vec{\nabla}\left(V_\mathrm{ext}+gn
  -\frac{\hbar^2}{2m\sqrt{n}}\vec{\nabla}^2\sqrt{n}
  +\frac{mv^2}{2}\right) = 0
\end{eqnarray}
where the velocity field $\vec{v}$ is proportional to the gradient of the phase,
$\vec{v}(x) = (1/{m})\vec{\nabla}S(x)$.
\TGasEq{ContEq} is the continuity equation expressing local number conservation while \TGasEq{QEuler} represents a quantum version of the Euler equation describing a frictionless fluid.
The quantum contribution adding to the classical Euler equation reflects the zero-point fluctuations encoded in the term proportional to $\hbar^2$ in \TGasEq{QEuler}:
A curved mean-field profile is subject to a quantum pressure which aims at flattening the density distribution.
We note that, since the velocity is a conservative or gradient field, the Euler equation describes, on a singly connected region of space, irrotational flow.
As an example which exhibits in a nice way the consequences of this irrotationality consider the motion of a Bose-Einstein condensate which resembles that of a scissors mode first discussed in the context of nuclear physics.
A condensed cloud trapped within an anisotropic, ellipsoidal potential which is excited by suddenly rotating the trap away from its prior position, starts oscillating around the new equilibrium orientation \cite{TGas:Marago2000a,TGas:GueryOdelin1999a,TGas:Jackson2001a}.
Although the oscillation of the density distribution resembles that of a rotation of the cloud, the velocity field shows that the flow pattern of particles is rather irrotational.

Irrotationality of the flow is one signature of the superfluidity present in a system with Bose-Einstein condensation.
A closely related and experimentally demonstrated property of such a system is the possibility of vortex formation.
The Gross-Pitaevskii equation possesses non-linear solutions describing a circular flow around a singularity at which the density $|\phi|^2$ vanishes, i.e., the phase $S(\vec{x})$ accumulates an integer multiple of $2\pi$ along one turn around the singular point (in 2 dimensions) or line (in 3D).
Vortices can be excited in trapped BECs \cite{TGas:Aftalion2006a} using circular polarised laser beams \cite{TGas:Madison2000a} and have been observed to form Abrikosov lattice structures \cite{TGas:Abo-Shaeer2001a} as known from liquid Helium \cite{TGas:Donnelly1991a}.
For a recent report on simulations see Ref.~\cite{TGas:Simula2008a}.

The GPE describes a colourful range of other nonlinear classical phenomena like solitons and nonlinear atom optics, phenomena which have been studied in many experiments and provide the frame of a research field in its own.
See, e.g., Refs.~\cite{TGas:Pethick2001a,TGas:Pitaevskii2003a}.
With the advent of the formation of bosonic pairs in ultracold Fermi gases classical nonlinear dynamics can be studied in even more systems.

\subsection{Beyond the GPE: Hartree-Fock-Bogoliubov mean-field theory}
\label{TGas:sec:HFB}
As discussed in the previous section, condensates exhibit superfluidity.
The Gross-Pitaevskii equation for the order parameter field $\phi(x)$ includes an Euler-like equation for a perfect fluid.
Superfluidity, in turn, does not require a non-vanishing condensate order parameter as is known from the low-temperature physics of helium which is a non-dilute and therefore strongly interacting system.
To describe BEC away from zero temperature and vanishing interactions, as well as away from thermal equilibrium, fluctuations need to be considered beyond the Gross-Pitaevskii approximation.
In \TGasSection{Obs} we identified $\phi(x)$ as a practically suitable measure for the asymptotic off-diagonal long-range order contained in the full two-point correlation function.
Describing a Bose gas beyond the GP approximation requires additional information about the more local properties and dynamics of the two-point function, i.e., about the connected correlation function $G(x,y)$ as well as about the back-reaction of this onto the evolution of $\phi$. 
In the GPE the terms accounting for this back-reaction were neglected when approximating $\langle\Phi^\dagger\Phi\Phi\rangle$ as a product of field expectation values.

\subsubsection{Time-dependent HFB equations}
\label{TGas:sec:TDHFB}
The dynamic equation for $G$ can be derived, as before, from the Liouville equation $i\hbar\partial_t\langle{\cal O}\rangle_t=\langle[{\cal O},H]\rangle_t$, where ${\cal O}$ is now to be replaced by the respective products of field operators $\Phi_a(x)$.
Rewriting all correlation functions appearing in the equations in terms of their cumulants, i.e., in terms of connected Green functions and neglecting all cumulants of order three and higher one obtains the equations
\begin{eqnarray}
\label{TGas:eq:HFB_phi}
   &&\big[i\hbar\partial_{t}-H_\mathrm{1B}(\vec{x})\big]\,\phi(\vec{x},t)
   = g\Big[
     \Big(\phi(\vec{x},t)^2+{\widetilde m}(\vec{x},\vec{x};t)\Big)\phi^*(\vec{x},t)
     + 2{\widetilde n}(\vec{x},\vec{x};t)\,\phi(\vec{x},t)
    \Big],
   \\
\label{TGas:eq:HFB_n}
  &&\big[i\hbar\partial_{t}-H_\mathrm{1B}(\vec{x})+H_\mathrm{1B}(\vec{y})\big]\,
   {\widetilde n}(\vec{x},\vec{y};t)
   = 
   g\,
  \Big[\Big(\phi(\vec{x},t)^2
   +{\widetilde m}(\vec{x},\vec{x};t)\Big){\widetilde m}^*(\vec{x},\vec{y};t)
  \nonumber\\
  &&\qquad\qquad
   +\ 2\Big(\phi^*(\vec{x},t)\phi(\vec{x},t)+{\widetilde n}(\vec{x},\vec{x};t)\Big) 
     {\widetilde n}(\vec{x},\vec{y};t)\Big]
   - g[\vec{x}\leftrightarrow\vec{y}]^*,
\\
\label{TGas:eq:HFB_m}
  &&\big[i\hbar\partial_{t}-H_\mathrm{1B}(\vec{x})-H_\mathrm{1B}(\vec{y})\big]\,
   {\widetilde m}(\vec{x},\vec{y};t)
  = 
   g\, \Big[\Big(\phi(\vec{x},t)^2
   +{\widetilde m}(\vec{x},\vec{x};t)\Big){\widetilde n}(\vec{y},\vec{x};t)
  \nonumber\\
  &&\qquad\qquad
   +2\Big(\phi^*(\vec{x},t)\phi(\vec{x},t) + {\widetilde n}(\vec{x},\vec{x};t)\Big)
    {\widetilde m}(\vec{x},\vec{y};t)
  + g[\vec{x}\leftrightarrow\vec{y}],
\end{eqnarray}
where $H_\mathrm{1B}(\vec{x})=-\hbar^2\nabla^2/2m+V_\mathrm{ext}(\vec{x})$ denotes the one-body Hamiltonian. 
The generalised GPE \TGaseq{HFB_phi} and the equations \TGaseq{HFB_n} and \TGaseq{HFB_m} for the connected propagator contain, through the normal and anomalous density matrices $\widetilde n$ and $\widetilde m$, respectively, only the statistical two-point function $F$, cf.~Eqs.~\TGaseq{GitoFrho}, \TGaseq{nmitoF}.
The above equations are commonly termed the (time-dependent) Hartree-Fock-Bogoliubov (HFB) equations \cite{TGas:Hartree1928a,TGas:Fock1930a,TGas:Bogoliubov1947a} which describe the mean-field dynamics beyond the GPE in leading order in the coupling $g$.
They form a closed system of partial differential equations which describe the coupled dynamics of the condensate and noncondensate components of an ultracold Bose gas.
The set of equations preserves important conservation laws such as the total number of particles and energy.
The exchange between the condensate and the noncondensed fractions is caused by the elastic direct and exchange collision processes between a condensate atom and an excited atom, as well as pair excitations out of the condensate, see \TGasFig{HFBdiagrams}.
\begin{figure}[tb]
\begin{center}
\resizebox{0.8\columnwidth}{!}{
\includegraphics{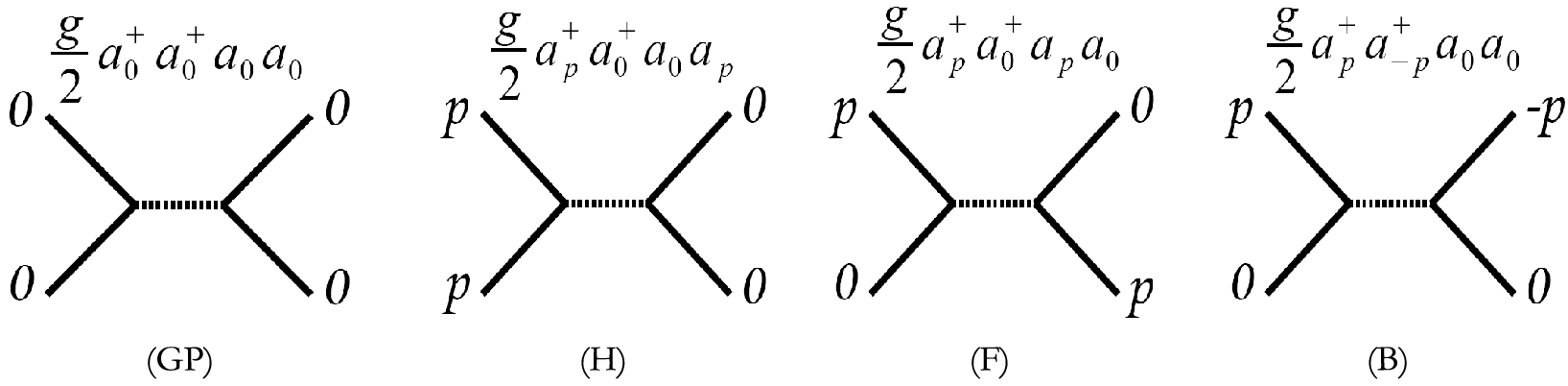}
}
\caption{
Schematic representation of the different scattering processes contributing to the dynamics of a homogeneous gas within the Hartree-Fock-Bogoliubov (HFB) mean-field approximation.
The left panel illustrates the Gross-Pitaevskii (GP) approximation, which includes scattering between atoms in the condensate ($p=0$) mode.
The right panel distinguishes between the three possible channels for elastic scattering between a condensate and an excited atom, the direct (Hartree), exchange (Fock), and pair production (Bogoliubov) processes.
In an operator language, these processes are described by the respective vertex operator contributions to the interaction Hamiltonian quoted above the diagrams.
All vertex operators are at most quadratic in excited-mode operators $a_p$ ($p\not=0$), such that the resulting Hamiltonian can be diagonalised, and it describes an effectively free system.
}
\label{TGas:fig:HFBdiagrams}
\end{center}
\end{figure}

\subsubsection{Linearised HFB equations}
\label{TGas:sec:LinHFB}
Let us finally consider the case that Eqs.~\TGaseq{HFB_phi}--\TGaseq{HFB_m} describe small-amplitude oscillations around their stationary solutions.
This approximation is, at first sight, irrelevant for the later discussion of far-from-equilibrium dynamics.
We will discuss it here since it has widely been used for ultracold gases, and in order to point to the difference between the so called Landau and Beliaev damping processes and the collisional damping we will discuss in the later sections.
See, e.g., Refs.~\cite{TGas:Giorgini2000a,TGas:Pitaevskii2003a} for a concise summary of the procedure outlined in the following.

On the right-hand side of \TGasEq{HFB_phi}, only diagonal elements of the normal and anomalous density matrices appear.
We can therefore focus on the closed set of equations for $\phi(x)$ and the diagonal elements $\widetilde n(\vec{x},t)\equiv\widetilde n(\vec{x},\vec{x},t)$ and $\widetilde m(\vec{x},t)\equiv\widetilde m(\vec{x},\vec{x},t)$.
To study small-amplitude deviations one linearises the generalised GPE \TGaseq{HFB_phi} as well as Eqs.~\TGaseq{HFB_n} and \TGaseq{HFB_m} in small displacements from the equilibrium values of $\phi$, $\widetilde n$, and $\widetilde m$,
\begin{eqnarray}
  \phi(\vec{x},t) 
  &=& \phi_0(\vec{x}) + \delta\phi(\vec{x},t),
  \nonumber\\
  \widetilde n(\vec{x},t) 
  &=& \widetilde n^0(\vec{x}) + \delta\widetilde n(\vec{x},t),
  \nonumber\\
  \widetilde m(\vec{x},t) 
  &=& \widetilde m^0(\vec{x}) + \delta\widetilde m(\vec{x},t),
\label{TGas:eq:SmallOscillations}
\end{eqnarray}
The time-independent part of the field expectation value is determined by the generalised stationary GPE
\begin{equation}
  \mu\phi_0(\vec{x})
  = \left[H_\mathrm{1B}(\vec{x})
    +g\left(n_0(\vec{x})+2\widetilde n^0(\vec{x})+\widetilde m^0(\vec{x})\right)\right]\phi_0(\vec{x}),
\label{TGas:eq:statGPE_HFB}
\end{equation}
with $n_0(\vec{x})=|\phi_0(\vec{x})|^2$.
We work in the grand canonical ensemble, with the Hamiltonian replaced by $K=H-\mu N$ which is equivalent to factor out a phase $\exp(-i\mu t/\hbar)$ from $\phi_0$ in order to make it time-independent.

In order to diagonalise the equations for the stationary densities $\widetilde n^0(\vec{x})$ and $\widetilde m^0(\vec{x})$ one transforms the fluctuation operators to a quasiparticle basis by means of the Bogoliubov transformation
\begin{equation}
\label{TGas:eq:BogTrafo}
  \left(\begin{array}{c}
  \widetilde\Phi(\vec{x},t)\\ \widetilde\Phi^\dagger(\vec{x},t)
  \end{array}\right)
  =\sum_j
  \left(\begin{array}{cc}
  u_j(\vec{x}) & v_j^*(\vec{x})\\
  v_j(\vec{x}) & u_j^*(\vec{x})
  \end{array}\right)
  \left(\begin{array}{c}
  \alpha_j(t)\\ \alpha_j^\dagger(t)
  \end{array}\right).
\end{equation}
Here, $\alpha_j$ and $\alpha_j^\dagger$ are quasiparticle operators which satisfy the Bose commutation relations $[\alpha_i(t),\alpha_j^\dagger(t)]=\delta_{ij}$, provided the mode functions $u_i$, $v_i$ are subject to the normalisation conditions $\int \mathrm{d}^dx\,[u_i^*(\vec{x})u_j(\vec{x})-v_i^*(\vec{x})v_j(\vec{x})]=\delta_{ij}$.

Defining the normal and anomalous quasiparticle density matrices $f_{ij}(t)=\langle\alpha_i^\dagger(t)\alpha_j(t)\rangle$ and $g_{ij}(t)=\langle\alpha_i(t)\alpha_j(t)\rangle$, respectively, the Bogoliubov-deGennes eigenvalue problem
\begin{equation}
\label{TGas:eq:BogdeGennes}
  \left(\begin{array}{cc}
  {\cal L}(\vec{x}) & g\,\left[n_0(\vec{x})+\widetilde m^0(\vec{x})\right]\\
  -g\,\left[n_0(\vec{x})+\widetilde m^0(\vec{x})\right]  & -{\cal L}(\vec{x})
  \end{array}\right)
  \left(\begin{array}{c}
  u_j(\vec{x})\\ v_j(\vec{x})
  \end{array}\right)
  =
  \epsilon_j
  \left(\begin{array}{c}
  u_j(\vec{x})\\ v_j(\vec{x})
  \end{array}\right),
\end{equation}
with ${\cal L}=H_\mathrm{1B}(\vec{x})-\mu+2gn(\vec{x})$, $n(\vec{x})=n_0(\vec{x})+\widetilde n^0(\vec{x})$, fixes the quasiparticle amplitudes $u_i$ and $v_i$, and yields a diagonal stationary part of the quasiparticle density matrix, $f^0_{ij}=f^0_i\delta_{ij}$ and a vanishing stationary anomalous quasiparticle density matrix $g^0_{ij}=0$.

The resulting linearised coupled equations for the time dependent variations $\delta\phi(\vec{x},t)$, $\delta f_{ij}(t)=f_{ij}(t)-\delta_{ij}f_i^0$ and $\delta g_{ij}(t)=g_{ij}(t)$ read
\begin{eqnarray}
\label{TGas:eq:GPE_HFB}
  i\hbar\partial_t\delta\phi(\vec{x},t) 
  &=& \left( H_\mathrm{1B}(\vec{x})-\mu
      +2g\, n(\vec{x})\right)\delta\phi(\vec{x},t) 
      + g\, n_0(\vec{x})\delta\phi^{\ast}(\vec{x},t) 
  \nonumber \\
  &&\quad+\ 2g\,\phi_0(\vec{x})\delta 
  \widetilde{n}(\vec{x},t) + g\,\phi_0(\vec{x})\delta \widetilde{m}(\vec{x},t),
\\
  i\hbar\partial_t\delta f_{ij}(t) 
  &=& (\epsilon_j-\epsilon_i)\delta f_{ij}(t)
  + 2g\,(f_i^0-f_j^0)\int \mathrm{d}^dx \;\phi_0(\vec{x}) 
  \big[ \big(\delta\phi(\vec{x},t)+\delta\phi^{\ast}(\vec{x},t)\big) 
  \big.
  \nonumber \\
  \big.
  &\times&\big(u_i(\vec{x})u_j^{\ast}(\vec{x})+v_i(\vec{x})v_j^{\ast}(\vec{x})\big)
  + \delta\phi(\vec{x},t)v_i(\vec{x})u_j^{\ast}(\vec{x}) 
  + \delta\phi^{\ast}(\vec{x},t)u_i(\vec{x})v_j^{\ast}(\vec{x}) \big] 
\label{TGas:eq:statHFB_n}
\\
  i\hbar\partial_t\delta g_{ij}(t) 
  &=& (\epsilon_j+\epsilon_i)\delta g_{ij}(t)
  + 2g\,(1+f_i^0+f_j^0)\int \mathrm{d}^dx \;
  \phi_0(\vec{x}) \big[ \big(\delta\phi(\vec{x},t)+\delta\phi^{\ast}(\vec{x},t)\big) 
  \big.
  \nonumber \\
  \big.
  &\times&\big(u_i^{\ast}(\vec{x})v_j^{\ast}(\vec{x})
  +v_i^{\ast}(\vec{x})u_j^{\ast}(\vec{x})\big)
  + \delta\phi(\vec{x},t)u_i^{\ast}(\vec{x})u_j^{\ast}(\vec{x}) 
  + \delta\phi^{\ast}(\vec{x},t)v_i^{\ast}(\vec{x})v_j^{\ast}(\vec{x}) \big].\qquad
\label{TGas:eq:statHFB_m}
\end{eqnarray}
The quantities $f_j^0=\langle\alpha_j^\dagger\alpha_j\rangle_0=[\exp(\beta\epsilon_j)-1]^{-1}$ are the equilibrium quasiparticle occupations, in terms of which the equilibrium non-condensate density is obtained as 
\begin{equation}
  \widetilde n^0(\vec{x})=\sum_j[(|u_j(\vec{x})|^2+|v_j(\vec{x})|^2) f_j^0+ |v_j(\vec{x})|^2].
\end{equation}
The variations of the normal and anomalous particle densities expressed in terms of the quasiparticle density variations read
\begin{eqnarray}
  \delta\widetilde n(\vec{x},t)
  &=& \sum_{ij}\left\{
      \left[u_i^*(\vec{x})u_j(\vec{x})+v_i^*(\vec{x})v_j(\vec{x})\right]\delta f_{ij}(t)
           +u_i(\vec{x})v_j(\vec{x})\delta g_{ij}(t)
	   +u_i^*(\vec{x})v_j^*(\vec{x})\delta g_{ij}^*(t)\right\},
  \nonumber\\
  \delta\widetilde m(\vec{x},t)
  &=& \sum_{ij}\left\{
           2v_i^*(\vec{x})u_j(\vec{x})\delta f_{ij}(t)
           +u_i(\vec{x})u_j(\vec{x})\delta g_{ij}(t)
	   +v_i^*(\vec{x})v_j^*(\vec{x})\delta g_{ij}^*(t)\right\}.
\label{TGas:eq:deltanmitofg}
\end{eqnarray}
Note that Eqs.~\TGaseq{statGPE_HFB}, \TGaseq{BogdeGennes}--\TGaseq{deltanmitofg} form a closed set of equations of motion.

Neglecting the cross terms coupling the condensate and noncondensate oscillations $\delta\phi$ and $\delta\widetilde n$, $\delta\widetilde m$, the equations for $\delta\phi$ and $\delta\phi^*$ can be disentangled by a Bogoliubov-like transformation.
The Bogoliubov frequencies are the resulting eigenvalues and therefore the frequencies of the BEC's elementary excitations.
For a translationally invariant gas, they read
\begin{equation}
\label{TGas:eq:BogEnergy}
  \epsilon_p=\left[\left(\frac{p^2}{2m}+g\,n_0\right)^2-g^2n_0^2\right]^{1/2},
\end{equation}
where the Popov approximation $\widetilde m^0=0$ has been chosen \cite{TGas:Popov1987a,TGas:Griffin1996b}.
This approximation ensures that the spectrum \TGaseq{BogEnergy} is gapless as required by the Hugenholtz-Pines \cite{TGas:Hugenholtz1959a} and Goldstone \cite{TGas:Goldstone:1961eq,TGas:Goldstone:1962es} theorems:
At low momenta, the energy \TGaseq{BogEnergy} is linear in the momentum $p=|\vec{p}|$.
In this limit the elementary oscillations are collective sound modes with dispersion $\omega_p=c_sp$, where $c_s=\sqrt{gn_0/m}$ is the sound velocity.
At high momenta, the Bogoliubov dispersion assumes the quadratic form of free particles, $\omega_p=p^2/2m+gn_0$.
The approximations made in deriving the above equations are valid if the gas is weakly interacting, i.e., for a small diluteness parameter $\eta=na^3$.
 
One can show that, with a linear dispersion, energy and momentum conservation restrict the possibilities for the excitation of particle modes in a flowing BEC when encountering obstacles, e.g., atoms at a cavity wall (see, e.g., Ref.~\cite{TGas:Pitaevskii2003a}). 
A linear dispersion implies a maximum critical velocity for frictionless flow, i.e., for superfluidity.
A weakly interacting BEC therefore obeys Landau's criterion for superfluidity in the same way as superfluid $^4$He which, in addition, shows a pronounced roton minimum at finite wave vectors.
Note that the Bogoliubov dispersion \TGaseq{BogEnergy} is already obtained in the Bogoliubov approximation where the time evolution of the excited modes as well as the back action of the static excitation numbers on the condensate oscillation frequencies are neglected. 

We emphasise that the HFB equations \TGaseq{statHFB_n}, \TGaseq{statHFB_m} are local in time and only involve single-time correlation functions.
The full HFB equations \TGaseq{HFB_n}, \TGaseq{HFB_m}, however, also determine the off-diagonal time dependence of $G_{ab}(x,y)$.
The fact that this does not feed back into the equations \TGaseq{HFB_n}, \TGaseq{HFB_m} for the density matrices reflects that the HFB approximation does not account for direct scattering required for collisional dissipation and thermalisation.
Hence, the HFB approximation is expected to be valid in the collisionless regime, where the mean free path is much larger than the scattering length.
Note, however, that the linearised equations \TGaseqs{GPE_HFB}{statHFB_m}, if the coupling of excitations of the condensate and noncondensed fractions account is taken into account, also describe one-to-two and two-to-one collision processes between the excitations, provided a BEC phase is present, i.e., $\phi_0\not0$.
These give rise to the so-called Landau and Beliaev damping caused by the mixing of superfluid and normal fluid phases, see, e.g., \cite{TGas:Giorgini2000a,TGas:Pitaevskii2003a} and Refs.~cited therein.
This damping is different in nature from the collisional dissipation obtained beyond the HFB approximation of the dynamic equations and discussed further in \TGasSect{NLO1N}.

\subsection{Beyond mean field}
\label{TGas:sec:BeyondMF}
The collisionless regime discussed so far is easily left behind in present-day experiments.
The preparation of ultracold atomic Bose and Fermi gases in various trapping environments allows to precisely study quantum many-body dynamics of strongly correlated systems, see, e.g., Refs.~\cite{TGas:Greiner2002a,TGas:Stoferle2004a,TGas:Kinoshita2006a,TGas:Sadler2006a}.
In particular, techniques exploiting zero-energy (magnetic and photoassociative, i.e., optical Feshbach) scattering resonances have helped to provide ultracold atomic gases with the importance they nowadays bear and the attraction they exert on physicists in most different areas of physics.
Such techniques allow, by means of external electromagnetic fields, to tune the scattering length freely between large negative and large positive numbers.
Special trapping configurations such as quasi one- and two-dimensional traps as well as optical lattices add to these possibilities and ask for descriptions beyond the mean-field level.

\begin{figure}[tb]
\begin{center}
\resizebox{1.0\columnwidth}{!}{
\includegraphics{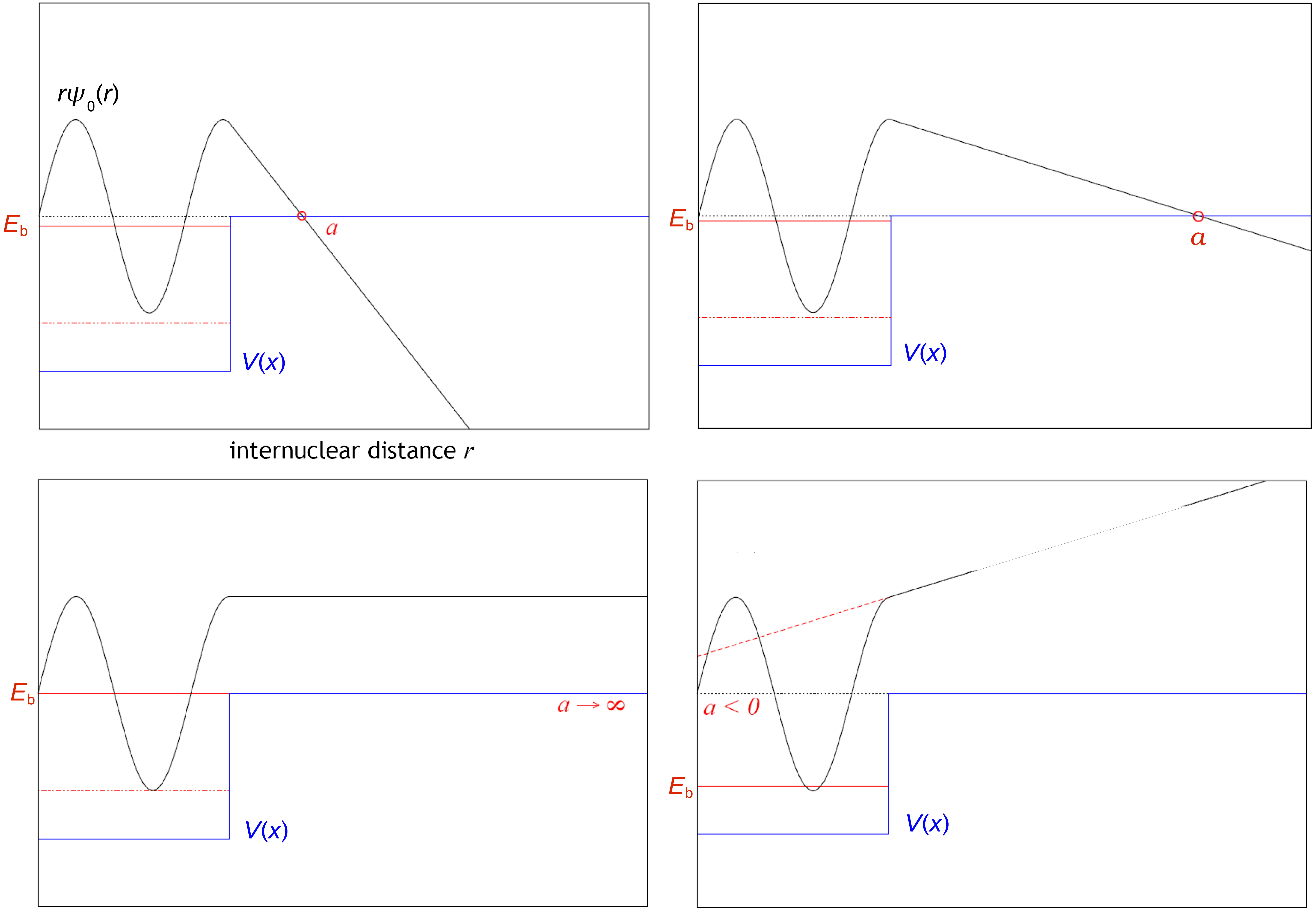}
}
\vspace*{2ex}\ \\
\caption{
(Color online)
Radial zero-energy scattering wave function $r\psi_0(r)=r\lim_{p\to0}\psi_{\vec{p}}(\vec{r})$ (black solid line) at small internuclear distances $r$, for the square-well potential $V(r)$ (drawn in blue).
Shown are four different potential depths $V(0)$, resulting in different $s$-wave scattering lengths $a$.
The scattering length is given by the intersection radius of the extrapolated wave function with the zero-energy axis, cf.~\TGasEq{ScattWFatZeroEnergy}.
(Red) horizontal lines in the potential wells indicate bound-state energy levels.
}
\label{TGas:fig:FBSqWell}
\end{center}
\end{figure}
\subsubsection{Feshbach resonances}
\label{TGas:sec:Feshbach}
Feshbach resonant scattering is most easily understood by realising that the value of the $s$-wave scattering length is directly related to the energy of the highest bound state the Born-Oppenheimer scattering potential supports below the zero-energy threshold.\footnote{%
Overviews and references to most of the relevant literature concerning magnetic and optical Feshbach resonances and photoassociation can be found in Refs.~\cite{TGas:Weiner1999a,TGas:Stwalley1999a,TGas:Williams2000a,TGas:Mies2000a,TGas:Burnett2002b,TGas:Weidemueller2003a,TGas:Gasenzer2004c,TGas:Kohler2006a}.}
To illustrate this consider the simple case of a square-well potential $V(r)$ which is non-zero only for $r\le r_0$ and tends to infinity at $r=0$, see \TGasFig{FBSqWell}.
The radial zero-energy scattering wave function $r\psi_0(r)=r\lim_{p\to0}\psi_{\vec{p}}(\vec{r})$ oscillates within the square well, with a frequency determined by the depth of the well, while its wave length outside the well is much larger, $\propto1/p$.
Hence, outside but close to the well, it is approximately linear in $r$.
As can be seen in \TGasFig{FBSqWell}, the continuity conditions for the scattering wave function at the edge of the well imply that $\psi_0(r)$ crosses the $r$-axis at $r=a$.
This can also be expressed in terms of the scattering amplitude approaching, for $p\to0$, a constant, the $s$-wave scattering length, $\lim_{p\to0}f_p(\Omega)=-a$:
\begin{equation}
\label{TGas:eq:ScattWFatZeroEnergy}
  r\psi_{\vec{p}}(\vec{r}) 
  ~~~\stackrel{r\to\infty}{\longrightarrow}~~~
  r\left(e^{i\vec{p}\cdot\vec{r}}+f_p(\Omega)\frac{e^{ipr}}{r}\right)
  ~~~\stackrel{p\to 0}{\longrightarrow}~~~
  r-a.
\end{equation}
The square-well example shows clearly that, if the potential depth is changed, the energy $E_b$ of the uppermost bound state shifts, and the scattering length goes through infinity when $E_b$ crosses zero. 
We remark that the number of nodes within the well corresponds to the number of bound states left in the potential well.
Very close to the resonance, the energy of the uppermost bound state with respect to threshold is proportional to the inverse of $a^2$,
\begin{equation}
\label{TGas:eq:FBBindingEnergy}
  E_\mathrm{b} = -\frac{\hbar^2}{ma^2}.
\end{equation}
The wave function of the uppermost bound state, as can be imagined from \TGasFig{FBSqWell}, is very similar to the wave function of the zero-energy scattering state for radii $r$ considerably smaller than $a$.
Close to $r=a$, however, since $E_b<0$, the bound-state wave function starts to differ from $\psi_0(r)$ and approaches zero for larger $r$.
As a consequence, the scattering length, if positive and larger than the extent of $V(r)$, measures the spatial extent of the bound state, i.e., the size of the respective dimer molecules.
Feshbach resonances have been exploited at large to produce degenerate molecular gases consisting of dimers of bosons, fermions, and of diatomic molecules of different species.
See, e.g., Refs.~\cite{TGas:Williams2000a,TGas:Weidemueller2003a,TGas:Gasenzer2004c,TGas:Kohler2006a,TGas:Doyle2004a,TGas:Dulieu2006a} for reviews and further references on cold molecules.
We remark that in these experiments, dimers could be identified at values of the scattering length on the order of one thousand Bohr radii \cite{TGas:Kohler2003b}.
These states are the largest and presumably most fragile molecules ever produced and measured in physics, see \TGasFig{phib}.

\begin{figure}[tb]
\begin{center}
\begin{minipage}{0.35\columnwidth}
\resizebox{0.85\columnwidth}{!}{
\includegraphics{TGas_fig3.eps}
}
\end{minipage}
\hspace*{0.05\columnwidth}
\begin{minipage}{0.57\columnwidth}
\resizebox{1.0\columnwidth}{!}{
\includegraphics{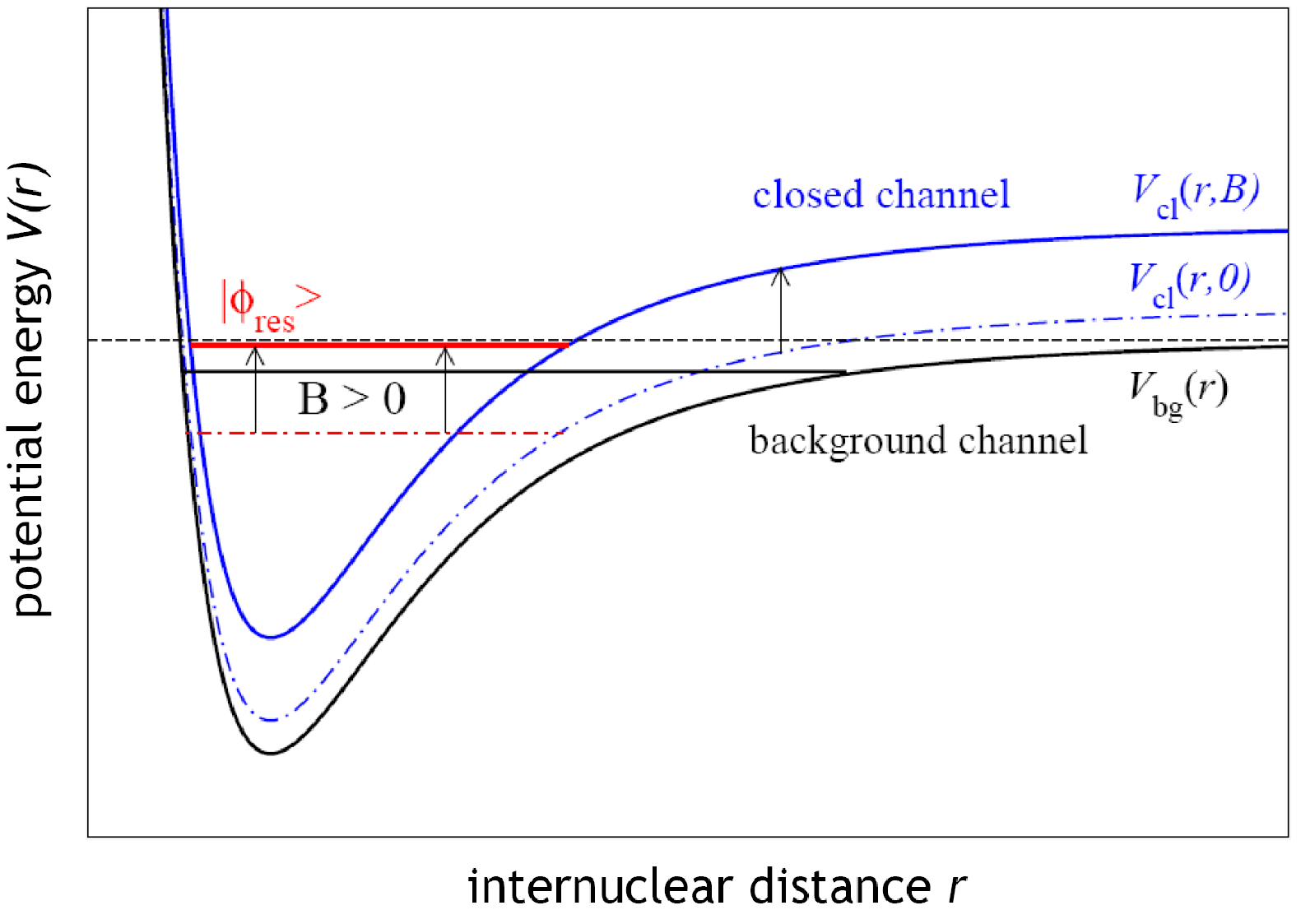}
}
\end{minipage}
\vspace*{2ex}\ \\
\begin{minipage}[t]{0.37\columnwidth}
\caption{
Coupled-channel bound states of $^{85}$Rb$_2$ at magnetic field strengths of $B_\mathrm{evolve}=16.0\ $mT and 
$B_\mathrm{min}=15.55\ $mT.
The dotted (dashed) curves indicate the closed ($\{-2,-2\}$ open) channel 
components. 
Note the extreme size at $B_\mathrm{min}$.
See text, Figs.~\TGasfig{pulse}, \TGasfig{MolOsc}, and Refs.~\cite{TGas:Kohler2003a,TGas:Kohler2003b} for more details.
}
\label{TGas:fig:phib}
\end{minipage}
\hspace*{0.01\columnwidth}
\begin{minipage}[t]{0.59\columnwidth}
\caption{
(Color online)
Coupled-channels picture of a magnetic Feshbach resonance.
The ultracold atoms collide, with an almost vanishing relative momentum, near the threshold of the background channel.
Intramolecular forces couple them to bound states of the closed-channel potential $V_\mathrm{cl}$ in which they are lacking the energy to separate to asymptotical distances.
By tuning the Zeeman shift between the asymptotic channel energies, a closed-channel bound state can be brought into resonance with the colliding atoms, causing a Feshbach-resonant increase of the $s$-wave scattering length $a$.
}
\label{TGas:fig:2ChannelFB}
\end{minipage}
\end{center}
\end{figure}
In order to meet the conditions for a magnetic Feshbach resonance, the effective Born-Oppenheimer potential of the gas atoms is modified by means of external magnetic fields coupling to the magnetic moment.
The atoms are usually trapped in a well-defined hyperfine state, such that the field causes a Zeeman shift relative to the energy of atoms in different polarisation states. 
By applying an external magnetic field, different scattering channels, corresponding to different asymptotic hyperfine states, can be shifted in energy relative to each other, see \TGasFig{2ChannelFB}.
Intramolecular electromagnetic forces couple these potentials, with a strength depending on the internuclear distance as well as on the energies of bound states supported by the system.
In this way, the effective scattering potential can be changed, and Feshbach resonances occur whenever a bound state of the coupled system crosses the energy of the asymptotically separated atom pair, see, e.g., Refs.~\cite{TGas:Mies2000a,TGas:Kohler2006a}.
We finally note, that photoassociation scattering is analogous to the magnetic Feshbach scattering described here.
There, the coupling between the channels is provided by polarised laser light.
In the most simple case of induced dipole transitions, which require asymptotically $P$-wave closed channels, spontaneous decay of the closed-channel bound state results in a complex scattering length 
\cite{TGas:Fedichev1996a,TGas:Bohn1996a,TGas:Gasenzer2004b}. 

\begin{figure}[tb]
\begin{center}
\begin{minipage}{0.42\columnwidth}
\resizebox{1.0\columnwidth}{!}{
\includegraphics{TGas_fig5.eps}
}
\end{minipage}
\hspace*{0.05\columnwidth}
\begin{minipage}{0.50\columnwidth}
\resizebox{0.9\columnwidth}{!}{
\includegraphics{TGas_fig6.eps}
}
\end{minipage}
\vspace*{2ex}\ \\
\begin{minipage}[t]{0.44\columnwidth}
\caption{
Scheme of a typical magnetic field pulse shape in the low
density ($n_0=3.9\times 10^{12}\ \mathrm{cm}^{-3}$) experiments in 
Ref.~\cite{TGas:Donley2002a}. The minimum magnetic field 
strength of the first and second pulse is 
$B_\mathrm{min}=15.55\ $mT. In the evolution 
period the field strength is chosen as $B_\mathrm{evolve}=16.0\ $mT.  
In the course of the experiments the evolution time $t_\mathrm{evolve}$
as well as $B_\mathrm{evolve}$ were varied. The dashed line 
indicates the position of the resonance at $B_0=15.49\ $mT.
}
\label{TGas:fig:pulse}
\end{minipage}
\hspace*{0.01\columnwidth}
\begin{minipage}[t]{0.52\columnwidth}
\caption{
The remaining fraction of condensate atoms, $n_{\rm c}(t_{\rm fin})$,
(solid line) together with the noncondensate fraction (dotted line), and 
the total density of unbound atoms (dashed line), 
as a function of the final time $t_{\rm fin}$, 
at the end of the magnetic-field pulse in \TGasFig{pulse}. All
densities are given relative to the initial density.
The figure shows the results of simulations of the HFB dynamic equations in the form described in Ref.~\cite{TGas:Kohler2003a}, for the experiment reported in Ref.~\cite{TGas:Donley2002a}.
See the main text for an outline of the experiment, and \cite{TGas:Kohler2003a} for more details.
}
\label{TGas:fig:MolOsc}
\end{minipage}
\end{center}
\end{figure}

\subsubsection{Ultracold gases near a Feshbach resonance}
\label{TGas:sec:FeshbachGases}
In a many-body system, the description in terms of binary scattering becomes unreliable close to the Feshbach resonance.
As follows from the above discussion, the scattering length can reach and exceed the mean atomic separation, and bound states are no longer binary but should be regarded as extended clusters involving a macroscopic number of particles.
For short evolution times, theoretical comparisons with experimental results for molecule formation in condensates, see, e.g., Refs.~\cite{TGas:Donley2002a,TGas:Kokkelmans2002a,TGas:Mackie2002b,TGas:Kohler2003a,TGas:Goral2004b}, indicate that the HFB dynamic equations discussed above can be applied even if the scattering length exceeds the mean interatomic spacing.
In this experiment, performed in the group of Carl Wieman at JILA, Boulder, Bose-Einstein condensed $^{85}$Rb atoms were, for the first time, observed to coherently bind to dimer molecules when the scattering length $a$ was tuned close to a Feshbach resonance \cite{TGas:Donley2002a}.
The setup worked as a Ramsey interferometer: 
In the first step, the magnetic field $B$ was tuned, for a few microseconds, close to $B_0$ such that $a(B)$ was on the order of $10^4$ Bohr radii $a_\mathrm{B}$.
This constituted the first (coupling) Ramsey pulse.
A longer evolution time $t_\mathrm{evolve}$ followed, during which $a(B)$ was ramped back to a few hundred $a_\mathrm{B}$, and in the end, a second Ramsey pulse, identical to the first one, was applied, see \TGasFig{pulse}.

The length of the pulses were sufficiently short such that the HFB mean-field equations yield results quantitatively close to measurements \cite{TGas:Kokkelmans2002a,TGas:Mackie2002b,TGas:Kohler2003a,TGas:Goral2004b}.
Before the effects of multiple collisions and higher correlations become important a certain time after the quench a description within mean-field approximation remains valid.
The calculations showed that the two Ramsey pulses coupled the colliding BEC atoms into the uppermost bound state of the two-channel system and coherently transferred atom pairs into molecules and vice versa.
Due to the short pulse times, the coupling evolution resembled that of the fraction of a Rabi oscillation between two energy eigenstates.
During the intermediate quasi free evolution, atoms and molecules could evolve such that a relative phase built up which was given by the binding energy relative to the free atoms multiplied by $t_\mathrm{evolve}$.
Depending on the value of this phase, the second Ramsey pulse lead to a further production of molecules or dissociated the previously formed dimers.
As a result, sinusoidal oscillations of the remaining fraction of atoms at the end of the pulse sequence were observed as a function of $t_\mathrm{evolve}$, see \TGasFig{MolOsc}.
The frequency of these oscillations precisely reproduced the expected binding energy of the Rubidium dimers \cite{TGas:Goral2004b}.

However, discrepancies between theory and experiment remained in other cases, in particular for longer evolution times under strong interactions \cite{TGas:Kohler2004a}, which may indicate that descriptions beyond mean field are required to interpret experimental data.
We will discuss such methods in the following sections.

Feshbach resonances have become, for experimenters, a versatile and convenient tool to control the collisional interactions.
The have, in particular, opened the door to the exploration of rich physics in parameter regimes never explored before.
For example, ultracold atomic Fermi gases can be manipulated such that they not only show the superconductor-like properties known from electron gases in solids but can cross from a BCS-like state containing Cooper pairs, over into a BEC of tightly bound molecules (For recent reviews cf.~Refs.~\cite{TGas:Chen2005a,TGas:Gurarie2007a,TGas:Inguscio2008a}).
As experiments are usually conducted in more than one spatial dimension and since the dynamics of Fermion gases can not be simulated by means of classical equations of motion, the functional field theoretical methods to be described in the following sections are expected to represent the most promising theoretical approach to the dynamics of strongly correlated Fermi gases beyond mean-field theory.

\subsubsection{Ultracold gases in lower dimensional traps and optical lattices}
\label{TGas:sec:LowDimTraps}
One- and two-dimensional traps \cite{TGas:Pitaevskii2003a,TGas:Stoferle2004a,TGas:Kinoshita2006a,TGas:Schmiedmayer2000a,TGas:Schreck2001a,TGas:Gorlitz2001a,TGas:Greiner2001a,TGas:Moritz2003a,TGas:Tolra2004a,TGas:Kinoshita2004a,TGas:Paredes2004a} as well as optical lattices \cite{TGas:Jaksch1998a,TGas:Bloch2004a,TGas:Morsch2006a} allow to realise strongly correlated many-body states of atoms. 
In an optical lattice, strong effective interactions can be induced by suppressing the hopping between adjacent lattice sites and thus increasing the weight of the interaction relative to the kinetic energy \cite{TGas:Jaksch1998a,TGas:vanOosten2001a}. 
This leads, in the limit of near-zero hopping or strong interactions, to a Mott-insulating state \cite{TGas:Greiner2002a}.
It is beyond the scope of this article to discuss in more detail the theory of ultracold gases trapped in such special configurations.
Some remarks concerning one-dimensional (1D) gases, though, are in order, as we will focus on such a system when applying the functional field-theory methods in later sections.

In special cases, the models describing 1D gases \cite{TGas:Lieb1963a,TGas:Girardeau1960a,TGas:McGuire1964a,TGas:Korepin1997a} allow to determine exact time-dependent solutions of the Schr\"odinger equation \cite{TGas:Girardeau2000b,TGas:Girardeau2003a} providing insight beyond various approximations, which is particularly important in strongly correlated regimes. 
These 1D systems are experimentally realized with atoms tightly confined in effectively 1D waveguides \cite{TGas:Stoferle2004a,TGas:Kinoshita2006a,TGas:Schreck2001a,TGas:Gorlitz2001a,TGas:Greiner2001a,TGas:Moritz2003a,TGas:Tolra2004a,TGas:Kinoshita2004a,TGas:Paredes2004a}, where nonequilibrium dynamics is considerably affected by the kinematic restrictions of the geometry \cite{TGas:Kinoshita2006a}, while quantum effects are enhanced \cite{TGas:Olshanii1998a,TGas:Petrov2000b,TGas:Dunjko2001a}. 
1D Bose gases are explored for various interaction strengths, from the Lieb-Liniger (LL) gas with finite coupling \cite{TGas:Schreck2001a,TGas:Gorlitz2001a,TGas:Greiner2001a,TGas:Kinoshita2006a} up to the so-called Tonks-Girardeau (TG) regime of ``impenetrable-core'' bosons \cite{TGas:Tonks1936a,TGas:Girardeau1960a,TGas:Kinoshita2004a,TGas:Paredes2004a,TGas:Kinoshita2006a}. 
The 1D gas enters the TG regime if the dimensionless interaction parameter $\gamma=g_\mathrm{1D}m V/(\hbar^2 N)$ is much larger than one.
Here, $g_\mathrm{1D}$ is the coupling parameter of the one-dimensional gas, e.g., $g_\mathrm{1D}=2\hbar^2a/(ml_\perp^2)$ for a cylindrical trap with transverse harmonic oscillator length $l_\perp$ \cite{TGas:Pitaevskii2003a}. 
In the Tonks-Girardeau limit $\gamma\to\infty$ the atoms can no longer pass each other and behave in many respects like a one-dimensional ideal Fermi gas \cite{TGas:Pitaevskii2003a}.

\subsubsection{Exact dynamics of an interacting 1D Bose gas}
\label{TGas:sec:ExactDyn}
Most theoretical studies of the exact time-dependence address the Tonks-Girardeau (TG) regime 
\cite{TGas:Girardeau2000b,TGas:Girardeau2000a,TGas:Ohberg2002a,TGas:Rigol2005a,TGas:Minguzzi2005a,TGas:DelCampo2006a,TGas:Rigol2007a,TGas:Pezer2007a}). 
In this limit, the complex many-body problem is considerably simplified due to the Fermi-Bose mapping property where dynamics follows a set of uncoupled single-particle Schr\" odinger equations \cite{TGas:Girardeau2000b}. 
A method for calculating the time-evolution of a LL gas with \textit{finite} interaction strength has recently been discussed in Refs.~\cite{TGas:Buljan2007a,TGas:Jukic2008a}. 
It has the potential to yield a valuable comparison to the results obtained from solving the dynamical field equations presented in the following sections.
We therefore devote this subsection to a brief excursion and outline the method which generalises Lieb and Liniger's exact diagonalisation method \cite{TGas:Lieb1963a} to time evolving $N$-particle quantum mechanical wave functions.

We consider the dynamics of $N$ indistinguishable $\delta$-interacting bosons in a 1D geometry \cite{TGas:Lieb1963a}.
The Schr\" odinger equation for this system is usually written as 
\begin{equation}
  i \partial_t\psi_B
  =-\sum_{i=1}^{N}\frac{\partial^2 \psi_B}{\partial x_i^2}+
    \sum_{1\leq i < j \leq N} 2c\,\delta(x_i-x_j)\psi_B, 
\label{TGas:LLmodel}
\end{equation}
where $\psi_B(x_1,\ldots,x_N,t)$ is the many-body wave function, and $c$ quantifies the strength of the interaction, which is related to the dimensionless 1D interaction parameter $\gamma=2c$ introduced above.
We do not impose any boundary conditions, i.e.~, the $x$-space is infinite which corresponds to a number of interesting experimental situations where the gas is initially localized within a certain region of space and then allowed to freely evolve \cite{TGas:Ohberg2002a,TGas:Rigol2005a,TGas:Minguzzi2005a,TGas:DelCampo2006a}. 

The idea is to construct exact solutions by differentiating a fully antisymmetric (fermionic) time-dependent wave function, which obeys the Schr\" odinger equation for a free Fermi gas \cite{TGas:Gaudin1983a}. 
The differential operator used for this depends on the interaction strength $c$ and the number of particles. 
When $c\rightarrow \infty$, the scheme reduces to Girardeau's time-dependent Fermi-Bose mapping \cite{TGas:Girardeau2000b}, valid for "impenetrable-core" bosons. 

Due to the Bose symmetry, it is sufficient to express the wave function $\psi_B$ in a single permutation sector of the configuration space, $R_1:x_1<x_2<\ldots<x_N$. 
Within $R_1$, $\psi_B$ obeys 
\begin{equation}
  i\partial_t \psi_B 
  = -\sum_{i=1}^{N} \partial^2 \psi_B/ \partial x_i^2,
\label{TGas:eq:free}
\end{equation} 
while interactions impose boundary conditions at the borders of $R_1$ \cite{TGas:Lieb1963a}:
\begin{equation}
  \left [
  1-\frac{1}{c}
  \left ( 
  \frac{\partial}{\partial x_{j+1}}-\frac{\partial}{\partial x_j}
  \right)
  \right]_{x_{j+1}=x_j}\psi_B=0.
\label{TGas:eq:interactions}
\end{equation}
This constraint creates a cusp in the many-body wave function when two particles touch, which should be present at any time during the dynamics. 
In the TG limit (i.e., when $c\rightarrow \infty$) the cusp condition is $\psi_B(x_1,\ldots,x_j,x_{j+1},\ldots,x_N,t)|_{x_{j+1}=x_j}$ $=0$ \cite{TGas:Girardeau1960a,TGas:Girardeau2000b}, which is trivially satisfied by an antisymmetric fermionic wave function $\psi_F(x_1,\ldots,x_N,t)$.
Hence, $\psi_B=\psi_F$ within $R_1$, which is the famous Fermi-Bose mapping \cite{TGas:Girardeau1960a,TGas:Girardeau2000b}. 
In many physically interesting cases, $\psi_F$ can be constructed as a Slater determinant 
\begin{equation}
  \psi_F(x_1,\ldots,x_N,t)
  =(N!)^{-\frac{1}{2}}
  \det[\phi_m(x_j,t)]_{m,j=1}^{N}.
\label{TGas:eq:Slater}
\end{equation}
Since $\psi_B=\psi_F$ within $R_1$, $\psi_F$ must obey $i\partial \psi_F /\partial t=-\sum_{j=1}^{N} \partial^2 \psi_F/ \partial x_j^2$, which implies that the (orthonormal) single-particle wave functions $\phi_m(x_j,t)$ evolve according to  
\begin{equation}
  i \partial \phi_m / \partial t
  =- \partial^2 \phi_m / \partial x^2;
\label{TGas:eq:master}
\end{equation}
$m=1,\ldots,N$. 
Thus, in the TG limit, the complexity of the many-body dynamics is reduced to solving a simple set of uncoupled single-particle equations, while the interaction constraint \TGaseq{interactions} is satisfied by the Fermi-Bose construction. 

The simplicity and success of this idea motivates us to choose an ansatz which automatically satisfies constraint \TGaseq{interactions} for any finite $c$ \cite{TGas:Gaudin1983a,TGas:Korepin1997a}. 
For this, define a differential operator 
\begin{equation}
  \hat O=\prod_{1\leq i < j \leq N} \hat B_{ij},
\label{TGas:eq:oO}
\end{equation}
where $\hat B_{ij}$ stands for 
\begin{equation}
  \hat B_{ij}
  =\left[
  1+\frac{1}{c}
  \left(
  \frac{\partial}{\partial x_{j}}-
  \frac{\partial}{\partial x_{i}}
  \right)
  \right]
\label{TGas:eq:Bij}.
\end{equation}
It can be shown that the wave function 
\begin{equation}
  \psi_{B}
  = {\mathcal N}_{c} \hat O \psi_F
  \mbox{ (inside $R_1$)},
\label{TGas:eq:ansatz}
\end{equation}
where ${\mathcal N}_{c}$ is a normalization constant, obeys the cusp condition \TGaseq{interactions} by construction \cite{TGas:Gaudin1983a,TGas:Korepin1997a}: 
Consider an auxiliary wave function 
\begin{eqnarray}
  \psi_\mathrm{AUX}(x_1,\ldots,x_N,t) 
  & = & \hat B_{j+1,j} \hat O \psi_F 
  \nonumber \\
  & = & \hat B_{j+1,j} \hat B_{j,j+1}\hat O'_{j,j+1} \psi_F,
\end{eqnarray}
where the primed operator $\hat O'_{j,j+1}=\hat O/\hat B_{j,j+1}$ omits the factor $\hat B_{j,j+1}$ as compared to $\hat O$. 
The auxiliary function can be written as 
\begin{equation}
  \psi_\mathrm{AUX}
  = \left [
  1-\frac{1}{c^2}
  \left ( 
  \frac{\partial}{\partial x_{j+1}}-\frac{\partial}{\partial x_j}
  \right)^2
  \right] \hat O'_{j,j+1} \psi_F.
\label{TGas:eq:AuxAsym}
\end{equation}
It is straightforward to verify that the operator $\hat B_{j+1,j} \hat B_{j,j+1}\hat O'_{j,j+1}$ in front of $\psi_F$ is invariant under the exchange of $x_j$ and $x_{j+1}$ \cite{TGas:Korepin1997a}. 
On the other hand, the fermionic wave function $\psi_F$ is antisymmetric with respect to the interchange of $x_j$ and $x_{j+1}$. 
Thus, $\psi_\mathrm{AUX}(x_1,\ldots,x_j,x_{j+1},\ldots,x_N,t)$ is antisymmetric with respect to the interchange of $x_j$ and $x_{j+1}$, which leads to
\begin{equation}
  \psi_\mathrm{AUX}(x_1,\ldots,x_j,x_{j+1},\ldots,x_N,t)|_{x_{j+1}
  =x_{j}}=0.
\end{equation}
This is fully equivalent to the cusp condition \TGaseq{interactions}, $\hat B_{j+1,j} \psi_B|_{x_{j+1}=x_{j}}=0$. 
Thus, the wave function \TGaseq{ansatz} obeys constraint \TGaseq{interactions} by construction. 

In order to exactly describe the dynamics of LL gases, the wave function \TGaseq{ansatz} should also obey \TGasEq{free} inside $R_1$. 
From the commutators $[\partial^2/\partial x_j^2,\hat O]=0$ and $[i\partial/\partial t,\hat O]=0$ it follows that if $\psi_F$ is given by \TGasEq{Slater} and the $\phi_m(x_j,t)$ obey \TGasEq{master}, then $\psi_B$ obeys \TGasEq{free}. 
Note that for $c\rightarrow\infty$, one recovers Girardeau's Fermi-Bose mapping \cite{TGas:Girardeau1960a,TGas:Girardeau2000b}, i.e., $\hat O=1$. 

The above procedure can now be used to calculate the time evolution of a bosonic wave function which at time $t=0$ is given in terms of the operator $\hat O_c$ acting on some localised fermionic function.
In order to calculate correlation functions, e.g., the single-particle density matrix $n$, \TGasEq{SPDM}, from this wave function, integrations over $N-2$ variables must be performed.
This can be achieved, at least for the diagonal elements of $n$, using Monte-Carlo techniques.
We point out that the above method has the advantage that it does not rely on calculating the LL energy eigenfunctions of the system nor on the determination of the projection of the initial wave function on these eigenfunctions.
The initial state is rather constructed from a localised Fermi wave function.
This approach suffices for making benchmark comparisons with approximative methods as those studied in the following.

We finally point out that for both, strongly correlated Bose and Fermi gases, methods inspired by quantum information theory have recently been developed, with much effort and success, for the description of nonequilibrium dynamics.
These time-dependent Density-Matrix Renormalisation-Group (tDMRG) methods draw from the empirical observation that the fraction of Hilbert space which becomes relevant during the time evolution of a generic many-body system scales only polynomially with the number of particle in contrast to the whole space which grows exponentially.
For details on these techniques see e.g.~Refs.~\cite{TGas:Vidal2004a,TGas:Daley2004a,TGas:White2004a,TGas:Manmana2005a}.

\section{Nonequilibrium quantum field theory}
\label{TGas:sec:NEqFT}
The following section intends to give a concise introduction to the functional integral formulation of nonequilibrium many-body dynamics.
The above brief discussion of mean-field dynamics has already given a taste of the formalism to be developed and made clear that the focus needs to be set on an efficient handling of the interactions which form the basis as well as the crucial difficulty of the whole theory.

Assuming basic knowledge about Feynman path integrals we will sketch the functional integral formulation of real-time quantum field theory (QFT) and, in particular, introduce the one-particle irreducible (1PI) effective action.
Eventually, to derive many-body dynamic equations beyond mean-field order which conserve crucial quantities like energy and particle number we will make use of an extended method which is based on the two-particle irreducible (2PI) effective action.
We will consider initial-value problems in QFT which naturally lead to the concept of the Schwinger-Keldysh closed time path (CTP).
The section will be closed with an application of the so derived equations of motion to describe equilibration of a Bose gas in one spatial dimension.

\subsection{Functional-integral approach}
\label{TGas:sec:FInt}
We briefly recall the picture of quantum mechanics as provided by Feynman's path integral formulation.
A common example is the time-evolution of a quantum mechanical state as exhibited by the transition amplitude from some initial state $|t_\mathrm{ini}\rangle$ to a final state $|t_\mathrm{fin}\rangle$.
Given a classical action depending on the generalised coordinate $\varphi(t)$ defined at each intermediate time step $t$ between $t_\mathrm{ini}$ and $t_\mathrm{fin}$, the transition matrix element can be expressed in the path integral form
\begin{equation}
\label{TGas:eq:QMPI}
  \langle t_\mathrm{fin}|t_\mathrm{ini}\rangle
  = \int{\cal D}\varphi\,e^{iS[\varphi]/\hbar},
\end{equation}
where ${\cal D}\varphi = \prod_{t=t_\mathrm{ini}}^{t_\mathrm{fin}}d\varphi(t)$, and the initial and final states are reflected in the fixed boundary values\footnote{We use the letter $\varphi$ instead of $x$ having in mind quantum mechanics as the $0+1$ dimensional special case of field theory.} $\varphi(t_{\mathrm{ini}})=\langle\varphi|t_\mathrm{ini}\rangle$, $\varphi(t_{\mathrm{fin}})=\langle\varphi|t_\mathrm{fin}\rangle$.
The classical action
\begin{equation}
\label{TGas:eq:SClitoL}
  S[\varphi] = \int_{t_\mathrm{ini}}^{t_\mathrm{fin}}dt\, L(\varphi,\dot\varphi)
\end{equation}
defines the dynamical process in terms of the time interval to be considered and the Lagrangian $L$ related to the Hamiltonian by a Legendre transform.
The classical dynamics is determined through Hamilton's principle
\begin{equation}
\label{TGas:eq:HamPrin}
  \delta S[\phi]=0,
\end{equation}
which leads to the Euler-Lagrange, i.e., the sought dynamic equation for $\varphi$.\footnote{In deriving the Euler-Lagrange equation the variation of the coordinate $\varphi$ is usually taken to vanish at the boundaries of the time interval $[t_\mathrm{ini},t_\mathrm{fin}]$.
This procedure applies to systems with differential evolution equations of second order in time. For dynamic equations of first order in time, as the GPE, care needs to be taken when using the path integral for initial value problems, see, e.g., Ref.~\cite{TGas:Ebert1994a}}
For instance, given particular initial values for the position and velocity of the child on the slide shown in \TGasFig{Slide} at $t=t_\mathrm{ini}$, this equation has the thick solid (red) path as solution.
Different paths require in general different initial conditions to be imposed.

On scales where quantum effects become relevant, the real world is somewhat more intricate.
Fluctuations around the classical path as depicted by the thin (black) solid lines in \TGasFig{Slide} imply the action $S[\varphi]$ to deviate from its classical extremal value, and, only if this deviation is larger than $\hbar$, the phase factor $\exp\{iS[\varphi]/\hbar\}$ suppresses the contributions of such paths to the integral through destructive interference.
Qualitatively new effects are in order like the ``quantum child'' which can tunnel through the edge of the slide as along the (yellow) path in \TGasFig{Slide}.

\begin{figure}[tb]
\begin{center}
\begin{minipage}[b]{0.47\columnwidth}
\resizebox{1.0\columnwidth}{!}{
\includegraphics{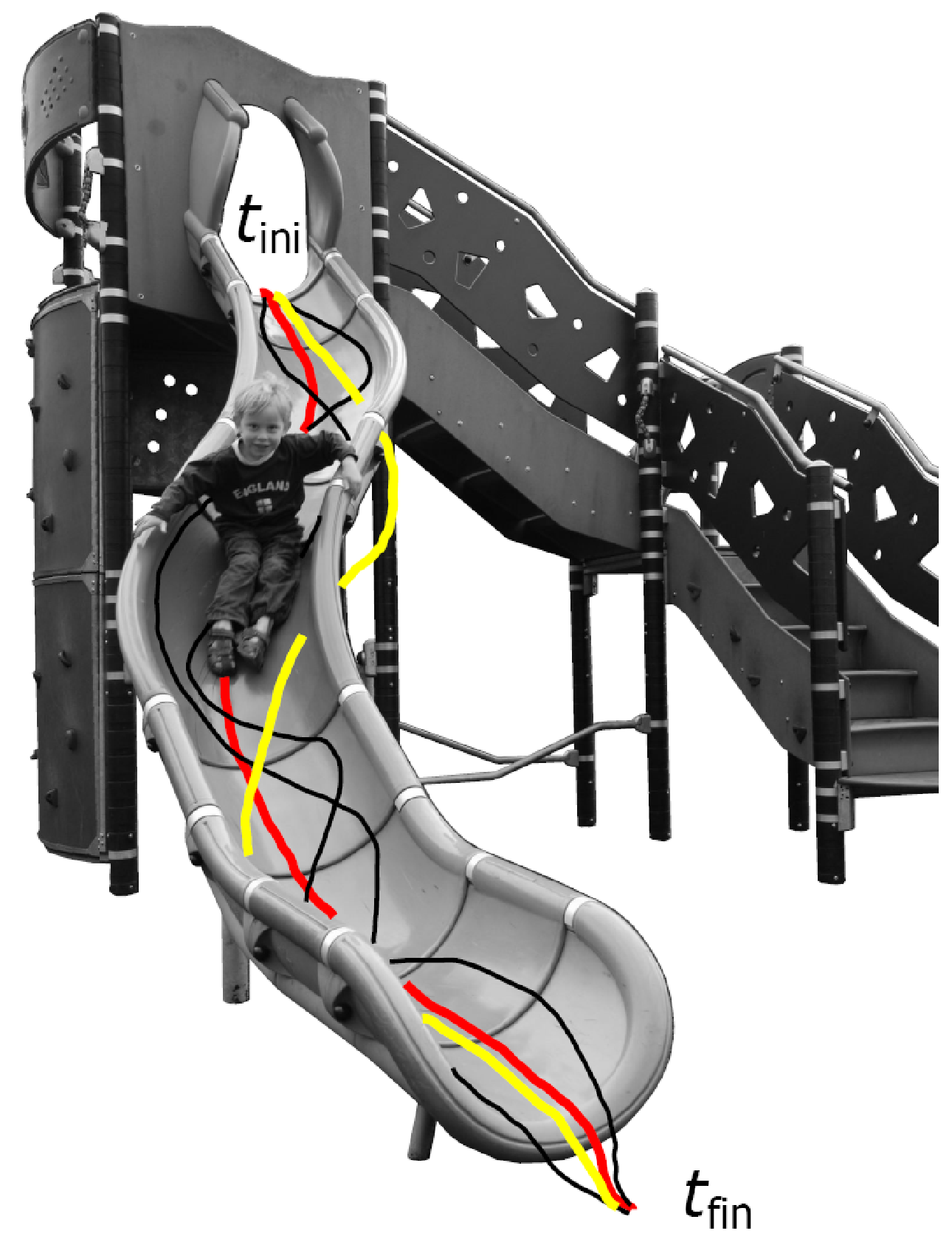}
}
\end{minipage}
\hspace*{0.05\columnwidth}
\begin{minipage}[b]{0.45\columnwidth}
\caption{
(Color online) Classical vs. quantum mechanics. The classical path for given boundary conditions at $t_\mathrm{ini}$ and/or $t_\mathrm{fin}$ is shown as thick (red) line.
The thin (black) paths would require, e.g., different initial values for $\varphi$, $\dot\varphi$.
In the microscopic world, the thin (black) paths add constructively to the path integral if their action $S[\varphi]$ deviates less than $\hbar$ from the extremal value corresponding to the classical path.
Also tunneling processes as indicated by the thick (yellow) line would add constructively to the integral.
}
\label{TGas:fig:Slide}
\end{minipage}
\end{center}
\end{figure}
We generalise this path-integral formulation to QFT, where the coordinates $\varphi$ become fields $\varphi(x)$ defined over time and space.
Moreover, we introduce external classical, i.e., non-fluctuating sources $J(x)$ to turn the path integral into a generating functional for correlation functions, similarly as in the (grand) canonical partition function in equilibrium physics. 
This generating functional reads  
\begin{equation}
\label{TGas:eq:ZJ}
  Z[J] = \int{\cal D}\varphi\, e^{i(S[\varphi]+\int J\varphi)}
\end{equation}
Here and in the following we shall use, if not explicitly stated otherwise, natural units, with $\hbar=1$.
We use the short-hand notation $\int J\varphi=\int_{\cal C} \mathrm{d}^{d+1}x J(x)\varphi(x)=\int_{t_\mathrm{ini}}^{t_\mathrm{fin}}\mathrm{d}x_0\int \mathrm{d}^dx J(x)\varphi(x)$, ${\cal C}=[t_\mathrm{ini},t_\mathrm{fin}]$.
For instance, it allows the field expectation value $\phi=\langle\Phi\rangle$ to be written as
\begin{equation}
\label{TGas:eq:phifromZ}
  \phi(x) = \left.\frac{\delta W[J]}{\delta J(x)}\right|_{J=0}
  = Z^{-1}\int{\cal D}\varphi\, \varphi(x)\,e^{iS[\varphi]},
\end{equation}
where $W[J]=-i\ln Z[J]$ is the Schwinger functional. 
We introduce the quantum effective action $\Gamma[\phi]$ by demanding that the full quantum dynamics of the field expectation value $\phi$ is given by Hamilton's principle applied to $\Gamma$, 
\begin{equation}
\label{TGas:eq:HamiltonGamma}
  \delta\Gamma[\phi]=0.
\end{equation}
This is equivalent to the functional measure containing a functional delta distribution which evaluates the fluctuating field to the expectation value $\phi$ which is implicitly defined by \TGasEq{phifromZ},
\begin{equation}
\label{TGas:eq:ZwithGamma}
  Z[J] = \int{\cal D}\varphi\,\delta[\varphi-\phi]\,
         e^{i(\Gamma[\varphi]+\int J\varphi)},
\end{equation}
with $\delta[\varphi-\phi]=\prod_{x}\delta(\varphi(x)-\phi(x))$. 
Evaluating the functional integral shows that the 1PI effective action defined as
\begin{equation}
\label{TGas:eq:Gamma1PILegTrafo}
  \Gamma[\phi] = W[J]-\int J\phi
\end{equation}
satisfies \TGasEq{ZwithGamma}, where it is implied that $J$, by inverting \TGasEq{phifromZ}, can be expressed in terms of $\phi$.
\TGasEq{Gamma1PILegTrafo} shows that the 1PI effective action is the Legendre transformation of the Schwinger functional with respect to the classical source field $J(x)$.
While \TGasEq{phifromZ} defines the classical field as the derivative of $W$, the inverse expression \TGaseq{HamiltonGamma} of $J$ as a derivative of $\Gamma$ constitutes the dynamic equation for $\phi$ in the presence of the source $J$.

$Z$ is the generating functional for time-ordered $n$-point correlation functions,
\begin{equation}
\label{TGas:eq:nCFfromZ}
  \langle{\cal T}_{\cal C}\Phi(x_1)\cdots\Phi(x_n)\rangle
  = \frac{1}{Z[J]}\left.\frac{\delta^nZ[J]}
                {i\delta J(x_1)\cdots i\delta J(x_n)}\right|_{J=0},
\end{equation}
The Schwinger functional, in turn, generates the connected correlation functions or $n$-point cumulants.
Hence, the connected propagator $G$ defined in \TGasEq{Gab} is obtained as 
\begin{equation}
\label{TGas:eq:GfromW}
  G(x,y) = -i\left.\frac{\delta W[J]}{\delta J(x)\delta J(y)}\right|_{J=0}.
\end{equation}

We briefly summarise the leading-order loop expansion of the 1PI effective action $\Gamma[\phi]$.
If the expansion of the classical action in powers of the field contains terms at most quadratic in $\varphi$ and its derivatives, i.e., if there are no interactions present, the functional integral is Gaussian and therefore can be evaluated analytically.
It is convenient to perform the shift $\varphi=\phi+\tilde\varphi$ and evaluate the Gaussian integrals over the fluctuation fields $\tilde\varphi$.
This yields the 1PI effective action up to one-loop order,
\begin{equation}
\label{TGas:eq:1PI1loop}
  \Gamma^\mathrm{(1loop)}[\phi]
  = S[\phi] - \frac{i}{2}\mathrm{Tr}\ln G_0,
\end{equation}
where the classical two-point function $G_0$ is the inverse of
\begin{equation}
\label{TGas:eq:G0}
  iG^{-1}_0(x,y)=i\frac{\delta^2 S[\phi]}{\delta\phi(x)\delta\phi(y)}.
\end{equation}
In the case that interactions, i.e., terms of cubic and higher order are present in the classical action $S[\varphi]$, \TGasEq{1PI1loop} represents the leading-order approximation to $\Gamma$ if the corrections resulting from the interactions are sufficiently small.
In this case the perturbative corrections can be calculated as a series of Feynman diagrams.
We finally note that the second derivative of $\Gamma[\phi]$ with respect to $\phi$,
\begin{equation}
\label{TGas:eq:DefSigma}
  -i\frac{\delta^2\Gamma[\phi]}{\delta\phi(x)\delta\phi(y)}
  = G^{-1}(x,y)
  = G_0^{-1}(x,y)-\Sigma(x,y),
\end{equation}
defines the proper self energy $\Sigma$ to which only 1PI diagrams contribute.

\subsection{The two-particle irreducible (2PI) effective action}
\label{TGas:sec:2PIEA}
The above steps set the framework of the functional approach to dynamics we choose in the following. 
We will develop it slightly further and use the two-particle irreducible (2PI) effective action instead of the 1PI one.
The reason is that we are interested, as in the mean-field theory, in the time evolution of the two-point Greens function $G(x,y)$.
It is therefore desirable to obtain the respective dynamic equation in the same way as that for $\phi$, as an Euler-Lagrange equation from Hamilton's principle applied to an action functional.
The 2PI effective action which is a functional of both, $\phi$ and $G$ fulfills these requirements.
Imagine it would be possible to calculate the 1PI and 2PI effective actions exactly, e.g. by evaluating their loop expansions to infinite order.
Results obtained from either of them would be identical. 
Their truncated expansions, however, are in general inequivalent.
The loop expansion of the 2PI effective action can be understood as the sum of the diagrams contributing to the 1PI effective action, reordered such that it can be expressed in terms of  full propagators $G$ which themselves can be represented as an infinite sum of diagrams involving only vertices and free propagators $G_0$.

The 2PI effective action has been introduced to solid-state theory (there called $\Phi$-functional) in the sixties \cite{TGas:Luttinger1960a,TGas:Baym1962a} and was later given its name in a relativistic quantum field theoretical formulation \cite{TGas:Cornwall1974a}.
See also Refs.~\cite{TGas:Vasiliev1998a,TGas:Kleinert1982a}.
The nonperturbative 2PI $1/{\cal N}$ approximation discussed in \TGasSect{NLO1N} below was introduced in Refs.~\cite{TGas:Berges:2001fi,TGas:Aarts:2002dj}.
Detailed studied and applications to scalar relativistic as well as gauge theories can be found in Refs.~\cite{TGas:Berges:2001fi,TGas:Berges:2002cz,TGas:Mihaila:2000sr,TGas:Cooper:2002qd,TGas:Arrizabalaga:2004iw,TGas:Berges:2002wr,TGas:Berges:2004ce,TGas:Berges:2004pu,TGas:Berges:2004yj,TGas:Arrizabalaga2004a,TGas:Aarts:2006pa}, to non-relativistic systems, in particular, ultracold gases in Refs.~\cite{TGas:Rey2004a,TGas:Baier:2004hm,TGas:Gasenzer:2005ze,TGas:Rey2005a,TGas:Temme2006a,TGas:Berges:2007ym,TGas:Branschadel2008a}.

To be specific, we consider a quantum field theory for a real $\cal N$-component scalar field $\varphi_a(x)$ ($a=1,...,{\cal N}$) with quartic interactions.
Its classical action reads
\begin{equation}
  S[\varphi]
  = \frac{1}{2}\int_{x y} \varphi_a(x) iD_{ab}^{-1}(x,y)\varphi_b(y)
    -\frac{g}{4{\cal N}} \int_{x} \varphi_a(x)\varphi_a(x)\varphi_b(x)\varphi_b(x) ,
\label{TGas:eq:Sclassphi4}
\end{equation}
with $\int_x \equiv \int \mathrm{d} x_0 \int \mathrm{d}^d x$.
As we shall focus on a single-species ultracold Bose gas described by a complex scalar field, we choose ${\cal N}=2$.
The free inverse classical propagator then reads, in the basis where the field indices number the real and imaginary parts, $\varphi=(\varphi_1+i\varphi_2)/\sqrt{2}$,
\begin{equation}
\label{TGas:eq:G0inv}
  iD^{-1}_{ab}(x,y)
  =\left.iG^{-1}_{0,ab}(x,y)\right|_{\phi=0}
  = \delta(x-y)\left[-i\sigma^2_{ab}\partial_{x_0}
   -H_\mathrm{1B}(x)\delta_{ab}
     \right],
\end{equation}
see \TGasEq{G0}.
Here $H_\mathrm{1B}(x)=-\partial^2_i/2m+V(x)$ denotes the
single-particle Hamiltonian with interaction potential $V(x)$ and
$
  \sigma^2 
$
the Pauli matrix in field-index space. 
Summation over double indices $a,b,c \ldots= 1,2$ is implied.
With this, the action \TGaseq{Sclassphi4} corresponds to the Hamiltonian \TGaseq{GPH}, for a contact interaction potential $V(r)=g\delta(r)$.
The Bose field commutators in this basis read
\begin{equation}
\label{TGas:eq:BoseCR12}
  [{\Phi}_a(t,\vec{x}),{\Phi}_b(t,\vec{y})]
  =-\sigma^2_{ab}\delta(\vec{x} -\vec{y}).
\end{equation}

The definition of the 2PI effective action is based on a generating functional
\begin{equation}
\label{TGas:eq:ZJR}
  Z[J,R] = \exp(iW[J,R])
         = \int{\cal D}\varphi\,
           e^{i(S[\varphi]+\int_x J_a(x)\varphi_a(x)+\frac{1}{2}\int_{xy}\varphi_a(x)R_{ab}(x,y)\varphi_b(y))}.
\end{equation}
depending on the one- and two-point sources $J_a(x)$ and $R_{ab}(x,y)$, respectively.
The field expectation value and the connected two-point function in the presence of sources are defined in accordance with the source-free case,
\begin{equation}
\label{TGas:eq:phiGfromZJR}
  \frac{\delta W[J,R]}{\delta J_a(x)}=\phi_a(x),~~~
  \frac{\delta W[J,R]}{\delta R_{ab}(x,y)}=\frac{1}{2}\Big(\phi_a(x)\phi_b(y)+G_{ab}(x,y)\Big).
\end{equation}
The additional source term $\propto R$ can be seen as a quadratic (mass) term modifying $S[\varphi]\to S^R[\varphi]=S[\varphi]+\int\varphi R\varphi$, such that the 1PI effective action in the presence of $R$ reads, to 1-loop order, cf.~\TGasEq{1PI1loop},
\begin{equation}
\label{TGas:eq:GammaR1loop}
  \Gamma^{R\mathrm{(1loop)}}[\phi]
  =S^R[\phi]+\frac{i}{2}\mathrm{Tr}\ln\left[G_0^{-1}(\phi)-iR\right].
\end{equation}

In order to arrive at the 2PI effective action one performs a further Legendre transform of $\Gamma^R[\phi]$ with respect to the source $R$,
\begin{eqnarray}
  \Gamma[\phi,G]
  &=&\Gamma^R[\phi]-\int_{xy}\frac{\delta\Gamma^R[\phi]}{\delta R_{ab}(x,y)}R_{ba}(y,x)
  \nonumber\\
  &=&\Gamma^R[\phi]-\int_{xy}\phi_a(x)R_{ab}(x,y)\phi_b(y)-\frac{1}{2}\mathrm{Tr}RG.
\label{TGas:eq:LTrafoGammaR}
\end{eqnarray}
Here we have used that
\begin{equation}
\label{TGas:eq:dGammadRequalsdWdR}
  \frac{\delta\Gamma^R[\phi]}{\delta R_{ab}(x,y)}
  =\frac{\delta W[J,R]}{\delta R_{ab}(x,y)}
   +\int_z\left(\frac{\delta W[J,R]}{\delta J_c(z)}-\phi_c(z)\right)\frac{\delta J_c(z)}{\delta R_{ab}(x,y)}
  =\frac{\delta W[J,R]}{\delta R_{ab}(x,y)},
\end{equation}
and its relation to $G$ in \TGasEq{phiGfromZJR}. 
From the double Legendre transform \TGaseq{LTrafoGammaR}, \TGaseq{Gamma1PILegTrafo}, one directly finds the stationary conditions for $\phi$ and $G$ which, in the absence of sources, will provide us with the dynamic equations we are heading for:
\begin{equation}
\label{TGas:eq:StatCondsphiG}
    \frac{\delta\Gamma[\phi,G]}{\delta\phi_a(x)}
  = -J_a(x)-\int_y R_{ab}(x,y)\phi_b(y),~~~
  \frac{\delta\Gamma[\phi,G]}{\delta G_{ab}(x,y)}
  = -\frac{1}{2}R_{ab}(x,y).
\end{equation}
Before proceeding to these we derive the loop expansion of $\Gamma[\phi,G]$ which provides a way to obtain approximated expressions for the action and to perform practical calculations.

Plugging \TGasEq{GammaR1loop} into \TGasEq{LTrafoGammaR} yields the one-loop part
\begin{equation}
\label{TGas:eq:Gamma2PI1loop}
  \Gamma^\mathrm{(1loop)}[\phi,G]
  = S[\phi]+\frac{i}{2}\mathrm{Tr}\left(\ln G^{-1}+G_0^{-1}G\right)+\mathrm{const.}
\end{equation}
of the 2PI effective action
\begin{equation}
\label{TGas:eq:Gamma2PI}
  \Gamma[\phi,G]=\Gamma^\mathrm{(1loop)}[\phi,G]+\Gamma_2[\phi,G],
\end{equation}
where we have used \TGasEq{DefSigma} in leading-order, $G^{-1}=G_0^{-1}-iR$.
Including all higher-loop terms in the ``rest'' $\Gamma_2$ in \TGasEq{Gamma2PI} one finds, varying $\Gamma$ with respect to $G$, that
\begin{equation}
\label{TGas:eq:SigmafromGamma2}
  \Sigma_{ab}(x,y;\phi,G)=2i\frac{\delta\Gamma_2[\phi,G]}{\delta G_{ab}(x,y)}.
\end{equation}
%
\begin{figure}[tb]
\begin{center}
\begin{minipage}{0.4\columnwidth}
\resizebox{1.0\columnwidth}{!}{
\includegraphics{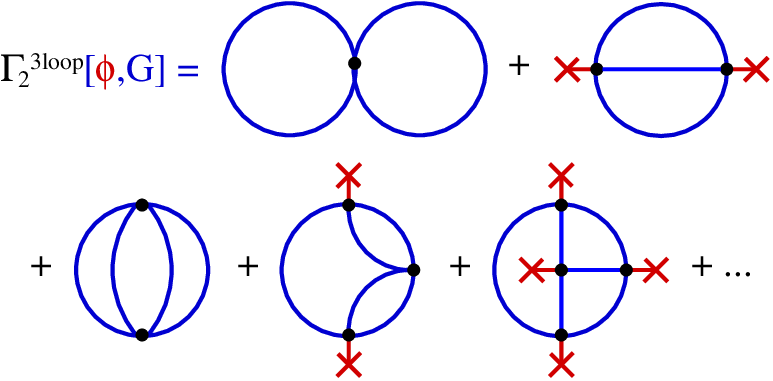}
}
\end{minipage}
\hspace*{0.05\columnwidth}
\begin{minipage}{0.4\columnwidth}
\resizebox{1.0\columnwidth}{!}{
\includegraphics{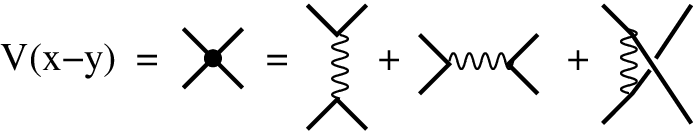}
}
\end{minipage}
\end{center}
\caption{
(Color online) (Left panel) Diagrammatic representation of the two- and three-loop diagrams contributing to the 2PI part $\Gamma_2[\phi,G]$ of the 2PI effective action, cf.~\TGasEq{Gamma2PI}.
The bare vertices are drawn as black dots.
Each such vertex is understood to represent a sum of the topologically different terms shown in the right panel.
At each vertex, it is summed over double field indices and double space-time variables according to the respective diagram in 
the right panel.
(Right panel)
The representation of the bare vertex in terms of a black dot stands for a sum of the three topologically different connections of the four `corners'.
The black lines do not represent propagators and are only drawn in order to illustrate the different possible connections of propagators and/or external fields at the vertices. 
}
\label{TGas:fig:Gamma23loop}
\end{figure}
Since the self energy $\Sigma$ is 1PI, and since taking the derivative with respect to $G$ corresponds to opening a propagator line, it follows that the rest term $\Gamma_2$ must consist of 2PI diagrams only.
This forms the central result that the 2PI effective action is given, besides the terms \TGaseq{Gamma2PI1loop}, by a series of all closed 2PI diagrams which can be formed from the full propagator $G$, the bare vertices defined by the classical action, and at most two external field insertions $\phi$.

The expansion of the 2PI part $\Gamma_2[\phi,G]$ up to 3-loop order, for the classical action defined in \TGasEq{Sclassphi4} is shown in \TGasFig{Gamma23loop}.
We emphasise that, although the diagrams in this expansion are proportional to a power of the bare coupling $g$, truncations of the series can not be regarded as perturbative in $g$ since the propagator $G$ itself represents an expansion to infinite order in the coupling.
The reason is that the stationarity condition for $G$, \TGasEq{StatCondsphiG} yields a perturbatively truncated expression for the inverse of $G$.

In order to arrive at a set of dynamic equations we need to discuss in more detail the implementation of the initial value problems we have in mind.

\subsection{Schwinger-Keldysh closed time path}
\label{TGas:sec:CTP}
We assume that the many-body state is initially, i.e., at time $t=t_0$, given by some general (mixed) density matrix $\rho(t_0)$
The time evolution of the expectation value of an operator ${\cal O}$ is then given as
\begin{equation}
\label{TGas:eq:Ot}
  \langle t|{\cal O}|t\rangle
  = \mathrm{Tr}\left[\rho(t_0)U^\dagger(t,t_0){\cal O}U(t,t_0)\right],
\end{equation}
where $U(t,t')={\cal T}\exp\{-i\int_{t'}^t \mathrm{d}t''\,H(t'')/\hbar\}$ denotes the time evolution operator as obtained from the Hamiltonian $H(t)$.

The operators ${\cal O}$ relevant for us, i.e., the $n$-point correlation functions, are products involving, in the Heisenberg picture, operators evaluated at different times. 
In the Schr\"odinger picture this implies additional time evolution operators between these factors.
Consider, for instance, the two-time Green function,
\begin{equation}
\label{TGas:eq:Gtwotime}
  \langle{\cal T}_{\cal C}\Phi_a(x)\Phi_b(y)\rangle_c
  =\mathrm{Tr}\left[\rho(t_0){\cal T}_{\cal C}\,U^\dagger(x_0)\Phi_a(x)U(x_0)\,U^\dagger(y_0)\Phi_b(y)U(y_0)\right]-\mathrm{disc.},
\end{equation}
where $U(t)\equiv U(t,t_0)$ and the operators are time-ordered in a way which leaves the ordering within the products $U^\dagger\Phi U$ invariant.
The disconnected part is denoted as `disc.'. 
\begin{figure}[tb]
\begin{center}
\resizebox{0.6\columnwidth}{!}{
\includegraphics{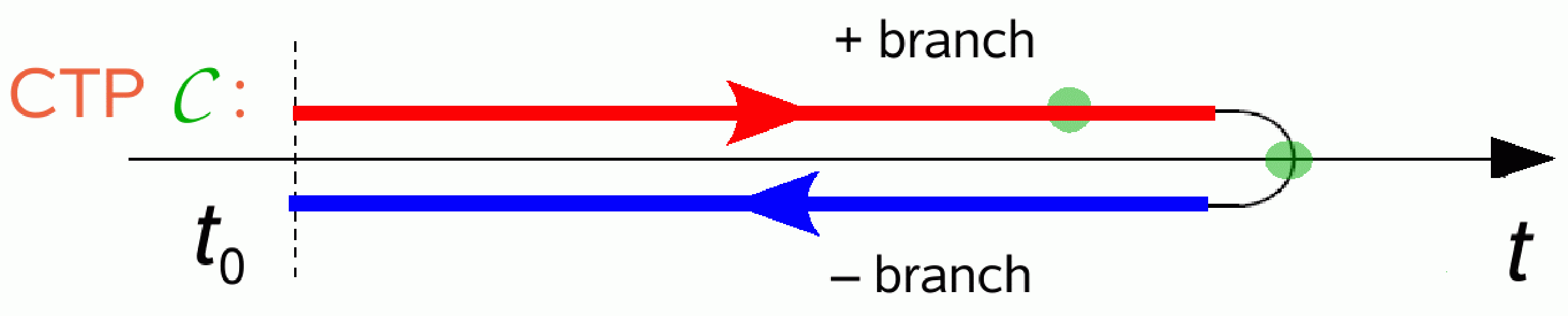}
}
\end{center}
\vspace*{-3ex}
\caption{
(Color online) Schwinger-Keldysh closed time path ${\cal C}$. 
The green dots indicate the times $x_0$ and $y_0$ for an example two-point function $G(x,y)$, see text.
The branches are drawn above and below the time axis only in order to make them separately visible.
}
\label{TGas:fig:CTP}
\end{figure}
The product of different time evolution operators and field operators can be visualised by means of the closed time path as shown in \TGasFig{CTP}.
Starting at time $t_0$, path sections leading to the maximum time appearing in the arguments of the field operators indicate time evolutions $U$.
One generically chooses all times to lie on the $+$ branch.
However, different time orderings can be handled simultaneously by allowing times on the $-$ branch as well and thereby doubling the range of possible times.
Clearly, the two-point Green functions, with times evaluated on either or both of the two branches, are not completely independent from each other, and one aim of the discussion in later sections will be to clarify the dependencies.
Here we only point out that the formalism to be developed naturally allows for two-time Green functions $G(x,y)$ and therefore for Fourier transforms over their relative time $x_0-y_0$.
These transforms, in turn, are functions of the frequency which, e.g., for a translationally invariant system, contain information about the spectral distribution of a particular momentum mode $\vec{p}$.
Beyond the mean-field approximation, collisions imply the redistribution of momentum between the particles.
These scattering effects emerge naturally as finite widths in the spectral distribution around the dispersion peak at $\omega(\vec{p})$.
How these properties emerge from the dynamical theory to be developed is the topic of \TGasSection{Kin}.

\subsection{Gaussian initial states}
\label{TGas:sec:GaussIniStates}
In the CTP formulation introduced in the previous section, the nonequilibrium generating functional for correlation functions can be split into a factor describing the initial conditions and one which contains all the ensuing quantum dynamics.
For this we insert unit operators written in the coherent state basis:
\begin{equation}
\label{TGas:eq:NonEqGenFunc}
  Z[J,R,\rho_0]
  =\int[\mathrm{d}\varphi_0^+][\mathrm{d}\varphi_0^-]\langle\varphi_0^+|\rho_0|\varphi_0^-\rangle
  \int_{\varphi_0^+}^{\varphi_0^-}{\cal D}'\varphi
  e^{i(S[\varphi]+\int J\varphi+\frac{1}{2}\int\varphi R\varphi)},
\end{equation}
The functional integrals over the fields at time $t_0$, $[d\varphi_0^\pm]=\prod_{\vec{x}}d\varphi^\pm(t_0,\vec{x})$, evaluated on the $+$ and $-$ branches of the CTP include the initial density matrix.
Their notion as limits of the dynamical functional integral implies that the primed measure ${\cal D}'\varphi=\prod_{x_0>t_0,\vec{x}}d\varphi^+(x)d\varphi^-(x)$ excludes these initial-time fields.
We emphasise that, due to causality, the CTP extends to the maximum time to be evaluated in a particular $n$-point function only.
At later times, the sources can be set to zero such that the time evolution operators on the corresponding $+$ and $-$ branches cancel by unitarity.
Note also that the CTP automatically arranges for the normalisation $Z[0,0]=1$.

The most general density matrix can be parametrised as 
\begin{equation}
\label{TGas:eq:ExpDM}
  \langle\varphi_0^+|\rho_0|\varphi_0^-\rangle
  ={\cal N}e^{if_{\cal C}[\varphi]},
\end{equation}
with the normalisation factor ${\cal N}$ and $f_{\cal C}[\varphi]$ expanded in powers of the fields:
\begin{equation}
\label{TGas:eq:fCexp}
  f_{\cal C}[\varphi]=\alpha_0+\sum_{n=1}^\infty\int_{x^{(1)},...,x^{(n)}}\alpha_n(x^{(1)},...,x^{(n)})\prod_{m=1}^{n}\varphi(x^{(m)}).
\end{equation}
Here, the coefficients $\alpha_n$ are non-zero only at the initial time $t_0$, at both ends of the CTP.
It is now clear that Gaussian initial density matrices, for which $\alpha_n\equiv0$ for $n\ge3$, can be absorbed into the integrand of the dynamical integral in \TGasEq{NonEqGenFunc} by a redefinition of the source fields $J$ and $R$.
The 2PI effective action approach then yields, once these modified sources are set to vanish, a closed set of dynamic equations for $\phi$ and $G$.
This set allows to specify initial values for these connected one- and two-point functions only, i.e., a Gaussian initial state implies that all higher-order connected $n$-point functions are assumed to vanish at $t=t_0$.\footnote{%
In the case that, at $t=t_0$, the $n$th-order connected correlation
function $\langle{\cal T}_{\cal C}\Phi_{a_1}(t_0,\vec{x}_1)\cdots$ $\Phi_{a_n}(t_0,\vec{x}_n)\rangle_c$ is non-zero, with all $m$th-order functions, $m>n$, vanishing, a straightforward generalisation of the approach involving the $n$PI effective action is at hand \protect\cite{TGas:Berges:2004pu}. }

\subsection{Dynamic equations}
\label{TGas:sec:DynEqs}
From the stationarity conditions \TGaseq{StatCondsphiG} one obtains the dynamic equation for the field expectation value $\phi$ which generalises the GPE to arbitrary order beyond the mean-field approximation, as well as a Schwinger-Dyson-like equation which, upon multiplication by $G$ has the form of a time evolution equation for the two-point Green function $G$,
\begin{equation}
\label{TGas:eq:EOMG}
  \int_zG^{-1}_{0,ac}(x,z)G_{cb}(z,y) 
  = \delta_{ab}\delta_{\cal C}(x-y)+\int_z\left[\Sigma_{ac}(x,z)+iR_{ac}(x,z)\right]G_{cb}(z,y),
\end{equation}
with $\delta_{\cal C}(x-y)=\delta_{\cal C}(x_0-y_0)\delta(\vec{x}-\vec{y})$, which is a first-order differential equation in $x_0$ for the non-relativistic case with free propagator \TGaseq{G0inv}.
Using the definition \TGaseq{GitoFrho} of $G$ in terms of the statistical and spectral functions as well as the commutation relations \TGaseq{BoseCR12} one derives the following set of dynamic equations:  
\begin{eqnarray}
&&   \Big(-i\sigma^2_{ab}\partial_{x_0}
    -g\,F_{ab}(x,x)\Big)\phi_b(x) -\Big(H_\mathrm{1B}(x)
    +\frac{g}{2}\big[\phi_c(x)\phi_c(x)
    +F_{cc}(x,x)\big]\Big)
    \phi_a(x) 
  \nonumber\\
  &&\phantom{\left[i\sigma^2_{ac}\partial_{x_0} + M_{ac}(x) \right]
    F_{cb}(x,y)}
   = \int_{t_0}^{x_0} \!\mathrm{d}y\,
   \Sigma^\rho_{ab}(x,y;\phi\equiv 0)\,\phi_b(y) ,
\label{TGas:eq:EOMphi}
  \\
  &&\left[i\sigma^2_{ac}\partial_{x_0} + M_{ac}(x) \right]
    F_{cb}(x,y)
  = - \int_{t_0}^{x_0} \! \mathrm{d} z\,
    \Sigma^{\rho}_{ac}(x,z;\phi) F_{cb}(z,y)
  + \int_{t_0}^{y_0} \! \mathrm{d}z\,
    \Sigma^{F}_{ac}(x,z;\phi) \rho_{cb}(z,y) ,
  \nonumber\\[1.5ex]
  &&
  \left[i\sigma^2_{ac}\partial_{x_0}
    +M_{ac}(x) \right] \rho_{cb} (x,y)  
  = -\int_{y_0}^{x_0} \! \mathrm{d}z\,
  \Sigma^{\rho}_{ac} (x,z;\phi)
  \rho_{cb}(z,y), \qquad
\label{TGas:eq:EOMFrho}
\end{eqnarray}
Here we employ the notation $\int_{t}^{t'}\mathrm{d}z = \int_t^{t'}\mathrm{d}z_0\int \mathrm{d}^dz$.
The ``mass'' matrix $M$ is defined as
\begin{equation}
  M_{ab}(x)
   = \delta_{ab}
  \Big[H_\mathrm{1B}(x)
   + \frac{g}{2}\Big(\phi_c(x)\phi_c(x)+F_{cc}(x,x)\Big)\Big]
   + g\Big(\phi_a(x)\phi_b(x)+F_{ab}(x,x)\Big)
\label{TGas:eq:MM}
\end{equation}
Moreover, the self energy has been decomposed into a part local in time and a nonlocal part written in terms of statistical and spectral components,
\begin{equation}
\label{TGas:eq:Sigma0Frho}
  \Sigma_{ab}(x,y)
  =\Sigma^{(0)}_{ab}(x)\delta_{\cal C}(x-y)+\Sigma^F_{ab}(x,y)-\frac{i}{2}\mathrm{sgn}_{\cal C}(x_0-y_0)\Sigma^\rho_{ab}(x,y).
\end{equation}
The local part, $\Sigma^{(0)}_{ab}(x)$, has been absorbed into the mass matrix $M$ while the non-local parts form the kernels for the memory integrals on the right-hand sides of the integro-differential dynamic equations \TGaseq{EOMphi}, \TGaseq{EOMFrho}. 
Note that only the two-loop, double-bubble term in $\Gamma_2$ contributes to $\Sigma^{(0)}$.

With $\Sigma^F$ and $\Sigma^\rho$ set to zero, which is equivalent to truncating the loop expansion of the 2PI effective action after the double-bubble diagram, one obtains a set of differential equations which are local in time and can be shown to be equivalent to the HFB dynamic equations introduced in \TGasSect{HFB}.
Note that the time derivatives in the equations for $F$ and $\rho$ only act on the respective first time variables $x_0$.
The corresponding equations in the second variable $y_0$ are obtained by symmetry considerations for $F$ and $\rho$.
The equations for different $y_0$ decouple, and only the diagonal functions with $x_0=y_0$ appear in $M$.
As initial conditions, $\phi$ and $F$ need to be specified at $x_0=y_0=t_0$.
Note that $\rho(t_0;\vec{x},\vec{y})$ is fixed by the commutation relations, \TGasEq{BoseCR12}.

Hence, in order to obtain Eqs.~\TGaseq{HFB_n} and \TGaseq{HFB_m} one needs to combine the equations \TGaseq{EOMFrho}, for $x_0=y_0$ with their counterparts with time derivatives acting on $y_0$.
The equation for $\rho$ which is decoupled from all other equation is thereby shown to be consistent with the conservation of the Bose commutators.
In summary, the HFB approximation is represented by the leading order (double-bubble) diagram contributing to $\Gamma_2$.

Setting, furthermore, $F(t_0,\vec{x};t_0,\vec{y})\equiv0$, $F$ remains identically zero at all times, and the equation for $\phi$ reduces to the Gross-Pitaevskii equation, cf.~\TGasEq{GPE}.
 
The non-Markovian integral terms on the right hand sides of Eqs.~\TGaseq{EOMphi} and \TGaseq{EOMFrho} open in a clear way the path to nonequilibrium dynamics beyond the mean-field approximation.
The higher-order loop diagrams, with one line opened yield contributions to the self energy which describe scattering processes involving the redistribution of particles between the different modes of the system.
As the corresponding contributions to $\Sigma^{F,\rho}$ are non-local in time they naturally imply the evolution of a non-trivial relative time dependence of the two-point correlation functions and therefore finite-width spectral distributions as discussed above.

Higher-order correlations formed in this way are not included explicitly in terms of higher-order $n$-point functions.
One may, however, imagine these functions to obtain non-zero values implicitly, according to their own dynamic equations which then have been integrated formally and reinserted into the equations for the one- and two-point functions.

Before we discuss in more detail the analytical implications of the non-Markovian terms and relate them to results obtained in the framework of kinetic theory we briefly discuss, in the following sections, specific truncations of the 2PI effective action and, in a particular truncation, an application to a weakly interacting ultracold Bose gas evolving from a nonequilibrium initial state in one spatial dimension.

\subsection{Truncations of the 2PI effective action}
\label{TGas:sec:2PITrunc}
In order to practically solve the dynamic equations \TGaseq{EOMphi} and \TGaseq{EOMFrho}, details about the self energy $\Sigma$ are required, and these are, in general, only available to a certain approximation. 
As discussed in \TGasSect{2PIEA}, the natural expansion of $\Gamma_2$ is in terms of 2PI closed loop diagrams involving only bare vertices and full propagators $G$.
This expansion can be truncated at any order, e.g., of powers of $g$ or the number of loops, without violating the most important conservation laws for energy and particle number which we briefly discuss in the following.

\subsubsection{Number conservation}
Particle number conservation is a consequence of the Noether theorem in conjunction with the invariance of the theory under orthogonal transformations and can be seen as follows.%
\footnote{Conservation of total particle number in a non-relativistic system corresponds to conservation of the difference of particles and antiparticles in a fully relativistic approach, i.e., to the conservation of charge. Neglecting the antiparticle sector of the Hilbert space in a non-relativistic system is equivalent to the constraint of a vanishing antiparticle number and, hence, amounts to the fact that the $O(2)$ symmetry of the Lagrangian ensures total number conservation.}
The stationarity conditions \TGaseq{StatCondsphiG} can be combined to the equation
\begin{equation}
\label{TGas:eq:StatCondCombiforNC}
  \sigma^2_{ab}\Big[
  \phi_a(x)\frac{\delta\Gamma[\phi,G]}{\delta\phi_b(x)}
  +2\int_y\frac{\delta\Gamma[\phi,G]}{\delta G_{cb}(y,x)}G_{ca}(y,x)\Big]
  =0,
\end{equation} 
with the elements $\sigma_{2,ab}$ of the Pauli 2-matrix.
From the specific expression \TGaseq{Gamma2PI} for the 2PI effective action follows that \TGasEq{StatCondCombiforNC} is equivalent to the relation
\begin{equation}
\label{TGas:eq:ContEq2PI}
  \partial_{x_0} n(x)-\vec{\nabla}\vec{j}(x)
   = -{2i}\sigma^2_{ab}\Big[
  \phi_a(x)\frac{\delta\Gamma_\mathrm{int}[\phi,G]}
                {\delta\phi_b(x)}
  +2\int_y \frac{\delta\Gamma_\mathrm{int}[\phi,G]}
                {\delta G_{cb}(y,x)}G_{ca}(y,x)
  \Big].
\end{equation} 
Here,
\begin{eqnarray}
\label{TGas:eq:totaldensity}
  n(x)
  &=& \phi_a(x)\phi_a(x)+G_{aa}(x,x),
  \\
\label{TGas:eq:totalcurrentdensity}
  \vec{j}(x)
  &=& \frac{1}{m}\big[\phi_2(x)\vec{\nabla}\phi_1(x)-\phi_1(x)\vec{\nabla}\phi_2(x)
  +\langle{\cal T}_{\cal C}(
   \hat\Phi_2(x)\vec{\nabla}\hat\Phi_1(x)-\hat\Phi_1(x)\vec{\nabla}\hat\Phi_2(x)
   )\rangle_c\big]
\end{eqnarray} 
are the total number and current densities, respectively.
Clearly, particle number is conserved locally if $n$ and $\vec{j}$ obey a continuity equation, i.e., if the right hand side of \TGasEq{ContEq2PI} vanishes identically.
We consider the specific structure of these terms:
The interaction part of the 2PI effective action occurring therein is defined as
\begin{equation}
\label{TGas:eq:Gamma2PIint}
   \Gamma_\mathrm{int}[\phi,G]
   = \Gamma[\phi,G] 
   - \frac{1}{2}\int_{xy}\,
      \phi_a(x)\,iD^{-1}_{ab}(x,y)\phi_b(y)
   \nonumber\\
   -\frac{i}{2}\mathrm{Tr}\left[D^{-1}G\right].
\end{equation} 
The 2PI effective action is, like the underlying classical action $S[\varphi]$ \TGaseq{Sclassphi4}, a singlet under $O(2)$ rotations.
It is parametrised by the fields $\phi_a$ and $G_{ab}$, where the number of $\phi$-fields has to be even in order to construct an $O(2)$-singlet. 
From the fields $\phi_a$ alone one can construct only one independent invariant under $O(2)$ rotations, which can be taken as $\mathrm{tr}(\phi\phi) \equiv \phi^2 = \phi_a \phi_a$.
All functions of $\phi$ and $G$, which are singlets under $O(2)$, can be built from the irreducible, i.e., in field-index space not factorisable, invariants~\cite{TGas:Berges:2001fi,TGas:Aarts:2002dj}
\begin{equation}
\label{TGas:eq:ONinvariants}
  \phi^2, 
  \quad\quad 
  \mathrm{tr} (G^n), 
  \quad\quad \mbox{and}  \quad\quad 
  \mathrm{tr} (\phi \phi G^{n}), 
\end{equation}
with $n=1,2,...$.
As before, the trace tr$(\cdot)$ only applies to the field-component indices while there is no integration over space-time, e.g., tr$(G^3)\equiv G_{ab}(x,y)G_{bc}(y,z)G_{ca}(z,x)$.

For contributions to $\Gamma_\mathrm{int}$ which contain only $\phi^2$ or tr$(G^n)$, the terms in square brackets in \TGasEq{ContEq2PI} either vanish separately or are symmetric under the exchange of $a$ and $b$.
Moreover, if a term contains an invariant of the form tr$(\phi\phi G^n)$, as, e.g., the contributions remaining in $\Gamma_\mathrm{int}$ from Tr$\{G_0^{-1}G\}$, the combination of the terms in square brackets in \TGasEq{ContEq2PI} is symmetric under transposition in field index space.
Hence, the total number density is conserved locally as a consequence of the $O(2)$ symmetry of the theory, and, more importantly, this is true for any set of approximate dynamic equations derived from a truncated but still $O(2)$-symmetric effective action.
Note, finally, that only terms in the action which contain mixed invariants tr$(\phi\phi G^n)$ induce exchange of particles between the condensate and the non-condensed fraction of the gas.

\subsubsection{Energy conservation}
Energy conservation follows from time translation invariance of $\Gamma$, cf., e.g., Ref.~\cite{TGas:Arrizabalaga:2005tf}.
Consider the general translations in continuous space and time which vanish at the boundary, $x^\mu\to x^\mu+\varepsilon^\mu(x)$, where $\varepsilon^\mu(x)$ is a time- and space-dependent infinitesimal 4-vector.
The mean field and 2-point functions transform, under these translations, to leading order in $\varepsilon$, as $\phi_a(x)\to\phi_a(x)+\varepsilon^\nu(x)\partial^x_\nu\phi_a(x)$, and $G_{ab}(x,y)\to G_{ab}(x,y)+\varepsilon^\nu(x)\partial^x_\nu G_{ab}(x,y)+\varepsilon^\nu(y)\partial^y_\nu G_{ab}(x,y)$, respectively.
Here, $\partial^x_\nu=\partial/\partial x^\nu$, etc.
One can show that under these transformations the variation of the 2PI effective action $\Gamma$ can be written as $\Gamma[\phi,G]\to\Gamma[\phi,G]+\delta\Gamma[\phi,G]$, with
\begin{eqnarray}
  \delta\Gamma[\phi,G] = \int_x T^{\mu\nu}(x)\,\partial^x_\mu\varepsilon_\nu(x).
\label{TGas:eq:deltaGamma}
\end{eqnarray}
Since, by virtue of the stationarity conditions \TGaseq{StatCondsphiG}, the variation $\delta\Gamma$ vanishes for all solutions of the equations of motion for $\phi_a$ and $G_{ab}$, an integration by parts shows that $T^{\mu\nu}$ is the conserved Noether current for the time-space-translations:
\begin{eqnarray}
  \delta\Gamma[\phi,G] = -\int_x \varepsilon_\nu(x)\,\partial^x_\mu T^{\mu\nu}(x) = 0.
\label{TGas:eq:EMTconservation}
\end{eqnarray}
$T^{\mu\nu}(x)$ is identified as the energy-momentum tensor, and the conservation law for total energy is expressed as $\partial^x_\mu T^{\mu0}(x)=0$ or $\partial_t\int \mathrm{d}^3x\,T^{00}(t,\vec{x})=0$.
Explicit expressions for the energy-momentum tensor have been calculated in Refs.~\cite{TGas:Arrizabalaga:2005tf,TGas:Temme2006a}

\subsubsection{NLO 2PI $1/{\cal N}$ expansion}
\label{TGas:sec:NLO1N}
Also the expansion of $\Gamma_2$ in terms of 2PI loop diagrams can be resummed to obtain alternative non-perturbative approximation schemes.
The most outstanding such scheme is the expansion in powers of the inverse number of field components $\cal N$ applied to the dynamical 2PI effective action first by Berges and collaborators \cite{TGas:Berges:2001fi,TGas:Aarts:2002dj,TGas:Berges:2004yj} and extensively studied since.
To next-to-leading order (NLO) this resummation scheme can be understood as the replacement of certain vertices in a loop expansion by a bubble-resummed vertex \cite{TGas:Aarts:2002dj,TGas:Gasenzer:2005ze}.
The result of the scheme has also been recovered using a functional renormalisation group inspired approach \cite{TGas:Gasenzer:2007za} where it results as a truncation in orders of proper $n$-point functions combined with an $s$-channel approximation of the equation for the proper four-vertex.

In the following sections we will employ the NLO $1/{\cal N}$ expansion scheme to calculate the dynamics of a single-species ultracold Bose gas described by a complex scalar field for which ${\cal N}=2$.
In the context of a non-relativistic Bose gas, this approximation has been discussed in detail in Refs.~\cite{TGas:Gasenzer:2005ze,TGas:Temme2006a,TGas:Berges:2007ym,TGas:Branschadel2008a}.
In this scheme, the contribution $\Gamma_2[\phi,G]$ to the 2PI effective action involves a leading (LO) and next-to-leading order (NLO) part which can be diagrammatically represented as shown in \TGasFig{DiagrExpGamma2NLO}.
\begin{figure}[tb]
\begin{center}
\resizebox{0.7\columnwidth}{!}{
\includegraphics{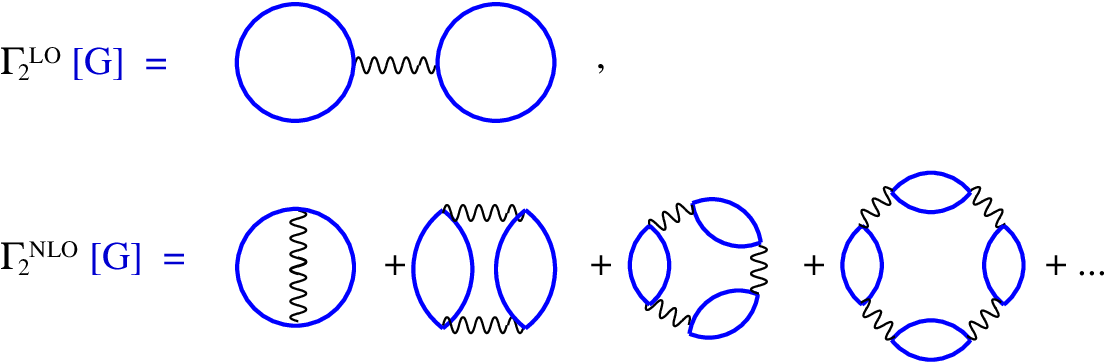}
}
\end{center}
\vspace*{-3ex}
\caption{
(Color online) Diagrammatic representation of the  leading order (LO) and next-to-leading order (NLO) contributions in the $1/\cal N$-expansion, to the 2PI part $\Gamma_2[\phi,G]$ of the 2PI effective action.
The thick blue lines represent 2-point functions $G_{ab}(x,y)$, the red crosses field insertions $\phi_a(x)$, and the wiggly lines a single vertex channel of those shown in the right panel of \TGasFig{Gamma23loop}.
At each vertex, it is summed over double field indices $a$ and integrated/summed over double time and space variables $x$.
}
\label{TGas:fig:DiagrExpGamma2NLO}
\end{figure}
While the leading-order contribution involves one diagram, in NLO a chain of bubble diagrams is resummed.
All of these diagrams are proportional to the same power of $1/\cal N$ since each vertex scales with $1/\cal N$, which is cancelled by the (blue) propagator loops which scale with $\cal N$ since they involve a summation over the field indices from $1$ to $\cal N$.
Note that the Hartree-Fock-Bogoliubov (HFB) approximation is given by an action $\Gamma_2$ which involves $\Gamma_2^\mathrm{LO}$ and the first diagram of $\Gamma_2^\mathrm{NLO}$ in \TGasFig{DiagrExpGamma2NLO} (b), see, e.g., Ref.~ \cite{TGas:Gasenzer:2005ze}.

From $\Gamma_2[\phi,G]=\Gamma_2^\mathrm{LO}[\phi,G]+\Gamma_2^\mathrm{NLO}[\phi,G]$ we obtain, using \TGasEq{SigmafromGamma2}, the self energies $\Sigma_{ab}(x,y)=\Sigma^F_{ab}(x,y)-(i/2)\mathrm{sgn}_{\cal C}(x_0-y_0)\Sigma^\rho_{ij}(x,y)$, with
\begin{eqnarray}
  \left(\begin{array}{r}
        \Sigma^F_{ab}(x,y) \\ -\frac{1}{2}\Sigma^\rho_{ab}(x,y)
	\end{array}\right)
   &=& -\frac{2g}{\cal N}\Bigg[\left(\begin{array}{r}
        I_F(x,y) \\ -\frac{1}{2}I_\rho(x,y)
	\end{array}\right)
        \phi_a(x)\phi_b(y) 
  \nonumber\\
  &&\qquad 
  +\ \left(\begin{array}{rr}
        \Delta_F(x,y) & 
	\frac{1}{2}\Delta_\rho(x,y) \\
       -\frac{1}{2}\Delta_\rho(x,y) & 
        \Delta_F(x,y) \end{array}\right)
    \left(\begin{array}{r}
        F_{ab}(x,y) \\ -\frac{1}{2}\rho_{ab}(x,y)
	\end{array}\right)\Bigg],
\label{TGas:eq:SigmaNLO1N}
\end{eqnarray}
where $\Delta_{F,\rho}(x,y)=I_{F,\rho}(x,y)+P_{F,\rho}(x,y;I_{F,\rho})$.
The resummation to NLO in $1/\cal N$ is expressed by the coupled integral equations for $I_{F,\rho}$ \cite{TGas:Berges:2004yj}:
\begin{eqnarray}
  \left(\begin{array}{r}
        I_F(x,y) \\ I_\rho(x,y)
	\end{array}\right)
  &=&
  \frac{g}{\cal N}\Bigg[
  \left(\begin{array}{c}
        F(x,y)^2-\frac{1}{4}\rho(x,y)^2 \\
	2F_{ab}(x,y)\rho_{ab}(x,y)
	\end{array}\right)
   -\ \int_{t_0}^{x_0}\mathrm{d}z\,I_\rho(x,z)
      \left(\begin{array}{c}
            F(z,y)^2-\frac{1}{4}\rho(z,y)^2 \\ 
            2F_{ab}(z,y)\rho_{ab}(z,y) 
	    \end{array}\right)
    \nonumber\\
   &&\qquad
   +\ \int_{t_0}^{y_0}\mathrm{d}z\,
      \left(\begin{array}{r}
            I_F(x,z) \\ 
            I_\rho(x,z) 
	    \end{array}\right)2F_{ab}(z,y)\rho_{ab}(z,y)\Bigg].
\label{TGas:eq:IFrho}
\end{eqnarray}
Here, $F^2=F_{ab}F_{ab}$, etc.
The functions $P_{F,\rho}$, which contribute to $\Delta_{F,\rho}$ in the self energies \TGaseq{SigmaNLO1N} and vanish if $\phi_i\equiv0$, read \cite{TGas:Berges:2007ym}
\begin{eqnarray}
  &&P_F(x,y;I_{F,\rho})
  = -\frac{2g}{{\cal N}}\Big\{
     H_F(x,y)
  +\int_{t_0}^{y_0}\mathrm{d}z\left[H_F(x,z)I_\rho(z,y)+I_F(x,z)H_\rho(z,y)\right]
  \nonumber\\
  &&\ 
  -\int_{t_0}^{x_0}\mathrm{d}z\left[H_\rho(x,z)I_F(z,y)+I_\rho(x,z)H_F(z,y)\right]
  -\int_{t_0}^{x_0}\mathrm{d}v\int_{t_0}^{y_0}\mathrm{d}w\, I_\rho(x,v)H_F(v,w)I_\rho(w,y)
  \nonumber\\
  &&\ 
  +\int_{t_0}^{x_0}\mathrm{d}v\int_{t_0}^{v_0}\mathrm{d}w\, I_\rho(x,v)H_\rho(v,w)I_F(w,y)
  +\int_{t_0}^{y_0}\mathrm{d}v\int_{v_0}^{y_0}\mathrm{d}w\, I_F(x,v)H_\rho(v,w)I_\rho(w,y)\Big\},
\label{TGas:eq:PF}
\end{eqnarray}
\begin{eqnarray}
  &&P_\rho(x,y;I_{F,\rho})
  = -\frac{2g}{{\cal N}}\Big\{
     H_\rho(x,y)
  -\int_{y_0}^{x_0}\mathrm{d}z\left[H_\rho(x,z)I_\rho(z,y)+I_\rho(x,z)H_\rho(z,y)\right]
  \nonumber\\
  &&\ 
  +\int_{y_0}^{x_0}\mathrm{d}v\int_{y_0}^{v_0}\mathrm{d}w\, I_\rho(x,v)H_\rho(v,w)I_\rho(w,y)\Big\},
\label{TGas:eq:Prho}
\end{eqnarray}
wherein the functions $H_{F,\rho}$ are defined as
\begin{eqnarray}
  H_F(x,y)
  &=& -\phi_a(x)F_{ab}(x,y)\phi_b(y),
  \nonumber\\
  H_\rho(x,y)
  &=& -\phi_a(x)\rho_{ab}(x,y)\phi_b(y).
\label{TGas:eq:HFHrho}
\end{eqnarray}
The technical procedure to solve the above dynamic equations in every time step requires the determination of the 
functions $I(x,y)$, before the actual propagation of the respective correlation functions.

\subsection{Functional renormalisation-group approach}
\label{TGas:sec:FRG}
We close our introduction to nonequilibrium quantum field theory with a short description of an alternative approach which is based on functional renormalisation group (RG) techniques. 
For more details on this approach see Refs.~\cite{TGas:Gasenzer:2007za,TGas:Gasenzer2008a}.
Dynamic equations will be derived which are similar in structure to the equations obtained from the 2PI effective action.
In a particular, ``$s$-channel'' approximation these equations are equivalent to the 2PI equations in NLO of the $1/{\cal N}$ approximation summarised above.
The functional renormalisation group (RG) techniques we will employ have been introduced and used extensively in the framework of equilibrium quantum field theory of strongly correlated systems.
See Refs.~\cite{TGas:Wetterich:1992yh,TGas:Bagnuls:2000ae,TGas:Litim:1998nf,TGas:Pawlowski:2005xe}, as well as \cite{TGas:Canet:2003yu} for non-equilibrium applications.

For a given initial-state density matrix $\rho_D(t_0)$, the renormalised finite quantum generating functional for time-dependent $n$-point correlation functions,
\begin{eqnarray}\label{TGas:eq:definingZneq}
  Z[J;\rho_D] 
  &=& {\rm Tr}\Big[\rho_D(t_0)\, {\cal T}_{\cal C}
  \exp\Big\{i \int_{x,{\cal C}}\! J_a(x) \Phi_a(x) \Big\}\Big],
\end{eqnarray}
(summation over double indices is implied) carries all the information of the quantum many-body evolution at times greater than the initial time $t_0$.  
${\cal T}_{\cal C}$ denotes, as before, time-ordering along the Schwinger-Keldysh closed time path
(CTP), see \TGasSect{CTP}, and $\int_{x,{\cal C}} \equiv\int_{\cal C}
\mathrm{d} x_0 \int \mathrm{d}^d x$.  
All connected Greens functions will be time-ordered along $\cal C$.

\subsubsection{Functional flow equation}
\label{TGas:sec:FFE}
The key idea of the approach to be described in the following is to first consider the generating functional for Green functions where all times are smaller than a maximum time $\tau$. 
This implies a time path ${\cal C}(\tau)$  which is closed at $t=\tau$, i.e., the maximum time in  \TGasFig{CTP} is set to $\tau$.
As a consequence, the generating functional $Z_\tau=Z_{{\cal C}(\tau)}$ has the source term 
\begin{equation}
 {\cal T}_{{\cal C}(\tau)}
  \exp\Big\{i \int_{x,{\cal C}(\tau)}\! J_a(x) \Phi_a(x) \Big\}\,. 
\end{equation}
At $\tau =t_0$, this results in a trivial $Z_{t_0}$ where all information is stored in the initial density matrix $\rho_D(t_0)$.
From this initial condition $Z_\tau$ can be computed by means of the time evolution $\partial_\tau Z_\tau$ for all times $\tau>t_0$.  

In the following, this evolution shall be derived by use of functional RG ideas. 
We note that $Z_\tau$ can be defined in terms of the full generating functional $Z_\infty$ in \TGasEq{definingZneq} by suppressing the propagation for times greater than $\tau$. 
This suppression is achieved by
\begin{eqnarray}
\label{TGas:eq:defZtau}
  Z_\tau 
  &=& \exp\Big\{-\frac{i}{2} \int_{x y,{\cal C}}\!
  \frac{\delta}{\delta J_a(x)}R_{\tau,ab}(x,y) \frac{\delta}{\delta
  J_b(y)}\Big\}Z,
\end{eqnarray}
where the function $R_\tau$ is chosen such that it suppresses the fields, i.e., $\delta/\delta J_a$, for all times $t> \tau$. 
This does not fix $R_\tau$ in a unique way, and a simple choice is 
\begin{equation}
\label{TGas:eq:Rchoice}
  -i R_{\tau,ab}(x,y)
  =\left\{\begin{array}{lcl} 
          \infty & \quad & \mathrm{for}\ x_0=y_0>\tau,\ \vec{x}=\vec{y},\ a=b  \\[1ex] 
          0  	 & 		 & \mathrm{otherwise} 
          \end{array} \right.,
\end{equation}
see \TGasFig{Rchoice}. 
\begin{figure}[tb]
\begin{center}
\begin{minipage}{0.35\columnwidth}
\resizebox{1.1\columnwidth}{!}{
\includegraphics{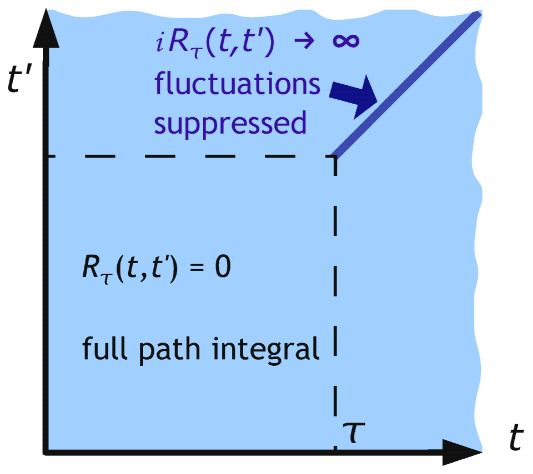}
}
\end{minipage}
\hspace*{0.05\columnwidth}
\begin{minipage}{0.57\columnwidth}
\resizebox{1.0\columnwidth}{!}{
\includegraphics{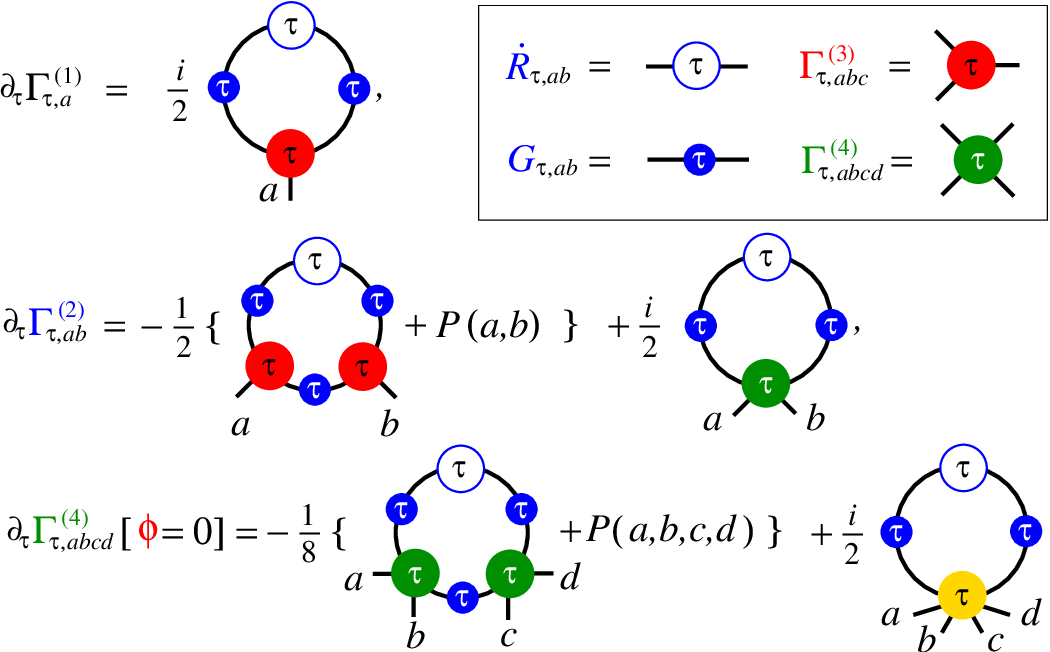}
}
\end{minipage}
\vspace*{2ex}\ \\
\begin{minipage}[t]{0.37\columnwidth}
\caption{
(Color online) The cutoff function $R_{\tau,ab}(x,y)$ in the time plane $\{x_0,y_0\}=\{t,t'\}$, $t,t'\ge t_0$. 
The function vanishes everywhere except for $t=t'>\tau$ where it tends to infinity and therefore implies a suppression of all fluctuations in the generating functional at times greater than $\tau$.
}
\label{TGas:fig:Rchoice}
\end{minipage}
\hspace*{0.01\columnwidth}
\begin{minipage}[t]{0.59\columnwidth}
\caption{(color online) Diagrammatic representation of
  the general flow equations for
  $\Gamma_\tau^{(1)}[\phi],\Gamma_\tau^{(2)}[\phi]$, and
  $\Gamma_{\tau,abcd}^{(4)}[\phi=0]$, for a $\phi^4$-theory.  Open
  circles with a $\tau$ denote $\partial_\tau R_{\tau,ab}$, solid
  lines with (blue) filled circles are $\tau$- and, in general,
  $\phi$-dependent two-point functions
  $G_{\tau,ab}=i[\Gamma_\tau^{(2)}+R_\tau]^{-1}_{ab}$.  All other
  filled circles denote proper field-dependent $n$-vertices
  $\Gamma^{(n)}_{\tau,abcd}$, $n=3,4,6$. $P$ implies a sum
  corresponding to all permutations of its arguments. }
\label{TGas:fig:FlowEqs}
\end{minipage}
\end{center}
\end{figure}

We emphasise that the cutoff $R_\tau$ in \TGasEq{defZtau} suppresses any time evolution at times greater than $\tau$.
Correlation functions derived from $Z_\tau$ vanish as soon as at least one of their time arguments is larger than $\tau$.  
Hence, the regularised generating functional \TGasEq{defZtau} is equivalent to a generating functional with a closed time path ${\cal C}(\tau)$ leading from $t_0$ to $\tau$ and back to $t_0$.  
Note that the CTP automatically arranges for the normalisation of $Z_\tau$. 

The restriction of the CTP to times $t_0\le t\le\tau$ implies that the differential equation for $Z_\tau$ describing the flow of the generating functional, and therefore that of the correlation functions, encodes the full time evolution of the system. 
Analogously, the time evolution of connected correlation functions is derived from that of the Schwinger functional $W_\tau=-i \ln Z_\tau$. 
It is more convenient, however, to work with the effective action
\begin{equation}
  \label{TGas:eq:effAction}
  \Gamma_\tau[\phi;R_\tau]
  = W_\tau[J;\rho_D]-\int_{\cal C}J_a\phi_a
      - \frac{1}{2}\int_{\cal C}\phi_a  R_{\tau,ab}\phi_b.
\end{equation}
Here, space-time arguments are suppressed, and $\phi_a(x)=\delta W_\tau/\delta J_a(x)|_{J\equiv0}$ is the usual field expectation value. 
From Eqs.~\TGaseq{defZtau} and \TGaseq{effAction} we derive the Functional RG or flow equation for the $\tau$-dependent effective action,
\begin{equation}
  \label{TGas:eq:flowGamma}
  \partial_\tau \Gamma_\tau
  = \frac{i}{2} \int_{{\cal C}}\!
  \left[\frac{1}{\Gamma^{(2)}_\tau+R_\tau}\right]_{ab}
  \partial_\tau R_{\tau,ab}\, ,
\end{equation}
where $\Gamma_\tau^{(n)}=\delta^n \Gamma_\tau/(\delta \phi)^n$.
Again, space-time arguments are suppressed which appear in analogy to the field indices $a,b$, see e.g.~Ref.~\cite{TGas:Pawlowski:2005xe}.
\TGasEq{flowGamma} is analogous to functional flow equations used extensively with regulators in momentum and/or frequency space to describe strongly correlated systems near equilibrium \cite{TGas:Wetterich:1992yh,TGas:Bagnuls:2000ae,TGas:Litim:1998nf,TGas:Pawlowski:2005xe}.
Its homogenous part relates to standard $\tau$-dependent renormalisation \cite{TGas:Pawlowski:2005xe}, and has been studied e.g.\ in \cite{TGas:Boyanovsky:1998aa,TGas:Ei:1999pk}.

\subsubsection{Flow equations for correlation functions}
\label{TGas:sec:FlowEqs}
To obtain a practically solvable set of dynamic equations, we derive the flow equation for the proper $n$-point Green function $\Gamma_{\tau}^{(n)}$ by taking the $n$th field derivative of \TGasEq{flowGamma}.  
\TGasFig{FlowEqs} shows a diagrammatic representation of the resulting equations for the $\tau$-dependent proper two- and four-point functions.
To be more specific, we consider, in the following, the special case of an ${\cal N}$-component scalar $\phi^4$ theory defined by the classical action \TGaseq{Sclassphi4}.

Our goal is to derive the full time-evolution of $\Gamma^{(n)}=\Gamma^{(n)}_{\infty}$, in particular, of the connected two-point function $G=i[\Gamma^{(2)}]^{-1}=i[\Gamma^{(2)}_{\infty}]^{-1}$. 
As a consequence of the above mentioned effective cut off of the CTP at  times greater than $\tau$, it will be sufficient, for the time evolution up to $t=\tau$, to determine the functions $\Gamma^{(n)}_{\tau}$ and thus the propagator 
\begin{equation}
\label{TGas:eq:Gtau}
  G_{\tau,ab}
  =i[\Gamma^{(2)}_{\tau}+R_\tau]_{ab}^{-1}.
\end{equation}

For conciseness we here only discuss the case $\phi_a\equiv0$, such that the action \TGaseq{Sclassphi4} implies that $\Gamma_{\tau}^{(3)}\equiv0$, and thus the flow of $\Gamma_\tau^{(1)}$ vanishes.  
Moreover, the equation for the proper two-point function involves, on the right hand side, only the term containing $\Gamma_\tau^{(4)}$,
\begin{equation}
\label{TGas:eq:flowGamma2}
  \partial_\tau \Gamma^{(2)}_{\tau,ab}
  = \frac{i}{2} \int_{{\cal C}}\!
  \Gamma^{(4)}_{\tau,abcd}
  (G_\tau\,[\partial_\tau R_\tau] G_\tau)_{dc}\, ,
\end{equation}
see also \TGasFig{FlowEqs}.
The term in parentheses stands for the regularised line.
We supplement \TGasEq{flowGamma2} with the flow equation for $\Gamma^{(4)}_\tau$, which, for $\phi_a\equiv0$, is drawn in \TGasFig{FlowEqs}.
This system of equations is still exact. 
For practical computations it needs to be closed which can be achieved by truncation or by supplementing it with equations for one or more higher $n$-vertices truncated at some higher order.  
Here we truncate by neglecting, in the equation for $\Gamma^{(4)}_\tau$, the term involving $\Gamma^{(6)}_\tau$,  
\begin{equation}
\label{TGas:eq:flowGamma4}
\partial_\tau \Gamma^{(4)}_{\tau,abcd} 
  = -\frac{1}{8}\int_{{\cal C}}\!  \Big\{
  \Gamma^{(4)}_{\tau,abef} G_{\tau,fg}
  \Gamma^{(4)}_{\tau,cdgh}
    \Big\}\,
  (G_\tau [\partial_\tau R_\tau] G_\tau)_{he} \,
  + P(a,b,c,d),
\end{equation}
where $P$ implies a sum corresponding to all permutations of its arguments.
In this way we obtain a closed set of integro-differential equations for the proper functions up to fourth order.  
As we will show in the following, they allow to derive, for a particular cutoff time $\tau$, a set of dynamic equations describing the time evolution of the two- and four-point functions up to time $t=\tau$.  
We emphasise that the only approximation here is the neglection of the six-point vertex, see \TGasFig{FlowEqs}.

\subsubsection{Dynamic equations}
\label{TGas:sec:RGDynEqs}
For the sharp cutoff $R_\tau$ chosen here the flow equations can be analytically integrated over $\tau$.
As pointed out above, our cutoff implies the connected two-point function to vanish at times greater than $\tau$, i.e., it can be written as
\begin{equation}
\label{TGas:eq:GtauSharp}
  G_{\tau,ab}
  =i\left[\Gamma^{(2)}_\tau\right]^{-1}_{ab}\,
  \theta(\tau-t_a)\,\theta(\tau-t_b)\, ,
\end{equation}
where $\theta(\tau)$ evaluates to $0$ for $\tau<0$ and to $1$ elsewhere, and where $t_a$ is the time argument corresponding to the field index $a$, etc.  Hence, the precise way in which the cutoff
$R_{\tau,ab}$ diverges at $t_a=t_b>\tau$ is chosen such that $\Gamma^{(2)}_\tau+R_\tau$ is the inverse of $-iG_\tau$ for all times $t_a,t_b$, see \TGasEq{Gtau}.  
Using \TGasEq{GtauSharp} one finds that
\begin{equation}
\label{TGas:eq:GRdotG}
  (G_\tau [\partial_\tau R_\tau] G_\tau)_{ab}
  =-iG_{\tau,ab}\partial_\tau[
  \theta(\tau-t_a)\,\theta(\tau-t_b)].
\end{equation}
Note that we have not used the specific choice \TGaseq{Rchoice} to derive \TGaseq{GRdotG} but simply the property \TGaseq{GtauSharp}, i.e., the suppression of any propagation for times $t>\tau$. 
After inserting \TGasEq{GRdotG} into Eqs.~\TGaseq{flowGamma2} and \TGaseq{flowGamma4} we can integrate over $\tau$ and obtain, after some algebra, the integral equations determining the flow of the proper functions from $t_0$ to some final time $t$,
\begin{eqnarray}
\label{TGas:eq:Gamma2t}
  &&\left.\Gamma^{(2)}_{\tau,ab}\right|_{t_0}^t 
  =\frac{1}{2}\int_{t_0,{\cal C}}^t\!
   \Gamma^{(4)}_{\tau_{cd},acbd} G_{\tau_{cd},dc},
  \\
\label{TGas:eq:Gamma4t}
  &&\left.\Gamma^{(4)}_{\tau,abcd}\right|_{t_0}^t 
  =\frac{i}{2}\int_{t_0,{\cal C}}^t\!
   \Gamma^{(4)}_{\tau_{efgh},abef}G_{\tau_{fg},fg}
   \Gamma^{(4)}_{\tau_{efgh},cdgh}G_{\tau_{eh},he}
   +\, (a\leftrightarrow c) + (a\leftrightarrow d) .
\end{eqnarray}
See Ref.~\cite{TGas:Gasenzer2008a} for more details of the derivation.
Double indices imply sums over field components, spatial integrals and time integrations over the CTP $\cal C$, from $t_0$ to $t$ and back to $t_0$.  
We furthermore introduced 
\begin{equation}
\label{TGas:eq:tauab}
  \tau_{ab}=\mathrm{max}\{t_a,t_b\},
  \qquad
  \tau_{abcd}
  =\mathrm{max}\{t_a,t_b,t_c,t_d\}.
\end{equation}
The brackets denote terms with the respective indices swapped.

From Eqs.~\TGaseq{Gamma2t} and \TGaseq{Gamma4t} it is clear that for the two- and four-point functions to be defined at $\tau=t$ we need to specify initial functions at $\tau=t_0$.  
We point out that, within the truncation scheme chosen above, we can  insert any set of proper two- and four-point functions, as long as  we set all $n$-vertices for $n=1,3$, and $n>4$ to vanish.  
Hence, the scheme corresponds to a Gaussian initial density matrix.
Here, we choose the respective classical proper functions defined by $S$ in \TGasEq{Sclassphi4}.  
Hence, the initial two- and four-point functions entering Eqs.~\TGaseq{Gamma2t} and \TGaseq{Gamma4t} read
\begin{eqnarray}
  \Gamma^{(2)}_{t_0,ab}
  &=&S^{(2)}_{ab}=iG^{-1}_{0,ab},
  \\
  \Gamma^{(4)}_{t_0,abcd}
  &=&S^{(4)}_{abcd}
  =-(2g/{\cal N})(\delta_{ab}\delta_{cd}+\delta_{ac}\delta_{bd}
     +\delta_{ad}\delta_{bc})
  \delta_{\cal C}(x_a-x_b)\delta_{\cal C}(x_b-x_c)\delta_{\cal C}(x_c-x_d).\ \ \ \ 
\end{eqnarray}
In order to arrive at a set of dynamic differential equations, we finally rewrite \TGasEq{Gamma2t} as
\begin{equation}
\label{TGas:eq:DynEqG2}
   iG_{0,ac}^{-1}G_{\tau_{cb},cb} 
   = i\delta_{{\cal C},ab}-\frac{1}{2}\int_{t_0,{\cal C}}^{\tau_{cb}}\!
   \Gamma^{(4)}_{\tau_{de},adce}\,G_{\tau_{de},ed}\,G_{\tau_{cb},cb},
\end{equation}
with $\delta_{{\cal C},ab}=\delta_{ab}\delta_{\cal
  C}(x_a-x_b)=\delta_{ab}\delta_{\cal
  C}(t_a-t_b)\delta(\mathbf{x}_a-\mathbf{x}_b)$.
\TGasEq{DynEqG2} is the dynamic (Dyson-Schwinger) equation for the connected two-point function $G_{\tau_{ab},ab}$.  
Since $\Gamma^{(2)}_{t_0,ab}=iG_0^{-1}$ represents a differential operator, the solution of \TGasEq{DynEqG2} finally requires another set of boundary conditions to be specified, depending on the form of the differential operator.  
In our case, this is the initial two-point function, fixed by the one-body density matrices as well as Bose statistics, see Sects.~\TGassect{GaussIniStates} and \TGassect{1DIniState}.
We also point out that we have set $t=\tau_{cb}$ since the flow of $G_{\tau,cb}$ stops at the maximum of the time arguments $t_b,t_c$.  
This can be proven from the structure of Eqs.~\TGaseq{Gamma4t}, \TGaseq{DynEqG2} but is more easily seen from the definition \TGaseq{defZtau}: 
Once the hard cutoff $\tau$ has passed the largest time appearing in a (connected) correlation function, the flow stops since the forward and backward parts of the CTP over greater times cancel identically in the functional integral.

Let us assume that $t=t_a$ denotes the present time, at which \TGasEq{DynEqG2} determines the further propagation of $G_{t_a,ab}$ for $t_b\le t_a$ (for $t_b>t_a$ the solution is then fixed by symmetry).  
All time arguments of the functions occurring on the right-hand sides of Eqs.~\TGaseq{Gamma2t}, \TGaseq{Gamma4t}, and therefore all cutoff times $\tau_{ab}$ and $\tau_{efgh}$ are evaluated at times $t'\le t$. 
Hence, in accordance with causality, Eqs.~\TGaseq{Gamma4t} and \TGaseq{DynEqG2} for a given initial $G_{t_0,ab}(t_0,t_0)$, can be solved iteratively.
\begin{figure}[tb]
\begin{center}
\resizebox{0.7\columnwidth}{!}{
\includegraphics{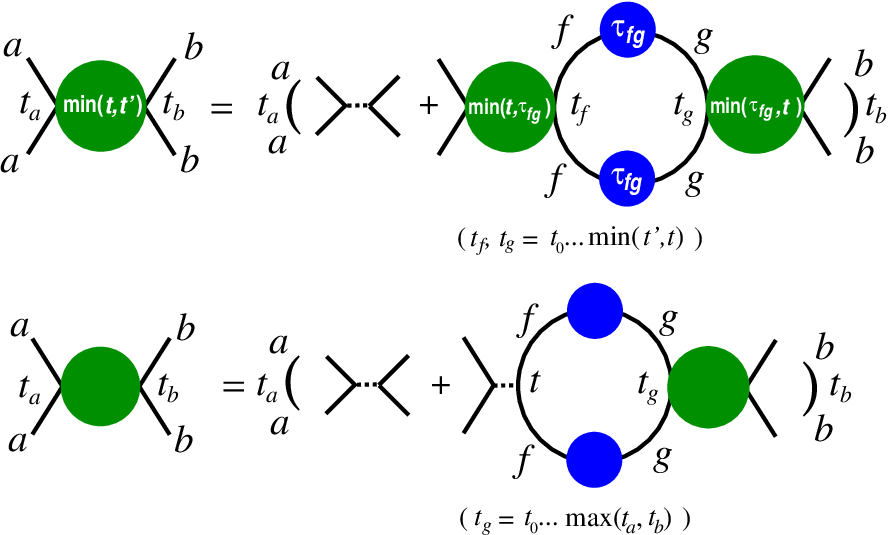}
}
\end{center}
\vspace*{-3ex} \caption{(color online) The upper equation is the
  $s$-channel projection of \TGasEq{Gamma4t}.  The second
  equation defines the resummed vertex appearing in the
  Dyson-Schwinger equation derived from the NLO $1/{\cal N}$
  approximation of the 2PI effective action
  \protect\cite{TGas:Berges:2001fi}.  The two definitions are identical in
  every order of a perturbative expansion (see text).  Dashed lines
  denote the $s$-channel part of the bare vertex $\Gamma_{t_0,ab}^s$,
  see text.  All other symbols correspond to those in
  \TGasFig{FlowEqs}.  Letters on internal lines indicate
  summation over field indices and integration over space and time
  (along the CTP from $t_0$ to $t$ and back).  The integration
  intervals are given in parentheses.  }
\label{TGas:fig:Bubblesum}
\end{figure}

This concludes the derivation of a closed set of dynamic equations for the two-point correlation function as obtained from a functional RG approach with a cutoff in real time.  
In the remainder of this section we will concentrate on rederiving the dynamic equations obtained from the 2PI effective action in NLO of the $1/{\cal N}$ expansion.

\subsubsection{From RG to 2PI next-to-leading order (NLO) $1/{\cal N}$}
\label{TGas:sec:Nonpert}
The mean-field approximation, i.e., the HFB equations, cf.~Sects.~\TGassect{HFB} and \TGassect{NLO1N}, are obtained by neglecting the flow of the four-point according to \TGasEq{Gamma4t}, $\Gamma^{(4)}_{t}\equiv\Gamma^{(4)}_{t_0}$.  
As a consequence, the flow parameter of $G_{\tau,ab}$ is, for the chosen cutoff, fixed to $\tau=\tau_{ab}$ and can thus be neglected.

As a first step beyond mean-field we consider the truncation in which the $s$-channel scattering diagram is included beyond the mean-field limit in the loop integral on the right-hand side of
\TGasEq{Gamma4t}.  
The result of this section will be that the obtained equations are equivalent to the 2PI equations in NLO of a $1/{\cal N}$ expansion.  
This truncation corresponds to keeping only one channel of each, the classical vertex and the one-loop integral term.  
The four-vertex then only depends on two space-time variables and field indices, $\Gamma^{(4)}_{t,acbd}=\Gamma^{(4)s}_{t,ab}\delta_{ac}\delta_{bd}$, and enters, in \TGasEq{DynEqG2}, as a cutoff-dependent self energy $\Sigma_{\tau_{ab},ab}\equiv\Gamma^{(4)s}_{\tau_{ab},ab}G_{\tau_{ab},ab}$.
We can always write $t=\mathrm{min}(t,t')$, with $t=t'=\tau_{ab}$, for the cutoff parameter in $\Gamma^{(4)s}_{t,ab}$, see \TGasFig{Bubblesum}, upper panel.  
Since the parameter in the (green) vertices on the right hand side is also the minimum of the maxima of integration times in the adjacent loop and the respective external time $t$ or $t'$, one can iterate the integral equation in order to obtain a perturbative series of bubble-chain diagrams consisting only of classical vertices and full, cutoff-dependent propagators.  
This procedure provides us with a proof that $\Sigma_{\tau_{ab},ab}$ is \emph{identical} to the NLO 2PI $1/{\cal N}$ self energy shown in the lower panel of \TGasFig{Bubblesum}.

A detailed formal proof of this result can be found in Ref.~\cite{TGas:Gasenzer2008a}.  
Here, we provide an argument that the above identity can be inferred in a comparatively easy way from the topology of the different terms in the flow equations for the two-, four- and six-point functions: 
Consider the untruncated set of equations as displayed in \TGasFig{FlowEqs}.  
First, non-$s$-channel contributions do not generate bubble-chains of the form shown in \TGasFig{Bubblesum}. 
Second, $\Gamma_\tau^{(6)}$ is one-particle irreducible. 
Its contribution to the flow of $\Gamma_\tau^{(4)}$ does not give rise to bubble-chains, even if inserted recursively into the first diagram on the right-hand side of the flow equation for $\Gamma_\tau^{(4)}$. 
In turn, by dropping the second diagram with the six-point function and using the $s$-channel truncation, the iterated flow equation generates only bubble-chain diagrams with full propagators as lines.  
Hence, a $\tau$-integration of this set of flow equations leads to dynamic equations which include all bubble-chain contributions and are therefore equal to those obtained in the NLO $1/{\cal N}$ expansion of the 2PI effective action, see \TGasSect{NLO1N}.
We emphasise that the above topological arguments are generally valid when comparing resummation schemes inherent in RG equations of the type of \TGasEq{flowGamma}, with those obtained from 2PI effective actions. 
This applies, e.g., to equilibrium flows \cite{TGas:Bagnuls:2000ae,TGas:Litim:1998nf,TGas:Pawlowski:2005xe} and thermal flows \cite{TGas:Litim:1998nf,TGas:Blaizot:2006rj}. 
For a comparison with 2PI results see Ref.~\cite{TGas:Blaizot:2006rj}, for the interrelation of 2PI methods and RG flows Ref.~\cite{TGas:Pawlowski:2005xe}.

\subsection{Dynamics of a one-dimensional Bose gas}
\label{TGas:sec:1DBoseEq}
In this section we apply the theoretical methods summarised above to describe the equilibration dynamics of a uniform ultracold gas of bosonic sodium atoms which are confined such that they can move in one spatial dimension only, see \TGasSect{LowDimTraps}. 
The gas is assumed to be initially far from thermal equilibrium. 
The ensuing equilibration process is found to happen on two different time scales. 
A fast dephasing period leads to a quasistationary state which shows certain near-equilibrium characteristics but is still far from being thermal. 
After this, the system approaches, within an at least ten times longer period, the actual equilibrium state.

\subsubsection{Initial conditions}
\label{TGas:sec:1DIniState}

The 2PI effective action approach is convenient for situations, where at time $t=t_0$ one has a Gaussian state, i.e., a state, for which all but the correlation functions of order one and two vanish.  
In the following we will consider a one-dimensional uniform system, for which the two-point functions $F_{ab}(x,y)$ and $\rho_{ab}(x,y)$ are spatially translation invariant. 
We therefore work in momentum space, where the kinetic energy operator is diagonal.
Moreover, we choose the field expectation value $\phi$ to vanish initially. 
Then, for reasons of number conservation, the equations of motion \TGaseq{EOMphi} and \TGaseq{EOMFrho} will conserve $\phi=0$ for all times, see \TGasSect{Obs}. 
We note that, since there is no spontaneous symmetry breaking in one spatial dimension at non-zero temperature, the field always approaches zero eventually, irrespective of its initial value.

Having prescribed initial values $F_{ab}(0,0;p)$ and $\rho_{ab}(0,0;p)$, the coupled system of integro-differential equations \TGaseq{EOMFrho} yields the time evolution of the two-point functions, in particular, of the momentum distribution
\begin{equation}
\label{TGas:eq:npoft}
  n(t,p) = \frac{1}{2}\Big(F_{11}(t,t;p)+F_{22}(t,t;p)-\alpha\Big),
\end{equation}
cf.~\TGasEq{nmitoF}.
We will later, in \TGasSect{QvsCl}, distinguish between quantum and classical statistical evolution.
For the quantum gas, one has $\alpha=1$ from the Bose commutation relations, while, for a gas following classical statistical evolution, $\alpha=0$.

We choose, at $t=0$, a Gaussian momentum distribution
\begin{equation}
\label{TGas:eq:npini}
  n(0,p) = \frac{n_1}{\sqrt{\pi}\sigma}e^{-p^2/\sigma^2}.
\end{equation}
which constitutes a far-from-equilibrium state if the interactions are non-zero and the corresponding interaction energy is much larger than the kinetic energy.

The initial pair correlation function vanishes,
$[F_{11}(t,t;p)-F_{22}(t,t;p)]/2+iF_{12}(t,t;p)=0$,
for $t=0$, in accordance with total atom number conservation at non-relativistic energies, see \TGasSect{Obs}.
Hence,
\begin{eqnarray}
\label{TGas:eq:IniValF}
  F_{11}(0,0;p)=F_{22}(0,0;p)
  =n(0,p)+\alpha/2,~~~~
  F_{12}(0,0;p)=F_{21}(0,0;p)
  \equiv0.
\end{eqnarray}
As far as the spectral functions are concerned, the Bose commutation relations \TGaseq{BoseCR12} imply:
\begin{eqnarray}
\label{TGas:eq:IniValrho}
  \rho_{11}(t,t;p)=\rho_{22}(t,t;p)
  \equiv0,~~~~
  -\rho_{12}(t,t;p)=\rho_{21}(t,t;p)
  \equiv1.
\end{eqnarray}

Let us investigate the dynamic evolution of a 1D Bose gas of sodium atoms with mass $m=22.99\,\TGasunit u$ ($^{23}\text{Na}$) in a box of length $L=N_sa_s$, with periodic boundary conditions. 
We choose the numerical grid such that it corresponds to a lattice of $N_s$ points in coordinate space with grid constant $a_s$, and the momenta on the Fourier transformed grid are $p_n=(2/a_s)\sin(n\pi/N_s)$ as explained in detail in Ref.~\cite{TGas:Berges:2004yj}.
The results presented in the following are obtained using $N_s=64$ modes on a spatial grid with grid constant $a_s=1.33\,\mu$m. 
We consider a line density of the atoms in the box of $n_1=10^7$ atoms$/$m. 
In this case the atoms are weakly interacting with each other, such that $g=g_\mathrm{1D}=\hbar^2\gamma n_1/m$, with the dimensionless parameter $\gamma=1.5\cdot10^{-3}$, see \TGasSect{LowDimTraps}. 
The width of the initial momentum distribution is chosen to be $\sigma=1.3\cdot10^5\,$m$^{-1}$.

\subsubsection{Equilibration of the quantum gas}
\label{TGas:sec:EquilQG}

To solve Eqs.~\TGaseq{EOMFrho}, with the self-energies given by Eqs.~\TGaseq{SigmaNLO1N} and the initial conditions in the previous section, a parallelised Runge-Kutta solver has been implemented and used on a cluster of $3\,$GHz dual processor PCs with up to one node per momentum mode. 
The correlation functions $F_{ab}(t,t^\prime;p)$ and $\rho_{ab}(t,t^\prime;p)$ were propagated, for fixed $t^\prime$, along $t$, using a second-order Runge-Kutta algorithm. 
After each Runge-Kutta step, the $I_{F,\rho}$ integrals were updated according to Eqs.~\TGaseq{IFrho}. 
The dynamic equations derived from the 2PI effective action are, by construction, number and energy conserving. 
While number conservation, by virtue of the O(2) symmetry of each diagram, is given exactly, energy conservation may be violated by the chosen discretization along the time axis.
Hence, in order to ensure optimal energy conservation numerically, a fourth-order Runge-Kutta algorithm was employed for the propagation of the correlation functions along the diagonal $t=t^\prime$.

\begin{figure}[tb]
\begin{minipage}[b]{0.50\columnwidth}
\begin{center}
\resizebox{1.0\columnwidth}{!}{
\includegraphics{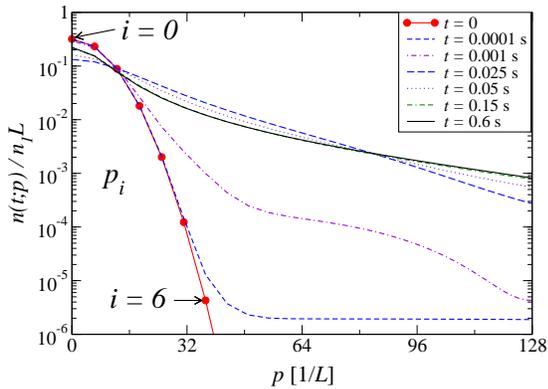}
}
\end{center}
\end{minipage}
\hspace*{0.03\columnwidth}
\begin{minipage}[b]{0.46\columnwidth}
\vspace*{-3ex} \caption{ (Color online) Momentum-mode distribution
$n(t;p)$ for the initial state (red filled circles, interpolated
by red solid line) and 6 subsequent times $t$ until no change can
be observed for $t>0.6\,$s. The interpolation of the final
distribution is shown as a black solid curve.
$n$ is normalised to the total number of
atoms, $n_1L=853$. The gas is in a far-from-equilibrium
state initially, characterised by a Gaussian distribution
$n(0;p)$, \TGasEq{npini}, with width
$\sigma=1.3\cdot10^5$m$^{-1}$. It is weakly interacting,
$\gamma=1.5\cdot10^{-3}$. Since we consider a homogeneous gas and
a symmetric initial state, the occupation numbers are invariant
under $p\to-p$. } 
\label{TGas:fig:n1ofp}
\end{minipage}
\end{figure}
\TGasFig{n1ofp} shows, as a (red) solid curve, the initial Gaussian momentum distribution of the gas, on a logarithmic scale, where it forms an inverted parabola. 
The filled circles indicate the numerically calculated modes $p_i$. 
In the same figure, the time evolution of the distribution is shown for different times between $t=0.1\,$ms and $0.6\,$s. 
For times greater than about $0.15\,$s, there is only very little change observed. 
As a function of time, the evolution of the single mode occupations is shown in \TGasFig{n1oft}. 
We observe that the system very quickly, after about $5\,\mu$s, evolves to a quasistationary state, and that the subsequent drift to the equilibrium distribution takes roughly ten times longer. 
In passing we note that the mean-field Hartree-Fock (HF) approximation, for which $\Sigma^F=\Sigma^\rho\equiv0$ in Eqs.~\TGaseq{EOMFrho}, conserves exactly all mode occupations and no equilibration is seen.

In order to estimate to which extent the final distribution approaches that of the actual equilibrium state of the gas, we fitted the distribution to the Bose-Einstein-like form $n(t;p)=[\exp\{(\omega(p)-\mu)/k_B\Theta(t;p)\}-1]^{-1}$, with a $p$-dependent temperature variable $\Theta(t;p)$. 
Here $\omega(p)$ was derived from the time-derivatives of the statistical function $F(t,t';p)$ at $t=t'$. 
If a Bose-Einstein distribution is approached the temperature can be obtained from the slope of $\log(n^{-1} + 1)$ and the chemical potential $\mu$ from its value at $\omega=0$. 
\TGasFig{betaofp} shows $\Theta(t;p)$ for $t=0...0.6\,$s. 
Obviously, during the quasistationary drift period, no temperature can be attributed to $n(t;p)$, while, for large $t$, $\Theta$ becomes approximately $p$-independent.
We deduce an approximate final temperature from $\Theta(0.6s;128/L)=T=0.35$nK with $\mu=1.08\,g_\mathrm{1D}n_1$ for the above given values of $g_\mathrm{1D}$ and $n_1$, which deviates from the HFB result $\mu=g_\mathrm{1D}n_1$ by about 8\%.

\begin{figure}[tb]
\begin{minipage}{0.44\columnwidth}
\begin{center}
\resizebox{1.0\columnwidth}{!}{
\includegraphics{TGas_fig15.eps}
}
\end{center}
\end{minipage}
\hspace*{0.05\columnwidth}
\begin{minipage}{0.49\columnwidth}
\begin{center}
\resizebox{1.00\columnwidth}{!}{
\includegraphics{TGas_fig16.eps}
}
\end{center}
\end{minipage}
\vspace*{2ex}\ \\
\begin{minipage}[t]{0.48\columnwidth}
\caption{
The normalised momentum-mode occupation numbers $n(t;p)/n_1L$,
corresponding to those shown in \TGasFig{n1ofp}, as
functions of time. Shown are the populations of the modes with
$p=p_i=2N_s/L\sin(i\pi/N_s)$, $i=0,1,...,N_s/2$, and one has
$n(t;-p)=n(t;p)$. A fast short-time dephasing period is followed
by a long quasistationary drift to the final equilibrium
distribution. Notice the double-logarithmic scale. }
\label{TGas:fig:n1oft}
\end{minipage}
\hspace*{0.01\columnwidth}
\begin{minipage}[t]{0.49\columnwidth}
\caption{Momentum and time dependent temperature variable $\Theta(t;p)$
obtained by fitting the distribution
$n(t;p)=[\exp\{(\omega(p)-\mu)/k_B\Theta(t;p)\}-1]^{-1}$ to the
distribution obtained from the results shown in
\TGasFig{n1oft}, for different, equally spaced times
between $t=0$ and $t=0.6\,$s. One observes that, during the
quasistationary period, $0.01\,$s$<t<0.1\,$s, no temperature can
be associated to the distribution. Only at very large times,
$\Theta$ becomes approximately $p$-independent.}
\label{TGas:fig:betaofp}
\end{minipage}
\end{figure}
In summary, the dynamic equations derived from the 2PI effective action expanded beyond the mean-field approximation can be applied to derive the evolution of a one-dimensional Bose gas starting with a momentum distribution far from equilibrium.
Their solutions show a non-secular evolution towards a state characterised by a thermal momentum distribution. 
The evolution is characterised by a fast initial dephasing followed by a long slow drift to thermal equilibrium.

A remark is in order:
As pointed out in \TGasSect{LowDimTraps}, a homogeneous 1D Bose gas with contact interactions represents an integrable system.
Hence, at large times, recurrence phenomena are expected which therefore prevent a thermalisation of the system.
This does not imply that the above results obtained for a finite-time evolution are invalid.
We point out that they were obtained for a weakly interacting gas for which the approximations made are expected to be justified.
Nonetheless, questions remain open --- whether in an actual system a quasi thermal state is reached and, if so, on which time scales and under which conditions the integrability of the system drives it away from there again \cite{TGas:Kinoshita2006a,TGas:Rigol2007a,TGas:Gangardt2008a}.

\section{From nonequilibrium to kinetic equations}
\label{TGas:sec:Kin}
Far-from-equilibrium dynamics is qualitatively different from evolution near equilibrium where distinct properties of the equilibrium state are still ``felt'' by the system.
We have mentioned a few times above that the statistical correlation function $F$ is, near equilibrium, no longer independent from the spectral function $\rho$.
Their interrelation is a manifestation of the fluctuation-dissipation relation well-known in nonequilibrium statistical physics.
Before we set our focus on how near-equilibrium behaviour emerges during the dynamical evolution starting far from equilibrium we illustrate the fluctuation-dissipation theorem in the framework of linear-response theory.

\subsection{Linear-response theory}
\label{TGas:sec:LinRes}
Analysing a simple quantum mechanical linear dissipative model, such a theorem can be easily derived from the properties of perturbation and response functions.
Let us consider a general many-body Hamiltonian $H$, perturbed in time around some equilibrium Hamiltonian $H_0$ by
\begin{equation}
\label{TGas:eq:LRH1}
  H_1(t) = - h e^{-i(\omega+i\epsilon)t} + \mathrm{h.c.},
\end{equation}
where $h$ is an operator inducing a small perturbation, and the regulator suppresses the perturbation for $t\to-\infty$.
In linear-response theory one studies the fluctuation
\begin{equation}
\label{TGas:eq:LRdeltaf}
  \delta\langle f^\dagger\rangle
  = \chi_{f^\dagger,h}(\omega)e^{-i(\omega+i\epsilon)}+\mathrm{c.c.},
\end{equation}
where $\chi_{f^\dagger,h}$ is commonly called the dynamic polarisability.
In perturbation theory, one finds
\begin{equation}
\label{TGas:eq:LRchifh}
  \chi_{f^\dagger,h}(\omega)
  =-Z^{-1}\sum_{m,n}e^{-\beta E_m}\left[
   \frac{\langle m|f^\dagger|n\rangle\langle n|h|m\rangle}
        {\omega-\omega_{mn}+i\epsilon}
  -\frac{\langle m|h|n\rangle\langle n|f^\dagger|m\rangle}
        {\omega+\omega_{mn}+i\epsilon}
   \right],
\end{equation}
with $\omega_{mn}=E_m-E_n$, $H_0|n\rangle=E_n|n\rangle$, $Z=\sum_n\exp\{-\beta E_n\}$, and where thermal equilibrium is assumed for $t\to-\infty$.
One also defines the \textit{dynamic structure factor}
\begin{equation}
\label{TGas:eq:LRSF}
   S_f(\omega)
   = Z^{-1}\sum_{m,n}e^{-\beta E_m}|\langle n|f|m\rangle|^2\delta(\omega-\omega_{nm}),
\end{equation}
which vanishes for $\omega<0$ and $T\to0$. 
In the case that $f=h$ one then finds that the dynamic structure factor is related as follows to the response function $\chi_f\equiv\chi_{f^\dagger,f}$.
\begin{equation}
\label{TGas:eq:LRchifSf}
  \chi_f(\omega)
  = -\int_{-\infty}^{\infty}\mathrm{d}\omega'\left[
   \frac{S_f(\omega')}{\omega-\omega'+i\epsilon}
  -\frac{S_{f^\dagger}(\omega')}{\omega+\omega'+i\epsilon}\right].
\end{equation}
Using the relation $(x-a\pm i\epsilon)^{-1}={\cal P}(x-a)^{-1}\mp i\pi\delta(x-a)$ involving the Cauchy principal part ${\cal P}$ one finds that the imaginary part of the response function is related to the dynamic structure factor as follows,
\begin{equation}
\label{TGas:eq:LRImchiSf}
  \mathrm{Im}\chi_f(\omega) = \pi(1-e^{-\beta\omega})S_f(\omega),
\end{equation}
where we have used that, at finite temperatures, one has $S_f(\omega)=e^{\beta\omega}S_{f^\dagger}(-\omega)$.
The imaginary part of $\chi$ is usually called the spectral density related to the operator $f$,
\begin{equation}
\label{TGas:eq:LRrhof}
  \rho_f(\omega)
  =Z^{-1}\sum_{m,n}(e^{-\beta E_m}-e^{-\beta E_n})
   |\langle n|f|m\rangle|^22\pi\delta(\omega-\omega_{nm})
  =2\mathrm{Im}\chi_f(\omega).
\end{equation}
In second-order perturbation theory one finds the rate of energy transfer to the system to be $dE/dt=2\omega\mathrm{Im}\chi_f(\omega)+\mathrm{terms}\propto\exp\{2i\omega\}$ which provides us with a physical interpretation of this quantity.

A sum-rule argument can now be used to relate the zeroth-order moment of the structure factor \TGaseq{LRSF} to the fluctuations of $f$,
\begin{equation}
\label{TGas:eq:LRSumRule}
  \int_{-\infty}^{\infty}\mathrm{d}\omega \left(S_f(\omega)+S_{f^\dagger}(\omega)\right)
  =\langle\{f,f^\dagger\}\rangle_\beta,
\end{equation}
where $\langle\cdot\rangle_\beta=Z^{-1}\sum_me^{-\beta E_m}\langle m|\cdot|m\rangle$, and where the completeness of the states $|m\rangle$ has been used. 
Using again $S_f(\omega)=e^{\beta\omega}S_{f^\dagger}(-\omega)$ one finds the fluctuation-dissipation relation
\begin{equation}
\label{TGas:eq:LRFlucDissInt}
  \langle\{f,f^\dagger\}\rangle_\beta
  =\frac{1}{\pi}\int_{-\infty}^{\infty}\mathrm{d}\omega\rho_f(\omega)
   \left(n_\mathrm{BE}(\omega)+\frac{1}{2}\right),
\end{equation}
with the Bose-Einstein distribution $n_\mathrm{BE}(\omega)=(\exp\{\beta\omega\}-1)^{-1}$.

\subsection{Equilibrium field theory}
\label{TGas:sec:FlucDiss}
In thermal equilibrium, correlation functions can be determined using the functional quantum field theoretical approach discussed in \TGasSect{NEqFT} if the real-time closed time path is replaced by the imaginary path ${\cal C}=[0,-i\beta]$.
$\beta$ denotes the inverse temperature.
Let us consider a spatially uniform system.
Since thermal equilibrium is time translation invariant, the two-point functions depend only on relative space and time coordinates.
In Fourier space one has, e.g.,
\begin{equation}
\label{TGas:eq:Feq}
  F^\mathrm{(eq)}_{ab}(x,y)
  =\int \frac{\mathrm{d}\omega\,d^dp}{(2\pi)^{d+1}}e^{-i\omega(x_0-y_0)+i\vec{p}(\vec{x}-\vec{y})}
  F^\mathrm{(eq)}_{ab}(\omega,\vec{p}).
\end{equation}
Since, in equilibrium, the path integral determines the expectation value with respect to the density matrix $\rho\propto\exp\{-\beta H\}$, the initial and final states must be set equal.
This leads to the periodicity condition for the imaginary-time propagator, $G_{ab}(x,y)|_{x_0=0}=G_{ab}(x,y)|_{x_0=-i\beta}$.
Inserting \TGasEq{GitoFrho} into this condition one obtains the relation
\begin{equation}
\label{TGas:eq:FlucDissFrho}
   F^\mathrm{(eq)}_{ab}(\omega,\vec{p})
   =-i\left(n_\mathrm{BE}(\omega)+\frac{1}{2}\right)
   \rho^\mathrm{(eq)}_{ab}(\omega,\vec{p})
\end{equation}
which, integrated over $\omega$, is identical to the fluctuation-dissipation relation \TGaseq{LRFlucDissInt} \cite{TGas:KadanoffBaym1962a,TGas:Aarts:1997kp}.
In deriving \TGasEq{FlucDissFrho} we have used that the time $x_0=0$ is the earliest on the imaginary path ${\cal C}$ while $x_0=-i\beta$ is the latest, such that $\mathrm{sgn}_{\cal C}(x_0-y_0)$ yields opposite signs in the two limits.
An equivalent fluctuation-dissipation relation can be derived, using the same arguments, for the equilibrium self-energy contributions $\Sigma_F$ and $\Sigma_\rho$.

We emphasise that in nonequilibrium field theory, the relation \TGaseq{FlucDissFrho} no longer holds.
Far-from-equilibrium dynamics in this respect allows for a greater variety of solutions.
Nonetheless, nonequilibrium field theory encompasses the near-equilibrium as well as vacuum theories, where $n_\mathrm{BE}(\omega)\to0$.

\begin{figure}[tb]
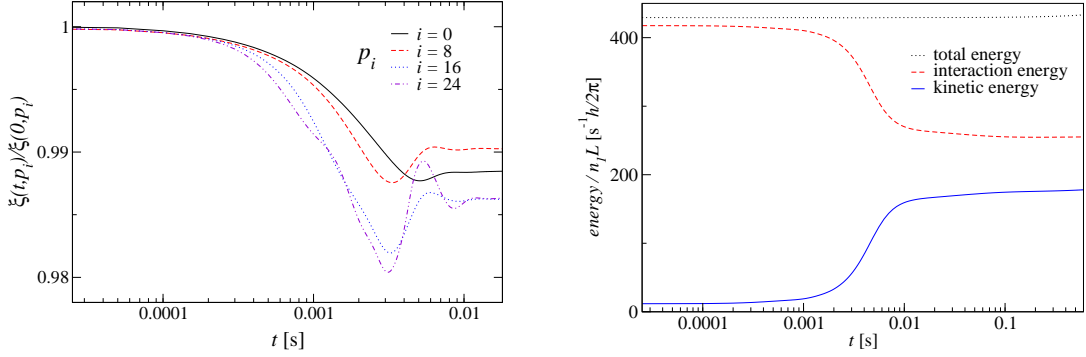

\begin{center}
\begin{minipage}{0.46\columnwidth}
\resizebox{1.0\columnwidth}{!}{
\includegraphics{TGas_fig17a.eps}
}
\end{minipage}
\hspace*{0.05\columnwidth}
\begin{minipage}{0.47\columnwidth}
\resizebox{1.0\columnwidth}{!}{
\includegraphics{TGas_fig17b.eps}
}
\end{minipage}
\vspace*{2ex}\ \\
\caption{
(Color online)
\textit{Left panel}:
Ratio of the envelopes of the unequal-time correlation functions, $\xi(t;p)=[(F_{11}(t,0;p)^2+F_{12}(t,0;p)^2)/(\rho_{11}(t,0;p)^2+\rho_{12}(t,0;p)^2)]^{1/2}/n(t;p)$, for four different momentum modes, as a function of time.
The evolution corresponds to that shown in \TGasFig{n1oft}, \TGasSect{EquilQG}.
Due to the normalization with respect to $n(t;p)$ all $\xi(t;p)$ are of the same order of magnitude.
$\xi$ is a measure of the interdependence of the statistical and spectral functions, and its settling to a constant value during the quasistationary drift period indicates that these functions become, as in thermal equilibrium, connected through a fluctuation-dissipation relation.
\textit{Right panel}:
Evolution of the kinetic and interaction contributions to the total energy of the gas.
During the quasistationary drift these contributions assume the same order of magnitude, calling in mind the virial theorem.
}
\label{TGas:fig:flucdiss}
\label{TGas:fig:energies}
\end{center}
\end{figure}

\subsection{Near-equilibrium time evolution of a 1D Bose gas}
\label{TGas:sec:NearEq1DBose}
We now study the equilibration process of the one-dimensional Bose gas presented in  \TGasSect{EquilQG} with respect to the characteristics of near-equilibrium evolution introduced in the previous section.
For this we consider the time dependence of the ratio of the envelopes of the unequal-time correlation functions, specifically, $\xi(t;p)=[(F_{11}(t,0;p)^2+F_{12}(t,0;p)^2)/(\rho_{11}(t,0;p)^2+\rho_{12}(t,0;p)^2)]^{1/2}/n(t;p)$.
Confer Refs.~\cite{TGas:Gasenzer:2005ze,TGas:Berges:2007ym} for more details.
Similar studies for relativistic gases have been presented in Refs.~\cite{TGas:Berges2001a,TGas:Aarts:2001qa}.
\TGasFig{flucdiss} (left panel) shows $\xi$, for four different momentum modes, as a function of time.
Due to the normalization with respect to $n(t;p)$ all $\xi(t;p)$ are of the same order of magnitude.
However, they show a distinct time evolution during the dephasing period, before they settle to a constant value during the quasistationary drift.
$\xi$ is a measure of the interdependence of the statistical and spectral functions, which, in thermal equilibrium, are connected through the fluctuation-dissipation relation given, in the momentum-frequency domain, in \TGasEq{FlucDissFrho} above. 
Hence, the time at which $\xi$ becomes stationary indicates that $F$ and $\rho$ are linked to each other long before the momentum distribution becomes thermal, as can be seen from \TGasFig{betaofp}.

A further signature is found when comparing the kinetic and interaction contributions to the total energy as shown in \TGasFig{energies}, right panel.
During the quasistationary drift, these contributions are constant and of the same order of magnitude, calling in mind the virial theorem.

In summary, during the drift period, as can be seen in Figs.~\TGasfig{n1ofp} and \TGasfig{betaofp}, the system is not yet in equilibrium as far as the momentum distribution and temperature are concerned.
The statistical and spectral correlation functions nevertheless are already locked to each other as shown in \TGasFig{flucdiss}, left panel.
Hence, one may expect that during this period the dynamics can be described using kinetic or transport equations, i.e., that it fulfills the requirements for the approximations implied by quantum Boltzmann equations.
We shall study this in more detail in the following sections.

\subsection{Transport equations}
\label{TGas:sec:TransportEq}
So far we have dealt with initial-value problems that describe the time evolution by means of in general coupled equations of motion for time dependent correlation functions, given specific values for them at the initial time.
Alternatively, the time-evolution of many-body systems is very often described in terms of kinetic or transport equations \cite{TGas:KadanoffBaym1962a}.
In general, the aim of quantum kinetic theory is to find evolution equations for distribution functions $f(\TGasF x, t)$ or $\tilde f(\TGasF p,t)$, interpreted, e.g., as particle number densities in $\TGasF x$- or $\TGasF p$-space.
These distribution functions can then be used to derive various transport properties of the many-particle system as, e.g., current of charge or energy. For that reason, the evolution equations for $f$ are generally referred to as transport equations.

Kinetic descriptions usually neglect the effect of correlations between different times of the evolution, i.e., they build, to a certain extent, on a Markovian approximation.
In particular, they neglect the initial dynamics directly after a change in the boundary conditions which drive the system out of equilibrium.
This shortcoming is cured in a dynamical approach as discussed above.
The buildup of correlations beyond the kinetic approximation, in these equations, is taken into account by means of non-Markovian integrations over the evolution history of correlation functions.

In the context of non-relativistic systems, in particular cold atomic gases cf., e.g., Refs.~\cite{TGas:KadanoffBaym1962a}, \cite{TGas:Proukakis1996a,TGas:Shi1998a,TGas:Gardiner1997a,TGas:Giorgini1997a,TGas:Proukakis1998a,TGas:Walser1999a,TGas:Gardiner2000a,TGas:Walser2000a,TGas:Rey2005a,TGas:Baier:2004hm}.
See also Refs.~\cite{TGas:Danielewicz1984a,TGas:Mrowczynski:1992hq,TGas:Boyanovsky:1996xx,TGas:Ivanov:1998nv,TGas:Lipavsky:2001di,TGas:Calzetta1988a,TGas:Blaizot:2001nr,TGas:Berges:2002wt,TGas:Prokopec:2003pj}, \cite{TGas:Konstandin:2004gy,TGas:Lindner:2005kv,TGas:Berges:2005md} 
in the context of relativistic physics.
A comparison of dynamic equations with their kinetic approximation for relativistic dynamics has been given in Refs.~\cite{TGas:Danielewicz1984a,TGas:KohlerHS1995a,TGas:Morawetz1999a,TGas:Juchem:2003bi,TGas:Lindner:2005kv,TGas:Berges:2005md}.
In this section, we derive transport equations from the 2PI dynamic equations \TGaseq{EOMFrho}, in leading order (LO) and next-to-leading order (NLO) of a gradient expansion.
See Refs.~\cite{TGas:Berges:2005md,TGas:Branschadel2008a} for more detailed discussions.
The field expectation value $\phi$ is set to zero as before.
If not stated explicitely otherwise, field indices $a,b,...$ are suppressed in the following, and all products of correlation functions are to be taken as matrix products.

We rewrite the exact dynamic equations \TGaseq{EOMFrho} with the self energy given in NLO $1/{\cal N}$ approximation as described in \TGasSect{NLO1N} in order to get evolution equations with respect to centre and relative coordinates, providing a starting point for a gradient expansion and transformation to Wigner space.
For simplicity we consider a system with spatially homogeneous initial conditions for the two-point function.
The initial values of the spectral function are fixed by the commutation relations, cf.~\TGasEq{IniValrho},
$
  \rho_{0,ab}(\mathbf{x},\mathbf{y})
  = -i\sigma_{2,ab}\delta(\mathbf{x}-\mathbf{y})
$.

To meet the requirements of the Fourier transformed equations in Wigner space on the one hand and to benefit from the spatial homogeneity of the system on the other hand, we rewrite the equations in terms of relative and centre space-time coordinates
\begin{equation}
    X = \frac{x+y}2, s = x-y ~~ \Leftrightarrow ~~ x = X+s/2,~ y = X-s/2
\end{equation}
where the two-point functions are to be transformed as, e.g.,
      $F(x,y) \mapsto F(X,s)$,
      $\Sigma^\rho(x,z) \mapsto \Sigma^\rho(X+s'/2,s-s')$,
      $F(z,y) \mapsto F(X+(s'-s)/2, s')$.
We additionally introduce
    $s' = z-y$
which later on serves as the integration variable in the memory integrals.

Using these definitions in Eqs.~\TGaseq{EOMFrho}, and adding these equations to the corresponding equations for the time derivatives with respect to $y_0$, one obtains differential equations with respect to the centre time coordinate $X_0$:
\begin{eqnarray}
   &&\TGaspdiff{X_0}F(X,s) 
     = i\sigma_2
         \Bigg\lbrace
           \Big[{\hat H}_{\mathrm{1B}}(X+\frac s 2)-{\hat H}_{\mathrm{1B}}(X-\frac s 2)
     +\Sigma^{(0)}(X+\frac s 2) -\Sigma^{(0)}(X-\frac s 2)\Big]F(X,s) 
     \nonumber \\
     && +\int_{s'}\theta(X_0+s'_0-\frac{s_0}2)\Big[
     \Sigma^R(X+\frac {s'} 2 , s-s') F(X-\frac{s-s'}2, s') 
     \nonumber \\
     && ~~~~~~~~~~ 
     - G^R(X+\frac{s'}2, s-s') \Sigma^F(X-\frac{s-s'}2,s') 
     +\Sigma^F(X+\frac {s'} 2 , s-s')G^A(X-\frac{s-s'}2, s') 
     \nonumber \\
     && ~~~~~~~~~~ 
     - F(X+\frac{s'}2, s-s') 
     \Sigma^A(X-\frac{s-s'}2,s')\Big]\Bigg\rbrace,
\label{TGas:eq:TE:FAbs}
\end{eqnarray}
\begin{eqnarray}
    &&\TGaspdiff{X_0}\rho(X,s)
     = i\sigma_2 \Big\lbrace
       \big[{\hat H}_{\mathrm{1B}}(X+\frac s 2)-{\hat H}_{\mathrm{1B}}(X-\frac s 2)
     +\Sigma^{(0)}(X+\frac s 2) 
     -\Sigma^{(0)}(X-\frac s 2) \big]\rho(X,s) 
     \nonumber \\
     && 
     +\int_{s'}\big[ \Sigma^R(X+\frac {s'}2,s-s') \rho(X-\frac{s-s'}2,s') 
     \nonumber \\
     && ~~~~~~~~~~ 
     -\rho(X+\frac{s'}2,s-s')\Sigma^A(X-\frac{s-s'}2,s') 
     +\Sigma^\rho(X+\frac{s'}2,s-s')G^A(X-\frac{s-s'}2, s') 
     \nonumber \\
     && ~~~~~~~~~~ 
     -G^R(X+\frac{s'}2,s-s')\Sigma^\rho(X-\frac{s-s'}2, s')
     \big]
    \Big\rbrace.
\label{TGas:eq:TE:rhoAbs}
\end{eqnarray}
Note that, introducing retarded and advanced Greens functions and self-energies,
\begin{eqnarray}
\label{TGas:eq:TE:rhoAR}
    G^R(x,y)
    &=&\theta(x_0-y_0)\rho(x,y),  ~~~~\,
    G^A(x,y)
    =-\theta(y_0-x_0)\rho(x,y), \\
\label{TGas:eq:TE:SigmaAR}
    \Sigma^R(x,y)
    &=&\theta(x_0-y_0)\Sigma^\rho(x,y),  ~~
    \Sigma^A(x,y)
    =-\theta(y_0-x_0)\Sigma^\rho(x,y), 
\end{eqnarray}
allowed us to send the integration limits of $s'_0$ in Eqs.~\TGaseq{TE:FAbs}, \TGaseq{TE:rhoAbs} to $\pm\infty$.
While the above equations have been obtained by adding the equations for $\partial_{x_0}F(x,y)$ and $\partial_{y_0}F(x,y)$, etc., a second set of equations for the derivatives of $F(X,s)$ and $\rho(X,s)$ with respect to $s_0$ results when subtracting the respective expressions.
These equations are provided and discussed further in Appendix A of Ref.~\cite{TGas:Branschadel2008a}.

\subsubsection{Approximations}
\label{TGas:sec:Approx}
With the aim to derive transport equations, we apply the following approximations:
\\

({\sl i})
The $\theta$-function is neglected in the evolution equations for $F$, Eqs.~\TGaseq{TE:FAbs}, taking into account that the correlations disappear for large relative times.
This corresponds to sending the initial time $t_0$ to the infinite past.
Note that, since an interacting system could have reached equilibrium at any finite time, transport equations are initialised by specifying $F$ and $\rho$ at a finite time using the equations with $t_0\to-\infty$ as approximate description \cite{TGas:Berges:2005md}.
\\

({\sl ii})
A \textit{gradient expansion} is applied with respect to the centre coordinates $X$.
In leading order this expansion corresponds to a Markovian approximation which neglects the variation of the correlation functions along the centre time coordinate.
The approximation is valid as long as this variation is small on the time scale set by the size of the memory kernel.
Physically, non-Markovian effects are expected to be small for systems not too far from thermal equilibrium.
In higher orders, this expansion is a derivative expansion of the correlation functions with respect to the centre time and takes into account non-Markovian effects.
\\

In leading order of the gradient expansion, one obtains from Eqs.~\TGaseq{TE:FAbs} and \TGaseq{TE:rhoAbs}, by keeping the dependence of all functions $\Sigma$ and $G$ on $X$ only and combining retarded and advanced functions to spectral functions, the evolution equations
\begin{eqnarray}
\label{TGas:eq:TE:motionF_LO}
   \TGaspdiff{X_0}F(X,s)
    &=& i\sigma_2
          \int_{s'}\Big[F(X, s-s')\Sigma^\rho(X,s')-\rho(X,s-s')\Sigma^F(X, s')\Big],
    \\
    \TGaspdiff{X_0}\rho(X,s) &=& 0.
\end{eqnarray}
Note that we have assumed homogeneous initial conditions such that all functions remain constant along the spatial centre coordinate.
In next-to-leading order in the gradient expansion with respect to the centre time coordinate, these equations receive additional corrections. 
They are given explicitely in Ref.~\cite{TGas:Branschadel2008a}, and we suppress them here for conciseness of the presentation.
The finally resulting NLO equations will be quoted below.

\subsubsection{Transformation to Wigner space}
\label{TGas:sec:TransWigner}
  The transformation to Wigner space involves a Fourier transformation with respect to the relative coordinate $s$
\begin{equation}
    \tilde F(X,p) = \int_{s} e^{i ps} F(X,s), \qquad
    \tilde \Sigma^R(X,p) = \int_{s} e^{i ps} \Sigma^R(X,s),
\end{equation}
etc., where $ps=p_0 s_0-\TGasF p\cdot\TGasF s$.

For a spatially homogeneous system one finds that the diagonal matrix elements of $\tilde F$ are purely real, while the off-diagonal matrix elements are purely imaginary.

We apply the transformation to the previously derived equations of motion and obtain
\begin{eqnarray}
\label{TGas:eq:TE:motionFWigner_LO}
      \TGaspDiff{X_0}[\tilde F(X,p)]
      &=& i\sigma_2
         \Big\lbrace
      \tilde F(X, p)\tilde\Sigma^\rho(X, p) 
      - \tilde \rho(X, p) \tilde \Sigma^F(X,p) \Big\rbrace
      \\
\label{TGas:eq:TE:motionRhoWigner_LO}
      \TGaspDiff{X_0}[\tilde \rho(X,p)]
      &=&\,0
\end{eqnarray}
as the leading-order transport equations.  
In deriving these one needs to take care of the integration limits when interchanging the derivative with respect to the centre time with the relative time integration.
This enters the definition of the derivative operator
\begin{eqnarray}
    &&\TGaspDiff{X_0}[\tilde F(X,p)]
    =
     \int_{-2X_0}^{2X_0} \mathrm{d} s_0\, \int\mathrm{d}^3 s \,  e^{i ps} \TGaspdiff{X_0} F(X,s) 
     \nonumber \\
    &&~ =~ \TGaspdiff{X_0} \tilde  F(X,p) -2\Big[e^{2i p_0 X_0}F(X, s_0=2X_0, \TGasF p) 
    + e^{-2i p_0 X_0}F(X, s_0=-2X_0, \TGasF p)\Big].
\label{TGas:eq:TE:limits1}
\end{eqnarray}
Assuming $\lim_{s_0\rightarrow\infty}F(s_0) = 0 = \lim_{s_0\rightarrow\infty}\rho(s_0)$ yields
$
    \lim_{X_0\rightarrow\infty}\TGaspDiff{X_0}=\TGaspdiff{X_0}.
$
The contributions from the integration limits account for the fact that the correlation functions are initialised at some time $x_0=y_0=0$. These contributions can be removed for sufficiently late times as the time correlations are expected to vanish for sufficiently large relative times.

In next-to-leading order of the gradient expansion, the transport equations receive additional contributions as follows.
  \begin{eqnarray}
\label{TGas:eq:TE:motionFWigner_NLO}
    \TGaspDiff{X_0}[\tilde F]
    &=&\,\mathrm{LO} +\sigma_2\Bigg\lbrace 
    \big[\TGaspdiff{X_0} \Sigma^{(0)}(X)\big] \TGaspdiff{p_0} \tilde F(X,p) 
    \nonumber\\
    && 
    - \frac 1 2\Big[
        \lbrace
          \tilde\Sigma^+(X,p),\tilde F(X,p)
        \rbrace_0
       +\lbrace
          \tilde \Sigma^F(X,p),\tilde \rho^+(X,p)
        \rbrace_0 \Big]\Bigg\rbrace,
    \\
    \TGaspDiff{X_0}[\tilde\rho]
    &=&\, \sigma_2\Bigg\lbrace \big[\TGaspdiff{X_0} \Sigma^{(0)}(X)\big] \TGaspdiff{p_0} \tilde\rho(X,p) \nonumber\\
    && - \frac 1 2\Big[
        \lbrace
          \tilde\Sigma^+(X,p),\tilde\rho(X,p)
        \rbrace_0
       +\lbrace
          \tilde \Sigma^\rho(X,p),\tilde \rho^+(X,p)
        \rbrace_0 \Big]\Bigg\rbrace,
  \end{eqnarray}
These contributions involve Poisson brackets with respect to $X_0$ and $p_0$:
  \begin{equation}
    \lbrace\tilde A,\tilde B\rbrace_0 = \TGaspdiffer{\tilde A}{p_0}\TGaspdiffer{\tilde B}{X_0}-\TGaspdiffer{\tilde A}{X_0}\TGaspdiffer{\tilde B}{p_0}.
  \end{equation}
%

\subsection{Comparison of dynamical and transport equations}
\label{TGas:sec:DynvsTrans}
It is desirable to compare the transport equations derived in the preceding section, in leading order, \TGasEq{TE:motionFWigner_LO}, and next-to-leading order, \TGasEq{TE:motionFWigner_NLO}, of the gradient expansion, with the equations of motion \TGaseq{EOMFrho}.
This is achieved in two steps.
First, one calculates $F$ and $\rho$ using equations \TGaseq{EOMFrho} in the NLO $1/\mathcal N$-approximation.
The results of this calculation are taken to calculate, after a Fourier transformation to Wigner space, both the left-hand and the right-hand sides of equations \TGaseq{TE:motionFWigner_LO} and \TGaseq{TE:motionFWigner_NLO}.
The left-hand side yields the time derivatives of the solutions of the 2PI dynamic equations with respect to the centre time coordinate, while the right-hand side generates the corresponding approximative derivatives in leading order and next-to-leading order of the gradient expansion.

\subsubsection{Physical setup}
We consider the same dilute homogeneous one-dimensional ($D=1$) gas of sodium atoms as defined in \TGasSect{1DIniState}.
For the results presented below the gas was chosen to be confined in a periodic box of length $L=N_s a_s\approx 43\,\TGasunit{\mu m}$.
The spacing of the numerical grid is $a_s=1.33\,\TGasunit{\mu m}$, and the number of lattice points is $N_s=32$.
Using different initial line densities $n_1$ between $10^5\,\TGasunit m^{-1}$ and $10^7\,\TGasunit m^{-1}$ the total particle number varies in the range from $\sim4$ to $\sim400$ particles.

Three different interaction strengths $\gamma$ are considered, see \TGasSect{LowDimTraps}, $\gamma=1.5\cdot10^{-3}$, $0.15$, $15$, with $\gamma n_1^2=mgn_1$ kept fixed by choosing the line densities accordingly.
As will be discussed in \TGasSect{QvsCl}, the effect of quantum fluctuations is, in this way, increased with $\gamma$ while the dynamics in the classical statistical approximation remains unchanged, see \TGasSect{QvsCl}.
The correlation functions are initialised as described in \TGasSect{1DIniState}, i.e., the initial momentum distribution $F(0,0,\TGasF p)$ is chosen to be Gaussian, peaked around $\mathbf{p}=0$, with a width of $\sigma=6.5\cdot 10^4\, \TGasunit m^{-1}$,
\begin{figure}[tb]
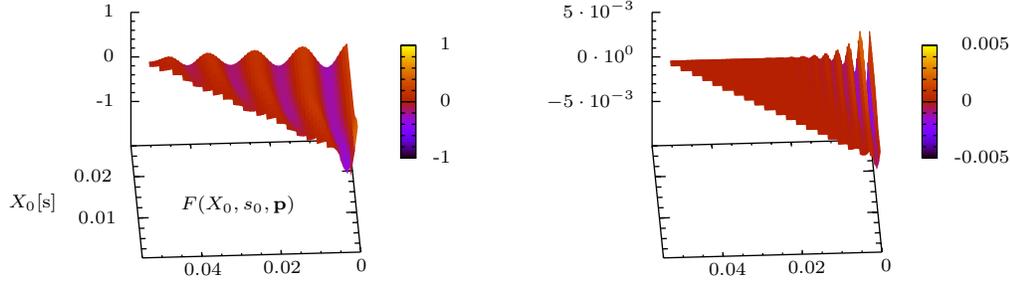

\begin{flushright}
     \ \\[1ex]
      \begin{scriptsize}
     \include{TGas_fig18}
      \end{scriptsize}
    \caption[]{The statistical two-point function $F$ for two different momentum modes as a function of the relative and centre time coordinates $s_0$ and $X_0$, respectively.
Left panel: momentum mode $n_p=0$, right panel: $n_p=8$.
The figures illustrate that the correlation function vanishes for sufficiently large relative times.
See the main text for the parameters chosen.
}
\label{TGas:fig:NR:Xs3d}
\end{flushright}
\end{figure}
and the Bose commutation relations fix the initial spectral function, $\rho(x_0,x_0,\TGasF p) = -i\sigma_2$.

\subsubsection{Numerical solution of the dynamic equations}
\label{TGas:sec:2PINumerics}
The 2PI dynamic equations are solved using the techniques described in \TGasSect{EquilQG} and their solutions transformed to Wigner space.
We expect the transport equations to reproduce, to a good approximation, the full dynamical evolution, if the two-point functions vary slowly with respect to the centre time coordinate $X_0$ as compared to the change with the relative time coordinate $s_0$.
Figure \TGasfig{NR:Xs3d} shows $F(X_0, s_0, \TGasF p)$ for two different momentum modes.
We see that for sufficiently late times the functions fall off to zero with increasing relative time $s_0$.
The spectral function $\rho$ shows a similar behaviour \cite{TGas:Branschadel2008a}, and the self energy contributions $\Sigma^{F,\rho}$ decay even faster in $s_0$ as they involve $F$ and $\rho$ at least to the third power. 
Compared to their oscillations along the relative time direction, only a weak dependence on the centre time coordinate is found.
The form of the correlators in the temporal plane shown in \TGasFig{NR:Xs3d} is to be compared with that of the correlators for an ideal gas which show an undamped oscillatory dependence in the $s_0$ direction, with the frequency given by the free dispersion $p_0={\TGasF p}^2/2m$.

As discussed in \TGasSect{FlucDiss}, the two-point functions $\tilde F$ and $\tilde\rho$ near equilibrium are expected to be related by the fluctuation dissipation relation \TGaseq{FlucDissFrho}
if the system is close to equilibrium \cite{TGas:KadanoffBaym1962a,TGas:Aarts:1997kp}.
As can be seen in  Figure \TGasfig{NR:fluctdisprel} the function $\tilde\rho_{21}(X_0,p)$ approaches $\tilde F_{11}(X_0,p)/(n(X_0, \TGasF p)+\frac 1 2)$ during the time evolution described by the 2PI dynamic equations \TGaseq{EOMFrho}, where $n(X_0, \TGasF p)= F_{11}(X_0, s_0=0, \TGasF p) -\frac 1 2$.
%
\begin{figure}[tb]
\begin{minipage}[b]{0.59\columnwidth}
\begin{center}
\resizebox{0.95\columnwidth}{!}{
\includegraphics{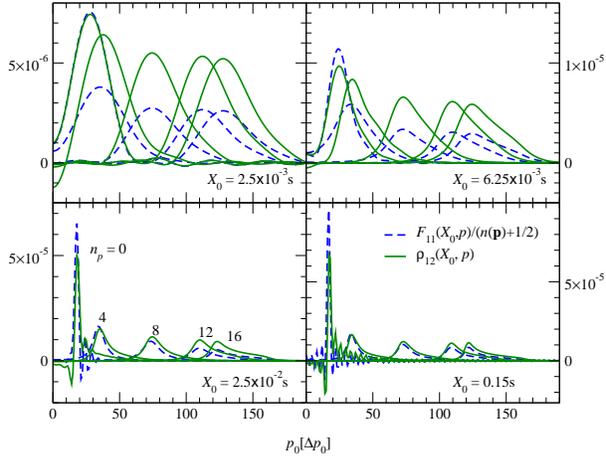}
}
\end{center}
\end{minipage}
\begin{minipage}[b]{0.39\columnwidth}
\caption[]{(color online) Spectral function $\tilde\rho$ (green solid lines), and statistical function $\tilde F$, normalised to the occupation number of the momentum mode $\TGasF p$ (blue dashed lines), as functions of $p_0$, for different times $X_0$ and different momentum modes $n_p$.
The difference between the respective spectral and normalised statistical functions decreases, indicating the emergence of a fluctuation-dissipation relation.
The good correspondence for the zero mode $n_p=0$ at the initial time is related to the choice of the initial condition.
}
\label{TGas:fig:NR:fluctdisprel}
\end{minipage}
\end{figure}

\subsubsection{2PI dynamical versus transport equations}
\label{TGas:sec:NR:em_vs_te}
We compare the time evolution as derived from the 2PI dynamic equations to their kinetic approximations following the procedure summarised in the introduction to \TGasSect{DynvsTrans}.
The absolute values of the left-hand side (LHS), the right-hand side in leading order (LO) and the right-hand side in next-to-leading order (NLO) of Eqns. \TGaseq{TE:motionFWigner_LO}, \TGaseq{TE:motionFWigner_NLO} are drawn in Figure \TGasfig{NR}.

As already seen in \TGasFig{n1oft} we observe three generic time regimes \cite{TGas:Berges:2001fi}:
Strong oscillations characteristic for early times, slow drifting at intermediate times, and a late-time approach to equilibrium characterised by vanishing time derivatives.

Close to the initial time, the strong oscillations of $F(X_0, s_0,p)$ and $\rho(X_0, s_0, p)$ in the centre time $X_0$ are due to the finite integration limits $s_0=\pm2X_0$ in the $s_0$ direction.
The intermediate drifting regime is reached (for $\gamma=1.5\cdot10^{-3}$ at $X_0\simeq0.003\,$s) when the contributions of the integration limits can be neglected.
Non-vanishing derivatives of $\tilde F$ and $\tilde\rho$ with respect to $X_0$ now solely result from the evolution of the correlations $F(X_0, s_0, \TGasF p)$ and $\rho(X_0, s_0, \TGasF p)$ in $X_0$.
As can be seen in Figure \TGasfig{NR:Xs3d}, the change with $X_0$ is slow compared to that with $s_0$.
Finally, equilibrium is approached when all time derivatives vanish.

We consider in more detail the cases corresponding to (A) weak, (B) moderate, and (C) strong effective interactions between the atoms:

\textsl{Case A}
corresponds to weak interactions, $\gamma=1.5\cdot 10^{-3}$.
The evolution is expected to be well described by Boltzmann-type equations after the initial oscillations have damped out.
Figure \TGasfig{NR}A (top row) shows the left-hand side (LHS, solid line) as well as the right-hand side of the transport equations in leading order (LO, dashed line) and next-to-leading order (NLO, dashed-dotted line) of the gradient expansion for two different momentum modes $n_p=4$ (left panel) and $n_p=8$ (right panel).
In the respective left columns, the frequency $p_0$ is that of the peak of the spectral function, cf. \TGasFig{NR:fluctdisprel} (`on-shell'), in the right columns, $p_0$ has been chosen two half-widths away from it (`off-shell').
One finds that the 2PI dynamic and the kinetic equations in general give the same results once the occupation numbers remain constant.
\begin{figure}[tb]
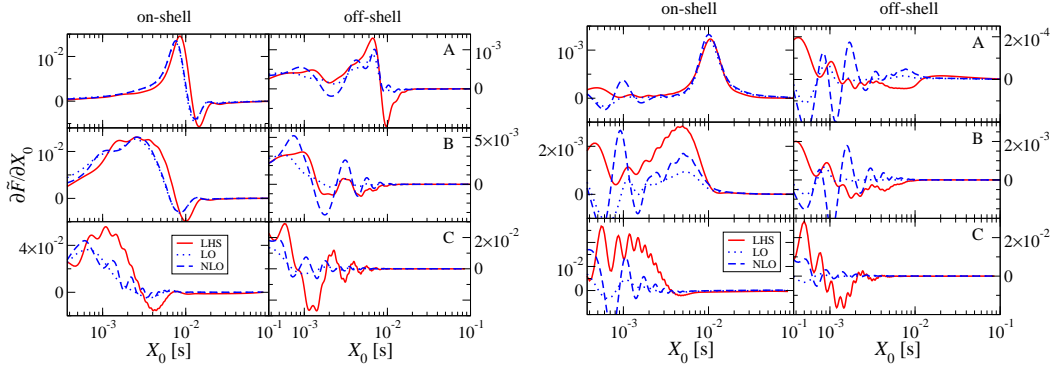

\begin{center}
\resizebox{0.48\columnwidth}{!}{
\includegraphics{TGas_fig20a.eps}
}
\resizebox{0.48\columnwidth}{!}{
\includegraphics{TGas_fig20b.eps}
}
\end{center}
\caption[]{(Color online)
Time-derivative of the statistical two-point function as a function of centre time $X_0$ for two different momentum modes, $n_p=4$ (left panel) and $n_p=8$ (right panel):
Comparison of results from 2PI dynamic equations and those from their kinetic approximation.
The (red) solid lines correspond to the left-hand side (LHS), the (blue) dashed(-dotted) lines to the the right-hand side in leading order (LO) (next-to-leading order, NLO) of the gradient expansion, Eqns. \TGaseq{TE:motionFWigner_LO} and \TGaseq{TE:motionFWigner_NLO}, respectively.
In the left column of each panel, the frequency $p_0$ is that of the peak of the spectral function, cf. \TGasFig{NR:fluctdisprel} (`on-shell'), in the respective right columns, $p_0$ has been chosen two half-widths away from it (`off-shell').
From top to bottom, the line density $n_1$ is rescaled as well as the interaction parameter $\gamma$, so that $m g n_1=\gamma n_1^2$ is kept fixed:
Top row: case A $n_1=10^7 \TGasunit m^{-1}$, $\gamma=1.5\cdot 10^{-3}$;
middle row: case B $n_1=10^6 \TGasunit m^{-1}$, $\gamma=0.15$;
bottom row: case C $n_1=10^5 \TGasunit m^{-1}$, $\gamma=15$.
}
\label{TGas:fig:NR}
\end{figure}

The NLO contributions depend on the $p_0$ derivatives of the correlation functions. 
The correlation functions in Wigner representation are non-zero for a continuum of frequencies $p_0$, see \TGasFig{NR:fluctdisprel}.
This results in non-vanishing NLO contributions.
However, as shown by the red dashed line, no significant contributions to the transport equations are found after the decay of the initial oscillations.

\textsl{Case B.}
We increase the interaction parameter to $\gamma=0.15$ while the initial line density is decreased to $n_1=10^6$.
In this way $g\propto\gamma n_1$ increases by a factor of $10$ such that quantum statistical correlations grow in importance, see \TGasSect{QvsCl}.
\TGasFig{NR} shows that equilibrium is reached faster, in particular for the higher momentum modes, while the intermediate drifting regime observed in case A is reduced.
For the lower momentum modes we get similar results as in case A while for the higher momentum modes there is an essential difference to the preceding case: correspondence between 2PI and transport equations is not reached until equilibration occurs.

\textsl{Case C.}
We finally choose strong coupling, $\gamma=15$, and decrease the line density to $n_1=10^5\TGasunit m^{-1}$.
Qualitatively similar results as in the preceding cases and a faster approach to an equilibrium configuration are found, see third line of \TGasFig{NR}.
\\

\begin{figure}[tb]
\begin{minipage}[b]{0.55\columnwidth}
\begin{center}
\resizebox{1.0\columnwidth}{!}{
\includegraphics{TGas_fig21.eps}
}
\end{center}
\end{minipage}
\hspace*{0.04\columnwidth}
\begin{minipage}[b]{0.40\columnwidth}
\caption[]{(Color online) 
Time dependent decay constant $\Gamma$ as defined in \TGasEq{TE:decayconst}, where the time derivative $\mathrm{d} \tilde F/\mathrm{d} X_0$ is replaced by the left-hand side (LHS, solid line), the right-hand side in leading order (LO, dotted line) and in next-to-leading order (NLO, dashed line) of the gradient expansion, Eqs. \TGaseq{TE:motionFWigner_LO} and (43) in \cite{TGas:Branschadel2008a}, for the on-shell and the off-shell case as a function of the centre time.
From top to bottom, the line density $n_1$ is rescaled as well as the interaction parameter $\gamma$ with $\gamma n_1^2=\text{const.}$, compare \TGasFig{NR}.
The three rows correspond to cases A--C as described in the text.
Colors indicate the momentum mode: $n_p=4$ (red), $n_p=8$ (blue), $n_p=12$ (green).
Non-constant (and negative) values for $\Gamma$ result from the fact that the time regime of exponential approach to equilibrium is not yet reached.
}
\label{TGas:fig:decayconst}
\end{minipage}
\end{figure}
To study in more detail the late-time behaviour, we assume that, at late times, the statistical correlation function decays exponentially to its equilibrium value with a decay constant $\Gamma(p)$,
\begin{equation}
  \tilde F(X_0, p)\approx \tilde F(X_0=\infty, p) + \Delta\tilde F(p)e^{-\Gamma(p) X_0},
\end{equation}
where $\Delta\tilde F(p)$ is some constant independent of $X_0$.
In \TGasFig{decayconst} we plot
\begin{equation}
  \Gamma(X_0, p)
  = - \frac{\partial^2 \tilde F(X_0, p)/\partial X_0^2}
      {\partial \tilde F(X_0, p)/\partial X_0},
\label{TGas:eq:TE:decayconst}
\end{equation}
where $\partial \tilde F/\partial X_0$ is given by the LHS, as well as by the LO and NLO  expressions on the RHS of \TGasEq{TE:motionFWigner_LO}.
We focus the range of times $X_0$ to those where $\Gamma(X_0, p)$ is settling to a constant, indicating the emergence of an evolution according to kinetic theory.
The top row of graphs in \TGasFig{decayconst} shows that the different mode evolutions are settling to an exponential decay at times between $0.07$ and $0.09$ seconds, deep in the drifting regime.
Hence, although a fluctuation dissipation relation is established almost an order of magnitude in time earlier, the kinetic approximation becomes strictly valid only at very late times.
\TGasFig{decayconst} also shows that it is in general not sufficient to take into account only the LO approximation in the gradient expansion.

Let us furthermore study the dependence of the decay constants $\Gamma$ on the line density $n_1$.
\TGasFig{Gammaofn1} shows $\Gamma$ for five different densities $n_1$.
We find an approximately linear dependence of $\Gamma$ on $n_1$ which indicates that the source of damping is rather an off-shell two-body than a three-body scattering effect.

\subsection{Summary}
\label{TGas:sec:KinSummary}

In this section we have studied further the dynamics of our equilibrating one-dimensional Bose gas, thereby focusing on the comparison between a fully dynamical approach on the basis of the 2PI effective action and the corresponding kinetic approximation in the form of Boltzmann-type transport equations.
Transport equations are derived from the 2PI dynamic equations for the two-point correlation functions by means of a gradient expansion with respect to centre coordinates and a subsequent Fourier transformation with respect to relative coordinates, i.e., a Wigner transformation.
Furthermore, the details of the initial state are neglected by sending the initial time to minus infinity.
We found that the time evolution of the correlation functions with respect to centre time occurs relatively slowly as compared to the oscillations with respect to relative time.

A comparison of the statistical and spectral correlation functions showed that they begin to be locked to each other as predicted by the equilibrium fluctuation-dissipation relation.
This occurs approximately at the same time as when the transport equations set in to be a good approximation to the dynamical evolution according to the full dynamic equations.
For weak couplings one observes good correspondence of dynamic and transport equations after an initial period of oscillations.
However, off-shell effects are not covered by transport equations before equilibration occurs.
Increasing the dimensionless interaction strength parameter $\gamma$ leads to significant differences between the 2PI dynamic and the transport equations.
Hence, the kinetic description of the time evolution in terms of an exponential decay with decay rate $\Gamma$ generically sets in to be valid only at very large times, when no essential change in the momentum profile of the system occurs any more.
The late-time evolution is predominantly due to two-body off-shell scattering effects.

\begin{figure}[tb]
\begin{minipage}[b]{0.57\columnwidth}
\begin{center}
\resizebox{1.0\columnwidth}{!}{
\includegraphics{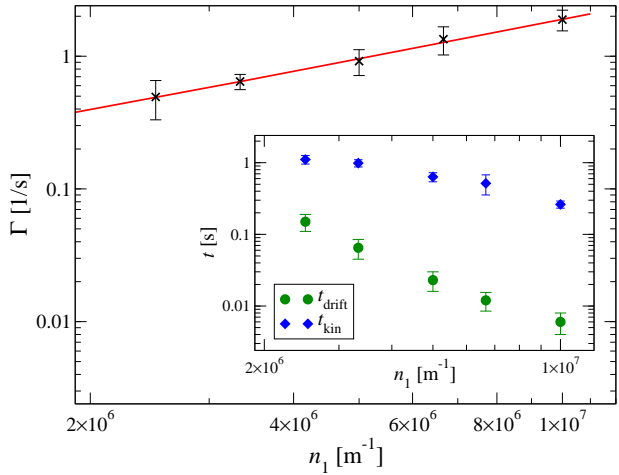}
}
\end{center}
\end{minipage}
\hspace*{0.04\columnwidth}
\begin{minipage}[b]{0.38\columnwidth}
\caption[]{(Color online) The decay constant $\Gamma$ as defined in \TGasEq{TE:decayconst} as a function of the particle density $n_1$, extracted from the solution of the 2PI dynamic equations at times $t_\mathrm{kin}$ as indicated with (blue) diamonds in the inset figure,
after which the evolution can be described to a good approximation by an exponential decay.
In the inset, the corresponding times $t_\mathrm{drift}$ at which the slow drift of the occupation number sets in are indicated with (green) circles.
The error bars indicate the variation of $\Gamma$, $t_\mathrm{kin}$, and $t_\mathrm{drift}$ over the different momentum modes.
The red line is a linear fit.
}
\label{TGas:fig:Gammaofn1}
\end{minipage}
\end{figure}

\section{Quantum vs.~classical statistical dynamics}
\label{TGas:sec:QvsCl}
Many experiments with Bose-Einstein condensates of dilute gases have shown that the Gross-Pitaevskii theory represents, in practice, a very good approximation to describe static and dynamic features of these systems.
Despite the fact that the first-order coherence reflected by this equation has its origin in the quantum nature of the Bose condensation phenomenon, the GPE arises as the classical field-theory approximation of the underlying quantum many-body problem. 
It thus neglects all quantum statistical fluctuations contributing to the dynamics of the scalar field. 
However, it is the role of these quantum statistical fluctuations which is of central importance for our quantitative understanding of quantum many-body dynamics, so far the least explored in the dynamical world. 

Two cases should be distinguished in this context: 
If the real-time dynamics of a Bose gas is dominated by classical statistical fluctuations then it can be well approximated by a large number of numerical integrations of the classical field equation (GPE) and Monte Carlo sampling techniques.
Such simulations are done in many different areas in physics and to some extent form a research field in its own.
In the context of ultracold Bose gases, see e.g., Refs.~\cite{TGas:Davis2002a,TGas:Davis2002b,TGas:Kohl2002a}.
Comparisons between simulations of classical field dynamics and the nonperturbative 2PI approach described in \TGasSect{NEqFT} have been presented in Refs.~\cite{TGas:Aarts:2001yn,TGas:Berges:2008wm}.
Simulations take into account nonperturbative dynamics, however, they neglect all quantum corrections. 
For fermions, a corresponding classical statistical description does not exist. 
The other case concerns dynamics where quantum fluctuations are relevant. 
In the following we quantitatively determine the role of quantum fluctuations for a time-evolving Bose gas.

Many experiments concerning ultracold Bose gases fall short of being sensitive to quantum statistical fluctuations and can be accurately described by classical field theory.
This encompasses mean-field approximations which are applicable if classical fluctuations are small as well as the case of strong classical fluctuations which requires simulations of classical field equations.
The importance of classical statistical fluctuations can rise if the gas is sufficiently dense. 
A combination of low densities and strong self-interaction can lead to enhanced quantum fluctuations as compared to classical statistical fluctuations. 
As discussed in \TGasSect{BeyondMF}, zero-energy scattering resonances, particularly the so called magnetic Feshbach resonances, as well as lower-dimensional gases so far have played a leading role in the creation of strong interactions in degenerate atomic quantum gases. 
Present-day experimental techniques allow for resonance-enhanced scattering lengths larger than the mean interatomic distance $n^{-1/3}$, such that the system is no longer in the collisionless regime. 

In the following we discuss the difference between the quantum and classical statistical theory which can be expressed in terms of interaction vertex terms for the quantum theory which are absent in the classical statistical theory. 
This implies that the classical generating functional is characterised by a reparametrisation property, which allows one to scale out the dependence of the dynamics on the scattering length $a$. 
As a consequence, for the classical dynamics the effects of a larger self-interaction can always be compensated by a smaller density. 
It is shown that quantum corrections violate this invariance property. 
They become of increasing importance with growing scattering length or reduced density. 
On this basis one derives a condition which experimenters may use to find signatures of quantum fluctuations when preparing and probing the dynamics of ultracold gases. 
To illustrate the differences we compare quantum and classical evolution for the example of a one-dimensional Bose gas equilibrating in one spatial dimension as it has already been the subject of Sects.~\TGassect{1DBoseEq} and \TGassect{DynvsTrans}.

\subsection{Functional-integral approach to classical statistical dynamics}
\label{TGas:sec:ClassStatFI}
The use of functional methods to describe the dynamics of classical correlations dates back to the work of Hopf in the context of statistical hydrodynamics \cite{TGas:Hopf1952a}.
A field theory for the description of classical fluctuations in terms of noncommutative classical fields was first suggested by Martin, Siggia, and Rose (MSR) \cite{TGas:Martin1973a} and has been extensively used in critical dynamics near equilibrium \cite{TGas:Hohenberg1977a}.
This theory has been reformulated later in terms of Lagrangian field theory employing functional methods \cite{TGas:Phythian1975a,TGas:DeDominicis1976a,TGas:Janssen1976a,TGas:Bausch1976a,TGas:Phythian1977a,TGas:DeDominicis1978a}.
In these field theoretical approaches to classical statistics, a doubling of the degrees of freedom occurs. 
For example, in the generating functional for Green functions, besides each field appearing in the fundamental Lagrangian, a second `response' field is integrated over.
The functional integral approach to quantum field dynamics developed by Schwinger and Keldysh employs the closed time path (CTP) contour \cite{TGas:Schwinger1961a,TGas:Keldysh1964a} in the time-ordered exponential integral as introduced in \TGasSect{CTP}. 
The doubling of fields in the MSR and Lagrangian approaches to classical dynamics corresponds to the fields evaluated separately on the two branches of the Schwinger-Keldysh CTP~\cite{TGas:Chou1980a,TGas:Chou1985a,TGas:Blagoev2001a}.
Implications of the differences between the classical and quantum vertices, similar to the case considered in this article, have been discussed, for other theories, e.g.\ in Refs.~\cite{TGas:Aarts1997a,TGas:Buchmuller1997a,TGas:Wetterich:1997rp,TGas:Aarts:1997kp,TGas:Cooper2001a,TGas:Blagoev2001a,TGas:Jeon2005a}.

\subsection{A quantum-mechanical example}
\label{TGas:sec:QMEx}
Let us begin with a simple quantum mechanical example, that of a quantum harmonic oscillator.
As is well known and was mentioned in our introduction to mean-field theory, the coherent states \TGaseq{CohState} formed from the energy eigenstates of the oscillator describe classical motion as it is, e.g., seen in the trajectory of the mean value of the position operator.
The wave packet shows a dispersionless motion in the potential.
As a reason for this special behaviour one can regard the absence of genuinely quantum effects like tunneling or quantum reflection.

In the context of field theory it is common to regard mean-field approximations, including the HFB theory as classical approximations in the statistical sense.
The reason for this is that the Hamiltonian is at most quadratic in the field operators and their derivatives.
Such a Hamiltonian can always be diagonalised and represents an essentially free system.

Before discussing these matters further let us shed a bit more light on the classicality of the quantum harmonic oscillator using the path-integral formulation of initial-value problems.
As discussed in detail in \TGasSect{CTP}, a path integral allowing to derive the time-evolution of operator expectation values from a given initial state involves a time-integration over a closed time path, leading from the initial time $t_0$ to the maximum time of interest and back to $t_0$.
Hence, the argument of the exponential in the measure, cf.~\TGasEq{NonEqGenFunc}, can be written as
\begin{equation}
\label{TGas:eq:SCTP}
  S_{\cal C}[\varphi]=S[\varphi^+]-S[\varphi^-],
\end{equation}
where $\varphi^\pm$ are the fluctuation fields evaluated on the outward ($+$) and backward ($-$) branches of the CTP, respectively.
Note that the relative minus sign stems from the reverted time integration on the `$-$' path.
The CTP provides a doubling of fluctuation coordinates $\varphi$ as compared to the path integral for the simple quantum mechanical transition matrix element \TGaseq{QMPI}.
In the case of the harmonic oscillator with frequency $\omega$ the difference \TGaseq{SCTP} can be explicitely written as
\begin{eqnarray}
  S_{\cal C}[\varphi]
  &=&\frac{1}{2}\int \mathrm{d}t\,\left[
      (\dot\varphi^+)^2-\omega^2(\varphi^+)^2
     -(\dot\varphi^-)^2+\omega^2(\varphi^-)^2\right]
  \nonumber\\
  &=& \frac{1}{2}\left\{\varphi_0^+\dot\varphi_0^+-\varphi_0^-\dot\varphi_0^-
   -\int \mathrm{d}t\,\left[
    \varphi^+\left(\partial_t^2+\omega^2\right)\varphi^+
   -\varphi^-\left(\partial_t^2+\omega^2\right)\varphi^-\right]\right\},
   \qquad
\label{TGas:eq:SCTPHO}
\end{eqnarray}
where $\dot\varphi^\pm=\partial_t\varphi^\pm$, and an integration by parts produced the boundary terms involving the coordinate $\varphi_0^\pm=\varphi^\pm(t_0)$ and velocity $\dot\varphi_0^\pm$ evaluated at $t=t_0$.
Expressing now the coordinates evaluated on the $+$ and $-$ branches of the CTP in terms of a ``centre'' coordinate $\varphi$ and a ``relative'' coordinate $\widetilde\varphi$, as $\varphi^\pm=\varphi\pm\widetilde\varphi/2$, we can rewrite $S_{\cal C}$ as 
\begin{equation}
\label{TGas:eq:STPHOitoClphi}
  S_{\cal C}[\varphi]=S[\varphi,\widetilde\varphi]
  = -\left\{\widetilde\varphi_0\partial_t\varphi_0+
     \int \mathrm{d}t\,
     \widetilde\varphi\left(\partial_t^2+\omega^2\right)\varphi\right\},
\end{equation}
where we have performed two further partial integrations in order to get rid of any terms involving derivatives of $\widetilde\varphi$ in the integrand.
In this way, also the boundary terms combine to the single term left in \TGasEq{STPHOitoClphi}.
Inserting \TGaseq{STPHOitoClphi} into the path integral \TGaseq{NonEqGenFunc} for initial-value problems one obtains
\begin{equation}
\label{TGas:eq:ClPIHO}
  \int{\cal D}\varphi^+{\cal D}\varphi^-\,
  \langle\varphi_0^+|\rho_0|\varphi_0^-\rangle\,
  e^{i(S[\varphi^+]-S[\varphi^-])}
  =
  \int{\cal D}\varphi{\cal D}
  \widetilde\varphi\,\rho[\varphi_0,\widetilde\varphi_0]\,
  e^{iS[\varphi,\widetilde\varphi]},
\end{equation}
with $\rho[\varphi_0,\widetilde\varphi_0]=\langle\varphi_0+\widetilde\varphi_0/2|\rho_0|\varphi_0-\widetilde\varphi_0/2\rangle$, since the Jacobian of the transformation to $\varphi$ and $\widetilde\varphi$ is equal to one.
This path integral is of the form discussed in the context of classical statistical physics and contains, besides the integration over the physical coordinate $\varphi$, an additional integration over an auxiliary coordinate $\widetilde\varphi$, see, e.g., Refs.~\cite{TGas:Phythian1975a,TGas:DeDominicis1976a,TGas:Janssen1976a,TGas:Bausch1976a,TGas:Phythian1977a,TGas:DeDominicis1978a}.
The final step is done by observing that the integral over $\widetilde\varphi$ can be analytically performed, which gives, as integrand of the remaining path integral, a functional delta distribution, 
\begin{equation}
\label{TGas:eq:ClPIHOdelta}
  \int{\cal D}\varphi{\cal D}\widetilde\varphi\,\rho[\varphi_0,\widetilde\varphi_0]\,
  e^{iS[\varphi,\widetilde\varphi]}
  =
  \int \mathrm{d}\varphi_0\mathrm{d}\pi_0\,W[\varphi_0,\pi_0]\,\int{\cal D}''\varphi\,
  \delta[\left(\partial_t^2+\omega^2\right)\varphi].
\end{equation}
Here, $\pi=\partial_t\varphi$ denotes the momentum canonically conjugate to the harmonic-oscillator coordinate $\varphi$, and the initial phase space distribution $W[\varphi_0,\pi_0]$ is defined as the Fourier transform of the density matrix in $\varphi_0$-$\widetilde\varphi_0$ representation,
\begin{equation}
\label{TGas:eq:Witorho}
  W[\varphi_0,\pi_0]
  = \int \mathrm{d}\widetilde\varphi_0\,\rho[\varphi_0,\widetilde\varphi_0]
  \exp\{-i\widetilde\varphi_0\pi_0\}.
\end{equation}
Note that the phase factor in \TGasEq{Witorho} originates from the boundary terms obtained in the partial integrations in Eqs.~\TGaseq{SCTPHO} and \TGaseq{STPHOitoClphi}.
The double-primed measure ${\cal D}''\varphi$ excludes the integrations over the initial-time coordinate $\varphi_0$ as well as over the next-to-initial-time coordinate which is rewritten as an integration over the initial velocity $\pi_0=\partial_t\varphi_0$. 

The integral obtained in \TGasEq{ClPIHOdelta} performs a classical average over a distribution which is given by the initial phase space distribution $W[\varphi_0,\pi_0]$ propagated to any later time of interest by means of the classical equation of motion.
For example, assume the path integral to be used for the evaluation of the mean value of $\Phi(t)$ at some time $t>t_0$,
\begin{eqnarray}
  \langle\Phi(t)\rangle
  &=& \int \mathrm{d}\varphi_0\mathrm{d}\pi_0\,W[\varphi_0,\pi_0]\,\int{\cal D}''\varphi\,\,\varphi(t)\,
  \delta[\left(\partial_t^2+\omega^2\right)\varphi]
  \nonumber\\
  &=& \int \mathrm{d}\varphi_0\mathrm{d}\pi_0\,W[\varphi_0,\pi_0]\,\int{\cal D}''\varphi\,\,\varphi(t)\,
  J[\varphi]\delta[\varphi-\varphi_\mathrm{cl}],
\label{TGas:eq:meanphi}
\end{eqnarray}
where $\varphi_\mathrm{cl}$ is the solution of the classical equation of motion $(\partial_t^2+\omega^2)\varphi(t)=0$ given initial values $\varphi_0$ and $\pi_0$ for coordinate and velocity, respectively.
The Jacobean $J[\varphi]=|\mathrm{det}(\delta^2 S[\varphi,\widetilde\varphi]/\delta\varphi\delta\widetilde\varphi)|$ is an irrelevant normalisation constant which has been discussed in detail in Ref.~\cite{TGas:Jeon2005a} and references therein.
The integral can be performed numerically if $W[\varphi_0,\pi_0]$ is a positive definite, normalised probability distribution in phase space, in the case discussed here for the initial position $\varphi_0$ and momentum $\pi_0$ of the classical oscillator.
Hereby, the functional delta distribution implies, that the initial coordinates sampling $W$ are evolved according to the classical oscillator equation from $t_0$ to time $t$.
The resulting distribution is used to perform the average over $\varphi$.

This implies that any correlation function describing the evolution of the quantum mechanical harmonic oscillator can be evaluated using classical statistical sampling given that the initial phase-space distribution is positive definite.
Nevertheless, the oscillator is of course inherently quantum since its ground state does not correspond to the classical ground state with $\varphi=\partial_t\varphi=0$.
A real initial state contains in any case these quantum fluctuation effects, though they may be negligible compared to the high-energy classical occupation.
Despite this, however, no further quantum fluctuations occur during the time evolution of the oscillator.

The reason for this classicality is, clearly, that the Hamiltonian is quadratic in $\varphi$.
Consider, e.g., the anharmonic oscillator, with classical action
\begin{equation}
\label{TGas:eq:SAHO}
  S[\varphi]
  =\int \mathrm{d}t\,\left[\frac{1}{2}\left(\dot\varphi^2-\omega^2\varphi^2\right)
   -\lambda\varphi^4\right].
\end{equation}
The quartic term no longer allows to rewrite the action on the close time path into a functional linear in $\widetilde\varphi$.
On rather obtains
\begin{equation}
\label{TGas:eq:SintAHOCTP}
  \int \mathrm{d}t\,\left({\varphi^+}^4-{\varphi^-}^4\right)
  = \frac{1}{2}\int \mathrm{d}t\,
    \left(\widetilde\varphi\varphi^3+\frac{1}{4}\widetilde\varphi^3\varphi\right).
\end{equation}
Neglecting the second term which is cubic in $\widetilde\varphi$ one can still integrate out $\widetilde\varphi$ and obtains the classical statistical integral
\begin{equation}
\label{TGas:eq:ClPIAHOdelta}
  \int \mathrm{d}\varphi_0\mathrm{d}\pi_0\,W[\varphi_0,\pi_0]\,\int{\cal D}''\varphi\,\,
  \delta[\left(\partial_t^2+\omega^2\right)\varphi+\lambda\varphi^3],
\end{equation}
again with the classical equation of motion in the argument of the functional delta distribution.
This procedure is, however, no longer possible if the second term in \TGasEq{SintAHOCTP} is kept, and the explicit path integration over $\widetilde\varphi$ must be kept during the evolution.
Hence, the quantum anharmonic oscillator indeed shows fluctuation effects beyond classical statistics.

We finally point out that the classical approximation neglecting the $\widetilde\varphi^3\varphi$ vertex is equivalent to the so-called Truncated Wigner Approximation known in quantum optics.
See Ref.~\cite{TGas:Polkovnikov2003a} for details as well an approach to include corrections caused by quantum fluctuations analogous to what is described above.
Closely related results were recently presented in 
Ref.~\cite{TGas:Plimak2008a}.

\subsection{Field theory}
\label{TGas:sec:ClvsQFT}
We will now summarise the difference between classical and quantum statistical evolution illustrated in the previous section for the case of field theory.
Specifically, we shall use, in this context, the 2PI functional approach introduced in \TGasSect{NEqFT} and study the dynamics of the one-dimensional Bose gas. 

Consider the scalar theory of a single-species interacting Bose gas as defined by the classical action \TGaseq{Sclassphi4}.
Introducing a linear transformation of the fields as
\begin{equation}
  \left(\begin{array}{c}
  \varphi_a \\
  {\tilde\varphi}_a
  \end{array}\right)
  \equiv R\,\left(\begin{array}{c}
  \varphi_a^+ \\
  \varphi_a^-
  \end{array}\right),
  ~~~\mathrm{with}~~
  R
  = \left(\begin{array}{rr} \frac{1}{2} & \frac{1}{2} \\
                             1           & -1
                             \end{array}\right), \qquad
\label{TGas:eq:RTransf}
\end{equation}
this action assumes the form
$S[\varphi,{\tilde\varphi}] = S_0[\varphi,{\tilde\varphi}] +
S_{\rm int}[\varphi,{\tilde\varphi}]$, with the free-field-theory action
\begin{equation}
  S_0[\varphi,{\tilde\varphi}]
  = \frac{1}{2}\int_{xy}\,
  \left( \varphi_a,{\tilde\varphi}_a \right)
  \left(\begin{array}{cc}
  0         & iG^{-1}_{0,ab}\\
  iG^{-1}_{0,ab}    & 0
  \end{array}\right)
  \left(\begin{array}{c} \varphi_b \\ {\tilde\varphi}_b \end{array}\right)
  \label{TGas:eq:freeS}
\end{equation}
and the interaction part
\begin{equation}
  S_{\rm int}[\varphi,{\tilde\varphi}]
  = -\frac{g}{2}\int_x {\tilde\varphi}_a(x)\varphi_a(x)\varphi_b(x)\varphi_b(x)
    -\frac{g}{8}\int_x {\tilde\varphi}_a(x){\tilde\varphi}_a(x)
                         {\tilde\varphi}_b(x)\varphi_b(x),
\label{TGas:eq:SqPhiXi}
\end{equation}
in full analogy to the anharmonic oscillator discussed above.

\begin{figure}[tb]
\begin{center}
\resizebox{0.5\columnwidth}{!}{
\includegraphics{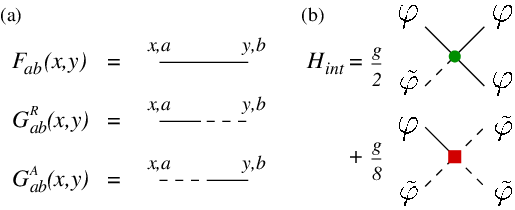}
}
\end{center}
\caption{(color online) (a) Diagrammatic
representation of the correlators in the $\varphi$-$\tilde\varphi$
basis. A full line indicates the $1$- or $\varphi$-component, a
broken line the $2$- or $\tilde\varphi$-component. $F_{ab}(x,y)$
is the statistical correlation function,
$G^R_{ab}(x,y)=\rho_{ab}(x,y)\theta(x_0-y_0)$ and
$G^A_{ab}(x,y)=-\rho_{ab}(x,y)\theta(y_0-x_0)$ the retarded and
advanced Green functions, respectively. Their representation in
terms of the real-valued spectral correlation function
$\rho_{ab}(x,y)$ exposes the $\theta$-functions which imply the
respective time ordering in $x_0$, $y_0$. (b) Diagrammatic
expansion of the quantum vertex contributing to the action as in \TGasEq{SqPhiXi}. 
The classical action is lacking the second contribution (red square).}
\label{TGas:fig:DiagramsFGRGAHI}
\end{figure}
It is beyond the scope of this article to present the derivation of quantum and classical versions of the 2PI effective action and the equations of motion obtained with this, which can be found in Ref.~\cite{TGas:Berges:2007ym}.
In there, also a more careful analysis of the boundary terms neglected above is given.
Here, we shall present a concise diagrammatic illustration of the differences emerging.
For this we note that the CTP propagator matrix is transformed, by $R$ from \TGasEq{RTransf}, to 
\begin{eqnarray}
\left(\begin{array}{cc}
    F_{ab} &
    -i G^\mathrm{R}_{ab} \\
    -i G^\mathrm{A}_{ab} &
    0 \end{array}\right)
    = R
   \left(\begin{array}{cc}
   G_{ab}^{++} & G_{ab}^{+-}\\
   G_{ab}^{-+} & G_{ab}^{--}
   \end{array}\right)
   R^{T},
   \label{TGas:eq:GR}
\end{eqnarray}
where $F_{ab}(x,y)$ is the statistical correlation function, \TGasEq{GitoFrho}, and $G^R_{ab}(x,y)=\rho_{ab}(x,y)\theta(x_0-y_0)$ and $G^A_{ab}(x,y)=-\rho_{ab}(x,y)\theta(y_0-x_0)$ are the retarded and advanced Green functions, respectively, which are directly related to the spectral function $\rho$.
$G^{st}_{ab}(x,y)$, with $s,t\in\{+,-\}$, denote the full propagator, with $x_0$ and $y_0$ evaluated on branch $s$ and $t$ of the CTP, respectively. 
Eqs.~\TGaseq{SqPhiXi} and \TGaseq{GR} suggest that the propagators and bare vertices forming the diagrams which occur in the effective action when written in the $R$-transformed basis can be diagrammatically represented as shown in \TGasFig{DiagramsFGRGAHI}.
See Ref.~\cite{TGas:Berges:2007ym} for details of the derivation.

We consider the possible one-loop as well as higher-loop, closed 2PI diagrams which can be formed from these constituents.
Obviously, solid lines only connect to solid lines, and dashed lines to dashed ones.
Since both vertices have an odd number of solid and dashed legs, no diagram up to the two-loop double-bubble diagram, i.e., up to first order in the coupling can contain a quantum vertex, depicted as a (red) diamond in \TGasFig{DiagramsFGRGAHI}.
Hence, up to the HFB mean-field approximation, the dynamic equations are of classical statistical nature, i.e., apart from quantum effects in the initial state, the ensuing dynamics can be simulated by Monte-Carlo methods.

\subsubsection{Rescaled fields}
\label{TGas:sec:RescaledFields}
The classical statistical generating functional, in which $S_\mathrm{int}$ is lacking the second term in \TGasEq{SqPhiXi}, exhibits an important reparametrisation property: 
If the fluctuating fields are rescaled according to
\begin{equation}
  \varphi_a(x) \to \varphi_a^\prime(x) 
  =  \sqrt{g}\,\varphi_a(x),
  ~~~
  \tilde{\varphi}_a(x) \to \tilde{\varphi}^\prime_a(x)
  = (1/{\sqrt{g}})\,{\tilde{\varphi}_a(x)} 
\label{TGas:eq:rescaling}
\end{equation}
then the coupling $g$ drops out of $S^\mathrm{cl}[\varphi,{\tilde \varphi}]=S_0[\varphi,{\tilde \varphi}]+S_{\rm int}^\mathrm{cl}[\varphi,{\tilde \varphi}]$ where $S^\mathrm{cl}$ denotes the term in \TGasEq{SqPhiXi} linear in $\widetilde\varphi$. 
The free part $S_0[\varphi,{\tilde \varphi}]$ remains unchanged and the interaction part becomes
\begin{equation}
  S_{\rm int}^\mathrm{cl}[\varphi^\prime,{\tilde \varphi}^\prime]
  = -\frac{1}{2}\int_x {\tilde\varphi}^\prime_a(x)
    \varphi^\prime_a(x)\varphi^\prime_b(x)\varphi^\prime_b(x).
\label{TGas:eq:SqPhiXiclRescaled}
\end{equation}
Moreover, the functional measure is invariant under the rescaling \TGaseq{rescaling}, and the sources can be redefined accordingly. 
Therefore, the classical statistical generating functional becomes independent of $g$, except for the coupling dependence entering the probability distribution fixing the initial conditions. 
Accordingly, as is well known for the Gross-Pitaevskii equation, the coupling does not enter the classical dynamic equations for correlation functions. 
All the $g$-dependence enters the initial conditions which are required to solve the dynamic equations.

In contrast to the classical case, this reparametrisation property is absent for the quantum theory.
After the rescaling \TGaseq{rescaling} one is left with  $S[\varphi^\prime,\tilde{\varphi}^\prime]$ whose coupling dependence is given by the interaction part
\begin{equation}
  S_{\rm int}[\varphi^\prime,\tilde{\varphi}^\prime]
  = -\frac{1}{2}\int_x {\tilde\varphi}^\prime_a(x)\varphi^\prime_a(x)\varphi^\prime_b(x)\varphi^\prime_b(x)
    -\frac{g^2}{8}\int_x {\tilde\varphi}^\prime_a(x){\tilde\varphi}^\prime_a(x)
                         {\tilde\varphi}^\prime_b(x)\varphi^\prime_b(x),
                         \quad
\label{TGas:eq:SqPhiXiRescaled}
\end{equation}
according to \TGasEq{SqPhiXi}. 
Comparing to \TGaseq{SqPhiXiclRescaled} one observes that the 'quantum' part of the vertex encodes all the $g$-dependence of the dynamics.

The comparison of quantum versus classical dynamics becomes particularly transparent using the above rescaling. 
The rescaled macroscopic field and statistical correlation function are given by
\begin{equation}
  \phi_a^\prime(x) 
   = \sqrt{g} \phi_a(x), 
  \quad F_{ab}^\prime(x,y)
   = g F_{ab}(x,y), 
\label{TGas:eq:phiFRescaled}
\end{equation}
while the spectral function $\rho_{ab}(x,y)$ remains unchanged as \TGasEq{GR} suggests and is shown in Ref.~\cite{TGas:Berges:2007ym}.
Similarly, we define for the statistical self-energy entering the dynamic equations \TGaseq{EOMFrho}
\begin{equation}
  \Sigma^{F\prime}_{ab}(x,y) = g \Sigma^F_{ab}(x,y).
\label{TGas:eq:SigmaRescaled}
\end{equation}

\subsubsection{Quantum vs. classical statistical self-energy}
\label{TGas:sec:QvsClSigma}
To identify the precise difference between the quantum and the classical time evolution, details about the self-energies are required. 
In the following, we quote the result for the self-energies in NLO of the 2PI $1/{\cal N}$ expansion introduced in \TGasSect{NLO1N}.
The self energies \TGaseq{SigmaNLO1N}, expressed in the rescaled variables \TGaseq{phiFRescaled} and \TGaseq{SigmaRescaled} read:
\begin{eqnarray}
  \left(\begin{array}{r}
        \Sigma^F_{ab}(x,y) \\ -\frac{1}{2}\Sigma^\rho_{ab}(x,y)
	\end{array}\right)
   &=& -\left(\begin{array}{r}
        I_F^\prime(x,y) \\ -\frac{1}{2}I_\rho(x,y)
	\end{array}\right)
        \phi_a^\prime(x)\phi_b^\prime(y) 
  \nonumber\\
  &&\qquad 
  -\ \left(\begin{array}{rr}
        \Delta_F^\prime(x,y) & 
	\frac{g^2}{2}\Delta_\rho(x,y) \\
       -\frac{1}{2}\Delta_\rho(x,y) & 
        \Delta_F^\prime(x,y) \end{array}\right)
    \left(\begin{array}{r}
        F_{ab}^\prime(x,y) \\ -\frac{1}{2}\rho_{ab}(x,y)
	\end{array}\right),
\label{TGas:eq:SigmaNLO1Nprimed}
\end{eqnarray}
with $\Delta_{F}^\prime(x,y)=I_{F}^\prime(x,y)+P_{F}^\prime(x,y)$, etc.
The functions $I_{F}^\prime$ and $I_{\rho}$ satisfy
\begin{eqnarray}
  \left(\begin{array}{r}
        I_F(x,y) \\ I_\rho(x,y)
	\end{array}\right)
  &=&
  -\left(\begin{array}{c}
        \Pi_F^\prime(x,y) \\
	\Pi_\rho(x,y)
	\end{array}\right)
   +\ \int_0^{x_0}\mathrm{d}z\,I_\rho(x,z)
      \left(\begin{array}{c}
            \Pi_F^\prime(z,y) \\ 
            \Pi_\rho(z,y) 
	    \end{array}\right)
   - \int_0^{y_0}\mathrm{d}z\,
      \left(\begin{array}{r}
            I_F^\prime(x,z) \\ 
            I_\rho(x,z) 
	    \end{array}\right)\Pi_\rho(z,y)\Bigg],
   \nonumber\\
\label{TGas:eq:IFrhoprimed}
\end{eqnarray}
with
\begin{eqnarray}
 \Pi_F^\prime (x,y) 
 & = & - \frac{1}{2}\Big[ F^\prime_{ab}(x,y) F^\prime_{ab}(x,y)
 - \frac{g^2}{4}\rho_{ab}(x,y) \rho_{ab}(x,y) \Big],
 \nonumber\\[1.5ex]
 \Pi_{\rho} (x,y) 
 & = & - F^\prime_{ab} (x,y) \rho_{ab} (x,y).
 \label{TGas:eq:piself}
\end{eqnarray}
%

\begin{figure}[tb]
\begin{minipage}[b]{0.54\columnwidth}
\begin{center}
\resizebox{1.0\columnwidth}{!}{
\includegraphics{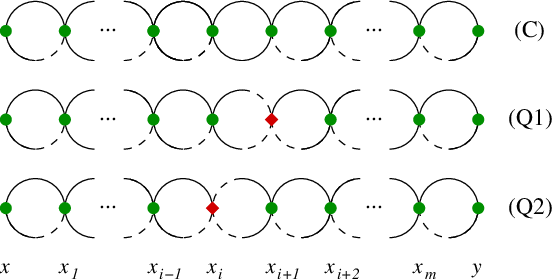}
}
\newline
\end{center}
\end{minipage}
\begin{minipage}[b]{0.45\columnwidth}
\caption{(color online) Bubble chains contributing to the functions $I_{F}^\prime$ and $I_\rho$ at NLO in the 2PI $1/\cal N$ expansion, see \TGasFig{DiagramsFGRGAHI} for definitions. 
The chains of type (Q1) and (Q2) only appear for a quantum system. 
Type (C) is present both in quantum and classical systems. 
These diagrams exhibit that in each term contributing to the functions $I_F^\prime$ and $I_\rho$, there is at most one loop involving two correlators $F'$ or $\rho$. 
In the classical limit, the loops
involving two $\rho$ are suppressed compared to
those with two $F'$.}
\label{TGas:fig:DiagramsIBubbleChains}
\end{minipage}
\end{figure}

Consider the bubble-chain diagrams occurring within the NLO $1/{\cal N}$ approximation as depicted in \TGasFig{DiagrExpGamma2NLO}.
Connecting the correlators through the respective vertices, in the $\varphi$-$\widetilde\varphi$ basis of \TGasFig{DiagramsFGRGAHI}, one finds which types of bubble chains appear. 
The classes of non-vanishing chains shown in \TGasFig{DiagramsIBubbleChains} confirm the structure of the functions $I_{F}^\prime(x,y)$ and $I_\rho(x,y)$ which are determined by the integral equation \TGaseq{IFrhoprimed}: 
In each term contributing to the diagrammatic expansion of these functions there is at most one loop containing two $F^\prime$ or two $\rho$ correlators, the latter resulting from either $G^\mathrm{R}$ or $G^\mathrm{A}$. 
In addition to this one finds that only the loop containing two $\rho$ correlators goes with the vertex which is present in the quantum case only. 
Hence, considering the classical dynamics, the $\sim g^2\rho^2$ terms are absent together with the `quantum' vertex. 
For the same reasons, the contribution $\sim (g^2/4) [I_\rho(x,y) + P_\rho(x,y) ]\rho_{ab}(x,y)$ to $\Sigma^F_{ab}(x,y)$, \TGasEq{SigmaNLO1N}, is absent for the classical statistical theory.
These arguments can be easily extended to the case that $\phi\not=0$ and the terms missing in the classical theory be determined \cite{TGas:Berges:2007ym}.

\subsection{Classicality criterion}
\label{TGas:sec:ClCrit}

Summarizing, one concludes that all equations \TGaseq{EOMphi}--\TGaseq{MM}, \TGaseq{PF}--\TGaseq{HFHrho}, and \TGaseq{SigmaNLO1Nprimed}--\TGaseq{piself} remain the same in the classical statistical limit except for differing expressions for the statistical components of the self-energy
\begin{eqnarray}
\Sigma^{F\prime}_{ab}(x,y) 
   & \stackrel{{\rm classical\,\,limit}}{\longrightarrow} 
   & - \Big\{
   I_F^\prime(x,y) \phi_a^\prime(x)\phi_b^\prime(y)
  + \left[ I_F^\prime(x,y)+P_F^\prime(x,y)\right] F_{ab}^\prime(x,y)\Big\},
  \nonumber\\
  \Pi_F^\prime(x,y) & 
      \stackrel{{\rm classical\,\,limit}}{\longrightarrow} & 
      - \frac{1}{2} F_{ab}^\prime(x,y)F_{ab}^\prime(x,y),
\label{TGas:eq:qmcl} \label{TGas:eq:SigmaNLO1NCl}
\end{eqnarray}
replacing the respective expression in Eqs.~\TGaseq{SigmaNLO1N}. 
The classical statistical self-energies can be obtained from the respective quantum ones by dropping two spectral ($\rho$-type) components compared to two statistical ($F$-type) functions. 
For vanishing $\phi$ where $P_{F,\rho}=0$ this corresponds to the result of Ref.~\cite{TGas:Aarts:2001yn}. 
After the rescaling \TGaseq{rescaling} the 'quantum' terms can be directly identified since they are the only $g$-dependent terms, which are absent in the classical statistical theory according to the above discussion. 
As a consequence, for the classical dynamics the effects of a larger coupling can always be compensated by changing the initial conditions such that $F g$, as well as $\phi \sqrt{g}$, remain constant. 
This cannot be achieved once quantum corrections are taken into account, since they become of increasing importance with growing coupling or reduced initial values for $F$ and $\phi$.

\TGasEq{SigmaNLO1NCl} describes the differences between quantum and classical statistical equations of motion. 
In turn one can ask under which conditions these differences are negligible.
In that case the quantum dynamics can be well approximated by classical statistical dynamics. 
Analysing the bubble one finds that a sufficient condition for the suppression of quantum fluctuations compared to classical statistical fluctuations is given by \cite{TGas:Berges:2007ym}
\begin{eqnarray}
  \left|F_{ab}^\prime(x,y)F_{cd}^\prime(z,w)\right|
  \gg\frac{3}{4} g^2 \left|\rho_{ab}(x,y)\rho_{cd}(z,w)\right| .
\label{TGas:eq:ClassCondNonEq}
\end{eqnarray}
This condition is not based on thermal equilibrium assumptions and holds also for far-from-equilibrium dynamics. 
In particular, it is independent of the value of the macroscopic field $\phi$. 
It can be applied, of course, also in thermal equilibrium, for which the statistical and spectral correlation functions are related by a fluctuation-dissipation relation, see \TGasEq{FlucDissFrho} in \TGasSect{FlucDiss}.
For large temperatures, $k_\mathrm{B} T\gg\omega-\mu$, one has $|F^{(\mathrm{eq})\prime}(\omega,\mathbf{p})|/g \gg |\rho^\mathrm{(eq)}(\omega,\mathbf{p})|$, i.e., the classicality condition is fulfilled for all modes whose occupation number $\sim F^{(\mathrm{eq})\prime}(\omega,\mathbf{p})/g$ is much larger than ${\cal O}(1)$. 

The equivalent statement also holds for nonequilibrium evolutions whenever it is possible to define a suitable 'occupation number' from a space-time or energy-momentum dependent proportionality between $F$ and $\rho$. 
Away from equilibrium the situation is often considerably more complicated. 
Strictly speaking the condition \TGaseq{ClassCondNonEq} must be valid at all times and for all space points, or momenta in Fourier space, for the classical and the quantum evolution to agree. 
In practice, however, it needs only be fulfilled for time and space averages. 
In \TGasSect{QvsClNumRes} we will demonstrate how quantum evolution can be approximated for not too late times by classical statistical dynamics, if the correlation functions satisfy \TGaseq{ClassCondNonEq} at initial time. 
In order to have quantum fluctuations playing a significant role, one needs, according to our above findings, to increase the interaction strength $g$ accordingly or change the phase-space structure by changing the external trapping potential. 
For example, in a one-dimensional trap, an effectively strong coupling and strong quantum fluctuations can be induced by reducing the line density of atoms while their interaction strength is kept constant.

\subsection{Quantum vs. classical evolution of an ultracold Bose gas}
\label{TGas:sec:QvsClNumRes}
In this section we apply the theoretical methods summarised above to the equilibrating 1D Bose gas studied already in Sects.~\TGassect{1DBoseEq}, \TGassect{NearEq1DBose}, and \TGassect{DynvsTrans}.
Along the lines discussed in the previous sections we compare the evolution involving only classical statistical fluctuations with that which also takes into account quantum corrections. 
Both, the classical and the quantum gas are assumed to be initially characterised by the same far-from-equilibrium initial conditions far. 
We solve Eqs.~\TGaseq{EOMFrho}, with the self-energies now given by Eqs.~\TGaseq{SigmaNLO1NCl}, for the initial conditions given in \TGasSection{1DIniState}, first with $\alpha=0$. 
For comparison we also consider the case $\alpha=1$ such that the classical and quantum initial correlation functions $F$ and $\rho$ are identical.

The time evolution of the initially Gaussian far-from-equilibrium momentum distribution of the classical gas is shown in \TGasFig{n1oftCl}. 
The mode occupations are shown as (black) solid lines, and for better comparison, we have quoted the quantum evolution from \TGasFig{n1oft} as (red) dashed curves.
\begin{figure}[tb]
\begin{minipage}[b]{0.59\columnwidth}
\begin{center}
\resizebox{1.0\columnwidth}{!}{
\includegraphics{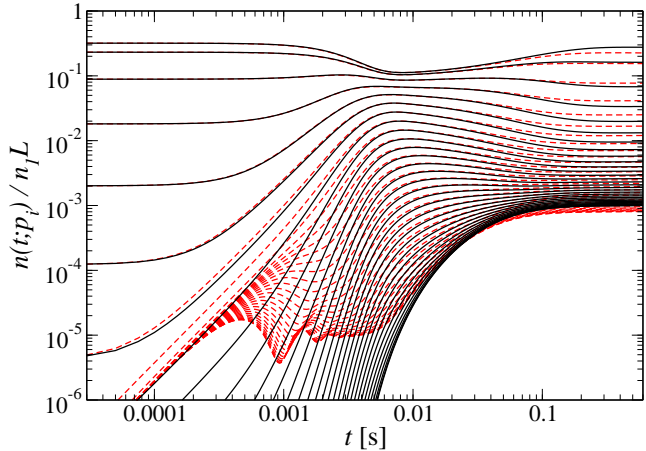}
}
\end{center}
\end{minipage}
\hspace*{0.03\columnwidth}
\begin{minipage}[b]{0.36\columnwidth}
\vspace*{-3ex} \caption{ (Color online) The normalised
momentum-mode occupation numbers $n(t;p)/n_1L$ for the classical
gas (black solid lines) compared to their quantum counterparts
from \TGasFig{n1oft} (red dashed lines), as functions
of time. Shown are the populations of the modes with
$p=p_i=2N_s/L\sin(i\pi/N_s)$, $i=0,1,...,N_s/2$, and one has
$n(t;-p)=n(t;p)$. In contrast to the quantum statistical evolution
there is no quick dephasing in the classical case, such that the
initially empty modes become only subsequently filled, as is
discussed in the main text. At large times, the classical and
quantum gases necessarily evolve to different distributions. }
\label{TGas:fig:n1oftCl}
\end{minipage}
\end{figure}
One observes that the time evolution of the modes with occupation number $n(t;p)>1$, i.e., $n(t;p)/n_1L>10^{-3}$, is, for most of the time, identical to that obtained in the quantum case, confirming condition \TGaseq{ClassCondNonEq}. 
As expected, for the strongly populated modes of a weakly interacting bosonic gas quantum fluctuations do not play a significant role for not too large times. 
Only when the evolution approaches the equilibrium state, the differences between quantum and classical statistics are expected to lead to a Bose-Einstein and classical distribution, respectively. 
We point out that, although we have chosen, for our comparison, the same initial occupation numbers in the two cases, the total energies are different since the initial correlation functions differ according to \TGasEq{IniValF}.
Hence, also the final-state occupation numbers of the low momentum modes can differ.

We finally point to the substantial differences in the short-time evolution of the weakly populated modes. 
While, in the quantum gas, the large-momentum mode populations are all growing at the same rate, the modes of the classical gas become populated much more gradually. 
The quantum-gas modes are occupied by $0.01$ and $1$ particle per mode already between $0.5$ and $1\,\mu$s, while the classical modes need up to ten times longer. 
The distinct quantum behaviour of the modes can be understood as follows. 
At the energies present in such an ultracold gas, the atomic interactions are essentially pointlike, i.e., the range of the potential is not resolved and the coupling function or scattering amplitude is constant over the range of relevant momenta. 
Hence, in a single scattering event, the distance of two atoms is localised to zero, such that the relative momentum of the atoms is completely unknown immediately after the collision. 
This means that the transfer probability of the atoms is the same for any final momentum mode, and this is observed as a quick, collective population in the respective, so far essentially unoccupied modes.

For comparison we repeated our calculations for a non-local interaction potential. 
In the momentum domain, this corresponds to a coupling function which is cut off at large momenta, and we chose the cutoff within the range of the momenta shown explicitly in \TGasFig{n1oftCl}. 
In this case we find that the quantum evolution is modified such that it becomes similar to that of the classical gas. 
In particular, all modes above the cutoff populate gradually one after each other.

\begin{figure}[tb]
\begin{minipage}[b]{0.59\columnwidth}
\begin{center}
\resizebox{1.0\columnwidth}{!}{
\includegraphics{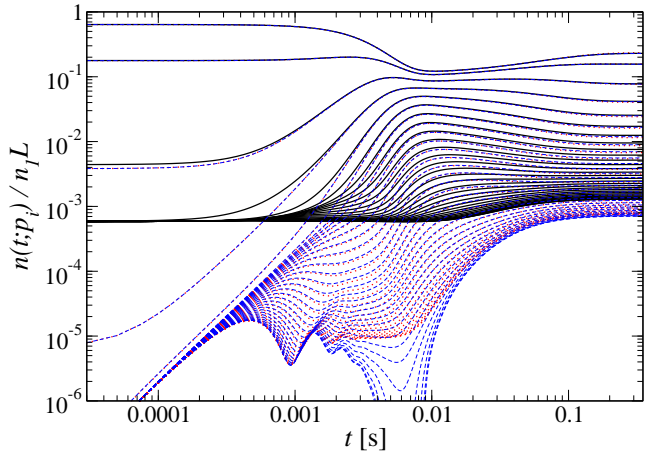}
}
\end{center}
\end{minipage}
\hspace*{0.03\columnwidth}
\begin{minipage}[b]{0.36\columnwidth}
\caption{
(Color online)
The normalised momentum-mode occupation numbers $n(t;p)/n_1L$ for the classical gas (black solid lines) compared to their quantum counterparts from \TGasFig{n1oft} (red dotted lines), as functions of time.
All parameters are chosen as in \TGasFig{n1oftCl}, except for $\sigma=6.5\cdot10^4\,$m$^{-1}$.
The dashed (blue) lines show $n(t,p)-1/2$, i.e., after subtracting the additional flat initial distribution which simulates the quantum ``zero point fluctuations''.
}
\label{TGas:fig:n1oftClqinic}
\end{minipage}
\end{figure}
The differences between the quantum and classical evolutions shown above depend, however, considerably on the choice of initial conditions. 
To compare the characteristics of the evolutions which are independent of the initial choice of $F$, we have repeated the classical calculations for an initial momentum distribution where, as compared to before, a constant occupation number $1/2$ has been added. 
Hence, we chose $\alpha=1$ in the initial values of $F$, \TGasEq{IniValF}, as in the quantum case, such that $F$ is identical for $x_0=y_0=t_0$ in the classical and quantum cases.
The results are shown in \TGasFig{n1oftClqinic}. 
The (red) dotted lines show, again, the quantum evolution, while the classical mode populations for the same initial conditions for $F$ and $\rho$ are shown as solid (black) lines. 
Subtracting $1/2$ from each $F_{aa}(t,t;p)$ gives the dashed (blue) lines. 
One finds, that during the initial period the evolution of the variation of the high-momentum modes with respect to their initial occupation is identical to the quantum evolution of the occupation numbers.
At intermediate times, however, there are deviations which lead to occupation numbers up to a percent lower than $1/2$. 
Although the chosen initial conditions which, in the classical case correspond to a base occupation of each mode with $1/2$ atom, seem unphysical, our results show that for the dilute, weakly interacting gas under consideration, there are differences only in those modes which, in the mean, are populated with less than one atom. 
Quantum statistical fluctuations play a role only for these modes.

\section{Summary}
\label{TGas:sec:Summary}
We have discussed functional field theoretical approaches to far-from-equilibrium quantum many-body dynamics.
The central topic were methods based on the real-time two-particle irreducible (2PI) effective action to derive coupled integro-differential equations of motion for correlation functions.
These equations were used to study, as a working example, the long-time evolution and equilibration of an ultracold Bose gas in one spatial dimension, starting in a state far away from equilibrium.
They allow to access phenomena which are beyond the reach of transport or kinetic theory, i.e., beyond that of (quantum) Boltzmann equations.
Non-perturbative approximation schemes were discussed which allow to gain access to the dynamics of strongly interacting systems and to long-time evolution and thermalisation.

By use of functional renormalisation-group (RG) techniques the 2PI dynamic equations were rederived in a particular truncation scheme of the RG flow equations, and an alternative non-perturbative approximation for use with strongly correlated systems was obtained.

The transition from initial to late-time nonequilibrium evolution and the emergence of transport theory were studied in detail and illustrated with the one-dimensional Bose gas example.
In the last part, we worked out the distinction between classical and quantum statistical fluctuations during the time evolution of a quantum system, first on the example of a quantum mechanical harmonic oscillator described by a Feynman path integral.
Extending the formulation to functional integrals allowed us to identify the effect of quantum fluctuations in terms of vertex contributions which are absent in the classical case.

Nonequilibrium quantum field theory has seen a vigorous development during the past decade, but many questions remain open leaving space for further progress.
Nonperturbative approximations to describe strongly correlated systems are at hand but need to be explored in more detail concerning their applicability to various specific configurations.
For such applications also numerical methods are desirable, as they are existing and used for classical statistical evolutions.
New field theoretical methods which provide the nonequilibrium time evolution, e.g., of equal-time correlation functions but do not require memory integrals over the history of the non-equal-time functions to be evaluated explicitly would help towards applications.

Experiments with ultracold gases have the potential to provide precise information about nonequilibrium dynamics of correlations.
The methods described in this article must be made accessible to these experiments, i.e., concrete applications need to be developed and optimal observables need to be found.
With these tasks in mind dynamical field theory is expected to continue to be and develop even stronger to an exciting research field.

\section*{Acknowledgments}
The author would like to thank R.~Alkofer, H.~Gies, B.-J.~Schaefer, H.~Latal, and L.~Mathe\-litsch for the organisation of, his invitation to, and their hospitality during the exciting 46.~Internationale Universit\"atswochen f\"ur Theoretische Physik at Schladming.
He furthermore thanks J.~Berges, A.~Bransch\"adel, H.~Buljan, K.~Burnett, S. Ke\ss ler, T.~K\"ohler, J.M.~Pawlowski, R.~Pezer, M.G.~Schmidt, M.~Seco, and K.~Temme for collaborations on part of the work presented here.
The author acknowledges support by the Deutsche Forschungsgemeinschaft and the Deutscher Akademischer Austausch Dienst.

\end{document}